\documentclass[final]{elsarticle}
\usepackage[T1]{fontenc}
\usepackage{courier}
\usepackage{amsmath,amssymb}
\usepackage{mathtools}
\usepackage[text={6.5in,9in},centering]{geometry}

\usepackage{textcomp}
\usepackage{wasysym}
\usepackage{tabularx}
\usepackage{multicol}
\usepackage{longtable,booktabs,colortbl}

\usepackage[vcentermath]{youngtab}
\usepackage{cancel}

\usepackage[pdftex,table,svgnames,hyperref]{xcolor}

\definecolor{darkred}{rgb}{0.5 0 0}
\definecolor{darkgreen}{rgb}{0.5 .5 0}
\definecolor{darkblue}{rgb}{0 0 .5}
\usepackage[pdftex,bookmarksopen]{hyperref}
 \hypersetup{%
    pdftitle={LieART - A Mathematica Application for Lie Algebras and Representation Theory}, %%
    pdfauthor={Robert Feger},%
    pdfsubject={},%
    pdfcreator={pdfTeX},%
    pdfproducer={Robert Feger},%
    pdfkeywords={},%
    colorlinks=true,%
  }

\usepackage{lieart}
\usepackage{mathematica}
\usepackage{lieartpaper}

\numberwithin{table}{section}

\journal{Computer Physics Communications}

\begin{document}

\begin{frontmatter}

\title{LieART -- A Mathematica Application for Lie Algebras and Representation Theory}

\author[WU]{Robert Feger\corref{author}}
\ead{robert.feger@physik.uni-wuerzburg.de}
\cortext[author] {Corresponding author}
\author[VU]{Thomas W. Kephart}
\ead{thomas.w.kephart@vanderbilt.edu}
\address[WU]{Universit\"at W\"urzburg, Institut f\"ur Theoretische Physik und Astrophysik, Emil-Hilb-Weg 22, 97074 W\"urzburg, Germany}
\address[VU]{Department of Physics and Astronomy, Vanderbilt University, Nashville, TN 37235}

\begin{abstract}
We present the Mathematica application ``LieART'' (\underline{Lie}
\underline{A}lgebras and \underline{R}epresentation \underline{T}heory) for
computations frequently encountered in Lie algebras and representation theory,
such as tensor product decomposition and subalgebra branching of irreducible
representations. LieART can handle all classical and exceptional Lie algebras.
It computes root systems of Lie algebras, weight systems and several other
properties of irreducible representations. LieART's user interface has been
created with a strong focus on usability and thus allows the input of
irreducible representations via their dimensional name, while the output is in the
textbook style used in most particle-physics publications. The unique Dynkin
labels of irreducible representations are used internally and can also be used
for input and output. LieART exploits the Weyl reflection group for most of the
calculations, resulting in fast computations and a low memory consumption.
Extensive tables of properties, tensor products and branching rules of
irreducible representations are included in the appendix.
\end{abstract}


\end{frontmatter}

\tableofcontents

\renewcommand{\descriptionlabel}[1]{\hspace{\labelsep}\emph{#1}}

\section*{Program Summary}

\begin{description}
\setlength{\itemsep}{0pt}
\item[Authors:]
	Robert Feger,  Thomas W. Kephart
\item[Program Title:] 
	LieART
\item[Licensing provisions:]
  GNU Lesser General Public License (LGPL)
\item[Programming language:]
	Mathematica
\item[Computer:]
	x86, x86\_64, PowerPC
\item[Operating system:]
	cross-platform
\item[RAM:] 
	$\geq 1\,\text{GB}$ recommended. Memory usage depends strongly on the Lie algebra's 
	rank and type, as well as the dimensionality of the representations in the 
	computation.
\item[Keywords:]
	Lie algebra; Lie group; representation theory; irreducible representation; 
	tensor product; branching rule; GUT; model building
\item[Classification:]
	4.2, 11.1
\item[External routines/libraries:]
	Wolfram Mathematica 8--10
\item[Nature of problem:]
	The use of Lie algebras and their representations is widespread in physics, 
	especially in particle physics. The description of nature in terms of gauge 
	theories requires the assignment of fields to representations of compact Lie 
	groups and their Lie algebas. Mass and interaction terms in the Lagrangian give 
	rise to the need for computing tensor products of representations of Lie 
	algebras. The mechanism of spontaneous symmetry breaking leads to the 
	application of subalgebra decomposition. This computer code was designed for the 
	purpose of Grand Unified Theory (GUT) Model building, where compact Lie groups 
	beyond the \U1, \SU2 and \SU3 of the Standard Model of particle physics are 
	needed. Tensor product decomposition and subalgebra decomposition have been 
	implemented for all classical Lie groups \SU{N}, \SO{N} and \Sp{2N} and the 
	exceptionals \E6, \E7, \E8, \F4 and	\G2.
\item[Solution method:]
	LieART generates the weight system of an irreducible representation (irrep) of a 
	Lie algebra by exploiting the Weyl reflection groups, which is inherent in all 
	simple Lie algebras. Tensor products are computed by the application of Klimyk's 
	formula, except for \SU{N}'s, where the Young-tableaux algorithm is used. 
	Subalgebra decomposition of \SU{N}'s are performed by projection matrices, which 
	are generated from an algorithm to determine maximal subalgebras as originally 
	developed by Dynkin \cite{Dynkin:1957um,Dynkin:1957dm}.
\item[Restrictions:]
	Internally irreps are represented by their unique Dynkin label. LieART's default 
	behavior in \texttt{TraditionalForm} is to print the dimensional name, which is 
	the labeling preferred by physicist. Most Lie algebras can have more than one 
	irrep of the same dimension and different irreps with the same dimension are 
	usually distinguished by one or more primes~(e.g.~$\irrep{175}$ and 
	$\irrep[1]{175}$ of \A4). To determine the need for one or more primes of an 
	irrep a brute-force loop over other irreps must be performed to search for 
	irreps with the same dimensionality. Since Lie algebras have an infinite number 
	of irreps, this loop must be cut off, which is done by limiting the maximum 
	Dynkin digit in the loop. In rare cases for irreps of high dimensionality in 
	high-rank algebras, if  the cutoff used  is too low, then the assignment of 
	primes will be incorrect, but the problem can be avoided by raising the cutoff. 
	However, in either case, this can only affect the display of the irrep because 
	all computations involving this irrep are correct, since the internal unique 
	representation of Dynkin labels is used.
\item[Running time:]
	From less than a second to hours depending on the Lie algebra's rank and type 
	and/or the dimensionality of the representations in the computation.
\end{description}
\newpage

\section{Introduction}

Lie groups are a key ingredient in modern physics, while smaller Lie groups like
\SU2 and \SO{3,1}, enter the quantum mechanics of elementary chemistry and
condensed matter physics, the full spectrum of Lie groups, i.e., the classical
groups \SU{N}, \SO{N} and \Sp{2N} and the exceptionals \E6, \E7, \E8, \F4 and
\G2, have all appeared with varying degrees of frequency in particle physics.
Lie groups have many other application e.g., to the theoretical physics of
gravity, string theory, etc. as well as applications to engineering and
elsewhere. Here we will focus on the Lie algebras of the compact forms of Lie groups that
are most useful for particle physics. Most of the results are
easily extended to the non-compact forms.

Shortly after the Standard Model was completed, Grand Unified Theories (GUTs) 
were proposed, where the Standard-Model gauge group
$\SU{3}_\text{C}{\times}\SU{2}_\text{L}{\times}\U{1}_\text{Y}$ is embedded in a
higher symmetry, typically \SU5 \cite{Georgi:1974sy}, \SO{10}
\cite{Georgi:1974xy,Fritzsch:1974nn} or \E6 \cite{Gursey:1975ki}, although other choices have
been tried. Major reviews appeared on the uses of Lie algebras \cite{Slansky,
McKay:99021}, including tables of irreducible representations (irreps) and their
invariants. There are also a number of useful textbooks that cover the topic
\cite{Georgi:1982jb, Ramond:2010zz, cahn1984semi}. While extensive tables
already exist for building GUT models, it has sometimes been necessary to go
beyond what is tabulated in the literature. Our purpose here is to give extended
tables that will satisfy most modern model-building requirements, but also
provide the software that allows one to go further as the situation may require.
In describing the software we will incorporate a review of most of the necessary
group-theory background. This includes root and weight systems, the associated
Weyl groups for all the classical and exceptional Lie algebras, orthogonal basis
systems, and Weyl-group orbits, which are used in our method of calculating
tensor products and irrep decompositions.

The theory of Lie algebras is in a mature state and many algorithms have been 
established to facilitate computations in representation theory. The 
correspondence of irreps to Young tableaux, especially for \SU{N}'s, with 
algorithms for decomposing tensor products and subalgebra decomposition, even 
allows complex calculations involving high-dimensional irreps by hand. 
Lie-algebra related computations have been implemented multiple times on the 
computer in many different programming languages. Popular programs with a 
similar aim as the software presented here are \cite{LiE,Schur,Simplie}. 
However, at the time we started the project no such implementation existed for 
the computer-algebra system Mathematica. (Meanwhile a package for computations 
in finite-dimensional and affine Lie algebras has been 
published~\cite{Nazarov:2011mv} that has a similar intention as our software, as 
well as a package for the calculation of the 2-loop renormalization-group 
equations of supersymmetric models based on gauge groups incorporating many Lie 
algebra related computations~\cite{Fonseca:2011sy}). 
Mathematica\textsuperscript{\textregistered} is a computer algebra software by 
Wolfram Research, Inc.\ which is widely used especially among particle 
physicists.

Originally intended as an in-house solution for a computerized grand-unified-model 
scan of \SU{N}'s in Mathematica \cite{Albright:2012zt}, we present here
the Mathematica application LieART (\underline{Lie} \underline{A}lgebras and
\underline{R}epresentation \underline{T}heory), that makes tensor products and
subalgebra branching of irreps of the classical and exceptional Lie algebras
available for this platform. LieART's code exploits the Weyl reflection group,
inherent in all simple Lie algebras, in many parts of the algorithms, which
makes computations fast and at the same time economical on memory.  We also
focused on the usability of LieART with a particle physicist as user in mind:
Irreps can be entered by their dimensional name, a nomenclature that physicists
prefer over the more unique Dynkin label. LieART displays results in textbook
style used in most particle-physics publications, e.g., \irrepbar{10} for the
conjugated 10-dimensional irrep of \SU5 instead of the corresponding Dynkin
label \dynkin{0,0,1,0}. The Dynkin label is used internally, but can also be
used as input and output. LieART can also display results in terms of \LaTeX\
commands, that are defined in a supplemental \LaTeX\ style file for the
inclusion of results in publications.

The paper is organized as follows: In Section~\ref{sec:DownloadAndInstallation} 
we give instructions for downloading and installing LieART, as well as locating 
its documentation integrated in Mathematica's help system. 
Section~\ref{sec:QuickStart} comprises a quick-start tutorial for LieART, 
introducing the most important functions for the most common tasks in an 
example-based fashion. Section~\ref{sec:TheoryAndImplementation} presents a 
self-contained overview of the Lie algebra theory used in LieART and gives notes 
on its implementation. Section~\ref{sec:Benchmarks} gives benchmarks for a few 
tensor-product decompositions and a subalgebra decomposition of a large irrep. 
In Section~\ref{LaTeXPackage} we present a \LaTeX\ style file included in LieART 
for displaying weights, roots and irreps properly. In 
Section~\ref{ConclusionsAndOutlook} we conclude and give an outlook on future 
versions. In the appendix we include an extensive collection of tables with 
properties of irreps, tensor products and branching rules. These tables follow 
\cite{Slansky} in selection and presentation style, but extend most of the 
results. We plan to maintain and further extend our tables, which can be used 
directly as lookup tables without the aid of LieART.

\section{Download and Installation}
\label{sec:DownloadAndInstallation}
\enlargethispage{2ex}

\subsection{Download}
LieART is hosted by Hepforge, IPPP Durham. The LieART project home page is

\href{http://lieart.hepforge.org/}{\texttt{http://lieart.hepforge.org/}}

and the LieART Mathematica application can be downloaded as tar.gz archive from

\href{http://www.hepforge.org/downloads/lieart/}{\texttt{http://www.hepforge.org/downloads/lieart/}}

\subsection{Automatic Installation}

Start Mathematica and in the front end select the menu entry
\newcommand\nextstep{$\mathtt{\;\to\;}$}

   \texttt{File}\nextstep\texttt{Install\ldots}

In the appearing dialog select \texttt{Application} as \texttt{Type of Item to
Install} and the tar.gz file in the open file dialog from \texttt{Source}.
(It is not necessary to decompress the tar.gz archive since Mathematica does this automatically.)
Choose whether you want to install LieART for an individual user or system wide. For
a system-wide installation you might be asked for the superuser password.

\subsection{Manual Installation}

The above procedure in Mathematica 7 only allows you to automatically install the
Mathematica package file (\texttt{LieART.m}) of LieART without the documentation. We therefore
suggest a manual installation of the LieART application in Mathematica 7 and
in Mathematica 8 through 10 if problems with the automatic installation occur.

Extract the archive to the subdirectory \texttt{AddOns/Applications} of the
directory to which \linebreak\texttt{\$UserBaseDirectory} is set for a user-only
installation. For a system-wide installation place it in the according
subdirectory of \texttt{\$InstallationDirectory}. Restart Mathematica to allow
it to integrate LieART's documentation in its help system.

\subsection{Documentation}

The documentation of LieART is integrated in Mathematica's help system. After
restarting Mathematica the following path should lead to LieART's documentation:

\texttt{Help}
\nextstep\texttt{Documentation Center}\newline
\hphantom{\nextstep}\nextstep\texttt{Add-Ons\:\&\:Packages} (at the bottom)\newline
\hphantom{\nextstep\nextstep}\nextstep\texttt{LieART}, Button labeled "\texttt{Documentation}"

(Alternatively, a search for ``LieART'' (with the correct case) in the Documentation Center leads to the same
page.) The displayed page serves as the documentation home of LieART and includes links
to the descriptions of its most important functions.

The documentation of LieART includes a \texttt{Quick Start Tutorial} for the impatient,
which can be found near the bottom of LieART's documentation home under the
section \texttt{Tutorials}.

Tables of representation properties, tensor products and branching rules
generated by LieART can be found in the section \texttt{Tables} at the bottom of
LieART's documentation home.

\subsection{\LaTeX\ Package}

LieART comes with a \LaTeX\ package that defines commands to display irreps, roots and weights properly.
The style file \texttt{lieart.sty} can be found in the subdirectory \texttt{latex/} of the LieART project tree.
Please copy it to a location where your \LaTeX\ installation can find it.
% \pagebreak

\section{Quick Start}
\label{sec:QuickStart}
\newcommand{\mmastring}[1]{\textcolor{gray}{#1}}

This section provides a tutorial introducing the most important and frequently used functions of LieART for Lie-algebra and representation-theory related calculations.
The functions are introduced based on simple examples that can easily be modified and extended to the user's desired application.
Most examples use irreducible representations (irreps) of \SU5, which most textbooks use in examples since it is less trivial than \SU3, but small enough to return results
almost instantly on any recent computer. Also, \SU5 frequently appears in unified model building since the Standard-Model gauge group is one of its maximal subgroups.
This tutorial can also be found in the LieART documentation integrated into the Mathematica Documentation Center as ``Quick Start Tutorial'' under the section ``Tutorials''
on the LieART documentation home.

This loads the package:
\begin{mathin}
<<\:LieART`
\end{mathin}\par
\stepcounter{outcount}

\subsection{Entering Irreducible Representations}

Irreps are internally described by their Dynkin
label with a combined head of \com{Irrep} and the Lie algebra.
\definition{
    \com{Irrep[\args{algebraClass}][\args{label}]} & irrep described by its \args{algebraClass} and Dynkin \args{label}.
}{Entering irreps by Dynkin label.}

The \args{algebraClass} follows the Dynkin classification of simple Lie algebras
and can only be \com{A}, \com{B}, \com{C}, \com{D} for the classical algebras
and \com{E6}, \com{E7}, \com{E8}, \com{F4} and \com{G2} for the exceptional
algebras. The precise classical algebra is determined by the length of the
Dynkin label.

Entering the \irrepbar{10} of \SU5 by its Dynkin label and algebra class:
\begin{mathin}
Irrep[A][0,0,1,0]//FullForm
\end{mathin}
\begin{mathout}
Irrep[A][0,0,1,0]
\end{mathout}

In \com{StandardForm} the irrep is displayed in the textbook notation of Dynkin labels:
\begin{mathin}
Irrep[A][0,0,1,0]//StandardForm
\end{mathin}
\begin{mathout}
\dynkin{0,0,1,0}
\end{mathout}

In \com{TraditionalForm} (default) the irrep is displayed by its dimensional name:
\begin{mathin}
Irrep[A][0,0,1,0]
\end{mathin}
\begin{mathout}
\irrepbar{10}
\end{mathout}
The default output format type of LieART is \com{TraditionalForm}. The
associated user setting is overwritten for the notebook LieART that is loaded in. For
\com{StandardForm} as output format type please set the global variable
\com{\$DefaultOutputForm=StandardForm}.

As an example for entering an irrep of an exceptional algebra, consider the \irrep{27} of \E6:
\begin{mathin}
Irrep[E6][1,0,0,0,0,0]
\end{mathin}
\begin{mathout}
\irrep{27}
\end{mathout}

Irreps may also be entered by their dimensional name. The package
transforms the irrep into its Dynkin label. Since the algebra of an irrep of a classical Lie algebra
becomes ambiguous with only the dimensional name, it has to be specified.
\definition{
    \com{Irrep[\args{algebra}][\args{dimname}]} & irrep entered by its \args{algebra} and dimensional name \args{dimname}.
}{Entering irreps by dimensional name.}
\pagebreak

Entering the \irrepbar{10} of \SU5 by its dimensional name specifying the algebra by its Dynkin classification \A4:
\begin{mathin}
Irrep[A4][Bar[10]]//InputForm
\end{mathin}
\begin{mathout}
Irrep[A][0,0,1,0]
\end{mathout}

The traditional name of the algebra \SU{5} may also be used:
\begin{mathin}
Irrep[SU5][Bar[10]]//InputForm
\end{mathin}
\begin{mathout}
Irrep[A][0,0,1,0]
\end{mathout}

Irreps of product algebras like $\SU3{\otimes}\SU2{\otimes}\U1$ are specified by
\com{ProductIrrep} with the individual irreps of simple Lie algebras as arguments.
\definition{
    \com{ProductIrrep[\args{irreps}]} & head of product \args{irreps}, gathering irreps of simple Lie algebras.
}{Product irreps.}

The product irrep $(\irrep{3}, \irrepbar{3})$ of $\SU3{\otimes}\SU3$:
\begin{mathin}
ProductIrrep[Irrep[SU3][3],Irrep[SU3][Bar[3]]]
\end{mathin}
\begin{mathout}
(\irrep{3},\irrepbar{3})
\end{mathout}
\begin{mathin}
\%//InputForm
\end{mathin}
\begin{mathout}
ProductIrrep[Irrep[A][1,0],Irrep[A][0,1]]
\end{mathout}
\begin{mathin}
ProductIrrep[Irrep[A][1,0],Irrep[A][0,1]]
\end{mathin}
\begin{mathout}
(\irrep{3},\irrepbar{3})
\end{mathout}

Take for example the left-handed quark doublet in the Standard-Model gauge group
$\SU3{\otimes}\SU2{\otimes}\U1$ (The \U1 charge is not typeset in bold face):
\begin{mathin}
ProductIrrep[Irrep[SU3][3],Irrep[SU2][2],Irrep[U1][1/3]]
\end{mathin}
\begin{mathout}
(\irrep{3},\irrep{2})\!($\mathtt{1/3}$)
\end{mathout}
\begin{mathin}
 \%//InputForm
\end{mathin}
\begin{mathout}
ProductIrrep[Irrep[A][1,0],Irrep[A][1],Irrep[U][1/3]]
\end{mathout}

\subsection{Decomposing Tensor Products}

\definition{
    \com{DecomposeProduct[\args{irreps}]} & decomposes the tensor product of several \args{irreps}.
}{Tensor product decomposition.}

Decompose the tensor product $\irrep{3}{\otimes}\irrepbar{3}$ of \SU3:
\begin{mathin}
DecomposeProduct[Irrep[SU3][3],Irrep[SU3][Bar[3]]]
\end{mathin}
\begin{mathout}
$\irrep{1}+\irrep{8}$
\end{mathout}

Decompose the tensor product $\irrep{27}{\otimes}\irrepbar{27}$ of \E6:
\begin{mathin}
DecomposeProduct[Irrep[E6][27],Irrep[E6][Bar[27]]]
\end{mathin}
\begin{mathout}
$\irrep{1}+\irrep{78}+\irrep{650}$
\end{mathout}

Decompose the tensor product $\irrep{3}{\otimes}\irrep{3}{\otimes}\irrep{3}$ of \SU3:
\begin{mathin}
DecomposeProduct[Irrep[SU3][3],Irrep[SU3][3],Irrep[SU3][3]]
\end{mathin}
\begin{mathout}
$\irrep{1}+2(\irrep{8})+\irrep{10}$
\end{mathout}

Decompose the tensor product $\irrep{8}{\otimes}\irrep{8}$ of \SU3:
\begin{mathin}
DecomposeProduct[Irrep[SU3][8],Irrep[SU3][8]]
\end{mathin}
\begin{mathout}
$\irrep{1}+2(\irrep{8})+\irrep{10}+\irrepbar{10}+\irrep{27}$
\end{mathout}
Internally a sum of irreps is represented by \com{IrrepPlus} and \com{IrrepTimes}, an analog of the built-in functions \com{Plus} and \com{Times}:
\begin{mathin}
\%//InputForm
\end{mathin}
\par
\medskip
\begin{mathout}
IrrepPlus[Irrep[A][0,0],\:IrrepTimes[2,\:Irrep[A][1,1]],\linebreak  Irrep[A][3,\,0],\:Irrep[A][0,3],\:Irrep[A][2,2]]
\end{mathout}
Results can be transformed into a list of irreps with \com{IrrepList}, suitable for further processing with Mathematica built-in functions like \com{Select} or \com{Cases}:
\begin{mathin}
\%//IrrepList
\end{mathin}
\begin{mathout}
\{\irrep{1},\irrep{8},\irrep{8},\irrep{10},\irrepbar{10},\irrep{27}\}
\end{mathout}
Decompose the tensor product $\irrep{4}{\otimes}\irrep{4}{\otimes}\irrep{6}{\otimes}\irrep{15}$ of \SU4:
\begin{mathin}
DecomposeProduct[Irrep[SU4][4],Irrep[SU4][4],Irrep[SU4][6],Irrep[SU4][15]]
\end{mathin}
\begin{mathout}
$2(\irrep{1})+7(\irrep{15})+4(\irrep[1]{20})+\irrep{35}+5(\irrep{45})+3(\irrepbar{45})+3(\irrep{84})+2(\irrep{175})+\irrep{256}$
\end{mathout}

The Mathematica built-in command \com{Times} for products is replaced by 
\com{DecomposeProduct} for irreps as arguments. E.g., decompose the tensor 
product $\irrepbar{10}{\otimes}\irrep{24}{\otimes}\irrep{45}$ of \SU5:
\begin{mathin}
	Irrep[SU5][Bar[10]]*Irrep[SU5][24]*Irrep[SU5][45]
\end{mathin}
\begin{mathout}
$3(\irrepbar{5})+6(\irrepbar{45})+3(\irrepbar{50})+5(\irrepbar{70})+2(\irrepbar{105})+\irrepbar[2]{175}+6(\irrepbar{280})+2(\irrepbar[1]{280})+\irrepbar{420}+\irrepbar[1]{450}+3(\irrepbar{480})+2(\irrepbar{720})+\irrepbar{1120}+\irrepbar{2520}$
\end{mathout}

For powers of irreps the Mathematica built-in command \com{Power} may be used. 
E.g., decompose the tensor product 
$\irrep{27}{\otimes}\irrep{27}{\otimes}\irrep{27}$ of \E6:
\begin{mathin}
Irrep[E6][27]\textasciicircum3
\end{mathin}
\begin{mathout}
$\irrep{1}+2(\irrep{78})+3(\irrep{650})+\irrep{2925}+\irrep{3003}+2(\irrep{5824})$
\end{mathout}

Decompose tensor products of product irreps $(\irrep{3},\,\irrepbar{3},\,\irrep{1}){\otimes}(\irrepbar{3},\,\irrep{3},\,\irrep{1})$ of $\SU3{\otimes}\SU3{\otimes}\SU3$:
\begin{mathin}
DecomposeProduct[\linebreak ProductIrrep[Irrep[SU3][3],Irrep[SU3][Bar[3]],Irrep[SU3][1]],\linebreak ProductIrrep[Irrep[SU3][Bar[3]],Irrep[SU3][3],Irrep[SU3][1]]]
\end{mathin}
\begin{mathout}
$(\irrep{1},\irrep{1},\irrep{1})+(\irrep{8},\irrep{1},\irrep{1})+(\irrep{1},\irrep{8},\irrep{1})+(\irrep{8},\irrep{8},\irrep{1})$
\end{mathout}

Decompose the tensor products $(\irrep{3},\,\irrep{2}){\otimes}(\irrepbar{3},\,\irrep{1})$ of $\SU3{\otimes}\SU2$:
\begin{mathin}
DecomposeProduct[\linebreak ProductIrrep[Irrep[SU3][3],Irrep[SU2][2]],\linebreak ProductIrrep[Irrep[SU3][Bar[3]],Irrep[SU2][1]]]
\end{mathin}
\begin{mathout}
$(\irrep{1},\irrep{2})+(\irrep{8},\irrep{2})$
\end{mathout}

\subsection{Decomposition to Subalgebras}

\definition{
    \com{DecomposeIrrep[\args{irrep},\,\args{subalgebra}]} & decomposes \args{irrep} to the specified \args{subalgebra}.\\
    \com{DecomposeIrrep[\args{pirrep},\,\args{subalgebra},\,\args{pos}]} & decomposes the product irrep \args{pirrep} at position \args{pos}.\\
}{Decompose irreps and product irreps.}

Decompose the \irrepbar{10} of \SU5 to $\SU3{\otimes}\SU2{\otimes}\U1$:
\begin{mathin}
DecomposeIrrep[Irrep[SU5][Bar[10]],ProductAlgebra[SU3,SU2,U1]]
\end{mathin}
\begin{mathout}
$(\irrep{1},\irrep{1})(6)+(\irrep{3},\irrep{1})(-4)+(\irrepbar{3},\irrep{2})(1)$
\end{mathout}
\pagebreak

\enlargethispage{10pt}
Decompose the \irrep{10} and the \irrepbar{5} of \SU5 to $\SU3{\otimes}\SU2{\otimes}\U1$ (\com{DecomposeIrrep} is \com{Listable}):
\begin{mathin}
DecomposeIrrep[\{Irrep[SU5][10],Irrep[SU5][Bar[5]]\},ProductAlgebra[SU3,SU2,U1]]
\end{mathin}
\begin{mathout}
$\{(\irrepbar{3},\irrep{1})(4)+(\irrep{3},\irrep{2})(-1)+(\irrep{1},\irrep{1})(-6),\,(\irrepbar{3},\irrep{1})(-2)+(\irrep{1},\irrep{2})(3)\}$
\end{mathout}

Decompose the \irrep{16} of \SO{10} to $\SU5{\otimes}\U1$:
\begin{mathin}
DecomposeIrrep[Irrep[SO10][16],ProductAlgebra[SU5,U1]]
\end{mathin}
\begin{mathout}
$(\irrep{1})(-5)+(\irrepbar{5})(3)+(\irrep{10})(-1)$
\end{mathout}

Decompose the \irrep{27} of \E6 to $\SU3{\otimes}\SU3{\otimes}\SU3$:
\begin{mathin}
DecomposeIrrep[Irrep[E6][27],ProductAlgebra[SU3,SU3,SU3]]
\end{mathin}
\begin{mathout}
$(\irrep{3},\irrep{1},\irrep{3})+(\irrep{1},\irrep{3},\irrepbar{3})+(\irrepbar{3},\irrepbar{3},\irrep{1})$
\end{mathout}

Decompose the \SU3 irrep \irrep{3} in $(\irrep{24},\irrep{3})(-3)$ of $\SU5{\otimes}\SU3{\otimes}\U1$ to
$\SU2{\otimes}\text{U}'(1)$,\newline i.e., $\SU5{\otimes}\SU3{\otimes}\U1\to\SU5{\otimes}\SU2{\otimes}\text{U}'(1){\otimes}\U1$:
\begin{mathin}
DecomposeIrrep[ProductIrrep[Irrep[SU5][24],Irrep[SU3][3],Irrep[U1][-3]], ProductAlgebra[SU2,U1],2]
\end{mathin}
\begin{mathout}
$(\irrep{24},\irrep{1})(-2)(-3)+(\irrep{24},\irrep{2})(1)(-3)$
\end{mathout}

The same decomposition as above displayed as branching rule:
\begin{mathin}
IrrepRule[\slot,DecomposeIrrep[\slot,ProductAlgebra[SU2,U1],2]]\&@
 ProductIrrep[Irrep[SU5][24],Irrep[SU3][3],Irrep[U1][-3]]
\end{mathin}
\begin{mathout}
$(\irrep{24},\irrep{3})(-3)\to(\irrep{24},\irrep{1})(-2)(-3)+(\irrep{24},\irrep{2})(1)(-3)$
\end{mathout}

Branching rules for all totally antisymmetric irreps, so-called basic irreps, of \SU6 to $\SU3{\otimes}\SU3{\otimes}\U1$:
\begin{mathin}
IrrepRule[\slot,DecomposeIrrep[\slot,ProductAlgebra[SU3,SU3,U1]]]\&/@\,\newline BasicIrreps[SU6]//TableForm
\end{mathin}
\begin{mathout}\hangindent=0ex%
$\irrep{6}\rightarrow(\irrep{3},\irrep{1})(1)+(\irrep{1},\irrep{3})(-1)$\newline
$\irrep{15}\rightarrow(\irrepbar{3},\irrep{1})(2)+(\irrep{1},\irrepbar{3})(-2)+(\irrep{3},\irrep{3})(0)$\newline
$\irrep{20}\rightarrow(\irrep{1},\irrep{1})(3)+(\irrep{1},\irrep{1})(-3)+(\irrep{3},\irrepbar{3})(-1)+(\irrepbar{3},\irrep{3})(1)$\newline
$\irrepbar{15}\rightarrow(\irrep{3},\irrep{1})(-2)+(\irrep{1},\irrep{3})(2)+(\irrepbar{3},\irrepbar{3})(0)$\newline
$\irrepbar{6}\rightarrow(\irrepbar{3},\irrep{1})(-1)+(\irrep{1},\irrepbar{3})(1)$\newline
\end{mathout}
\vspace{-15pt}

\subsection{Young Tableaux}
\Yautoscale0
\Yboxdim13pt

The irreps of \SU{N} have a correspondence to Young tableaux, which can be displayed by \com{YoungTableau}.
\definition{
    \com{YoungTableau[\args{irrep}]} & Displays the Young tableau associated with an \SU{N} \args{irrep}.\\
}{Young tableaux.}

Young tableau of the \irrep{720} of \SU5:
\begin{mathin}
YoungTableau[Irrep[A][1,2,0,1]]
\end{mathin}
\begin{mathout}\label{out:YoungTableau720SU5}
\large\yng(4,3,1,1)
\end{mathout}

Display Young tableaux of \SU4 irreps with a maximum of one column per box count:
\begin{mathin}
Row[Row[\{\textcolor{DarkGreen}{\#},"{:}\ ",YoungTableau[\textcolor{DarkGreen}{\#}]\}]\&/@\newline
SortBy[Irrep[A]@@@Tuples[\{0,1\},3],Dim],Spacer[10]]
\end{mathin}
\newcommand{\irrepandtableau}[2]{\irrep{#1}{:}\:#2\quad}
\begin{mathout}
$\irrepandtableau{\irrep{1}}{\bullet}
\irrepandtableau{\irrepbar{4}}{\yng(1,1,1)}
\irrepandtableau{\irrep{4}}{\yng(1)}
\irrepandtableau{\irrep{6}}{\yng(1,1)}
\irrepandtableau{\irrep{15}}{\yng(2,1,1)}
\irrepandtableau{\irrep{20}}{\yng(2,2,1)}
\irrepandtableau{\irrepbar{20}}{\yng(2,1)}
\irrepandtableau{\irrep{64}}{\yng(3,2,1)}$
\end{mathout}
% \pagebreak

\section{Theoretical Background and Implementation}
\label{sec:TheoryAndImplementation}

In this section we give a self-contained overview of the Lie algebra theory used
and implemented in LieART. It is subdivided into parts discussing basic
properties of Lie algebras, roots, weights, Weyl orbits, representations and decompositions.
Every subsection begins with a list of the
relevant LieART functions followed by text that introduces the necessary theory
with reference to the functions and notes on their implementation. This section
is not intended as a pedagogical introduction to Lie algebras and we refer the
reader to the excellent literature serving this purpose \cite{Slansky,
Georgi:1982jb, cahn1984semi}.

\subsection{Algebras}
\label{ssec:Algebras}

\definition{
    \com{Rank[\args{expr}]}                             & gives the rank of the algebra of \args{expr}, which can be an irrep, a weight a root or an algebra itself.\\
    \com{Algebra[\args{algebraClass}][\args{rank}]}     & represents a classical algebra of the type \args{algebraClass}, which can only be \com{A}, \com{B}, \com{C} or \com{D}, with rank \args{rank}.\\
    \com{Algebra[\args{expr}]}                          & gives the algebra (classical or exceptional) of \args{expr}, which may be an irrep, a weight or a root in any basis.\\
    \com{OrthogonalSimpleRoots[\args{algebra}]}         & gives the simple roots of \args{algebra} in the orthogonal basis.\\
    \com{CartanMatrix[\args{algebra}]}                  & gives the Cartan matrix of \args{algebra}.\\
    \com{OmegaMatrix[\args{algebra}]}                   & gives the matrix of fundamental weights of \args{algebra} as rows.\\
    \com{OrthogonalFundamentalWeights[\args{algebra}]}  & gives the fundamental weights of \args{algebra} in the orthogonal basis.\\
    \com{OrthogonalBasis[\args{expr}]}                  & transforms \args{expr} from any basis into the orthogonal basis.\\
    \com{OmegaBasis[\args{expr}]}                       & transforms \args{expr} from any basis into the $\omega$-basis. \\
    \com{AlphaBasis[\args{expr}]}                       & transforms \args{expr} from any basis into the $\alpha$-basis.\\
    \com{DMatrix[\args{algebra}]}                       & gives a matrix with inverse length factors of simple roots on the main diagonal.\\
    \com{ScalarProduct[\args{weight1},\args{weight2}]}  & gives the scalar product of \args{expr1} and \args{expr2} in any basis. \args{expr1} and \args{expr2} may be weights or roots.\\
    \com{MetricTensor[\args{algebra}]}                  & gives the metric tensor or quadratic-form matrix of \args{algebra}.\\
}{Basic Algebra Properties.}

\begin{figure}[t]
    \begin{center}
        \includegraphics{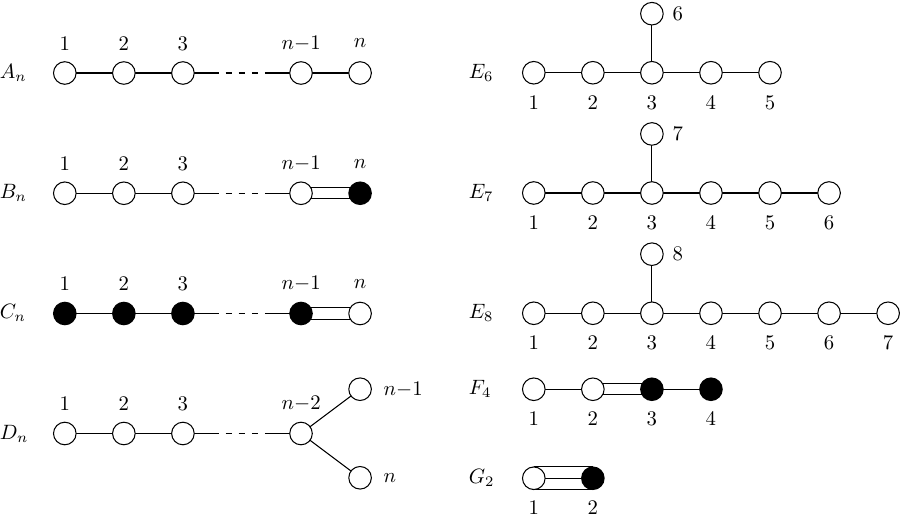}
        \caption{\label{fig:DynkinDiagrams} Dynkin Diagrams of classical and exceptional simple Lie algebras.}
    \end{center}
\end{figure}

\newcommand{\liebracket}[2]{\left[#1, #2\right]}

\subsubsection{Definition}

A \emph{Lie Algebra} is a vector space $g$ over a field $F$ with the \emph{Lie
bracket} $\left[\cdot,\cdot\right]$ as binary operation, which is bilinear,
alternating and fulfills the Jacoby identity. The Lie bracket is often referred
to as the commutator. The Lie brackets of the generators $t_i$ of the Lie
algebra are
\begin{equation}\label{eq:DefinitionStructureConstants}
    \liebracket{t_i}{t_j} = f_{ijk} t_k
\end{equation}
with the so-called \emph{structure constants} $f_{ijk}$, that fully determine
the algebra. A Lie algebra is called \emph{simple} when it contains no
non-trivial ideals. A \emph{semi-simple} Lie algebra is a sum of simple ones.

\subsubsection{Roots}

The generators $t_i$ of a simple Lie algebra in the Cartan-Weyl basis fall into
two sets: The so-called \emph{Cartan subalgebra}, $H$, contains all
simultaneously diagonalizable generators $h_i$, i.e., the generators are
Hermitian and mutually commute (the Cartan subalgebra is abelian):
\begin{equation}
    h_i=h_i^\dagger, \qquad\liebracket{h_i}{h_j} = 0, \qquad  i,j=1,\ldots,n.
\end{equation}
The number of simultaneously diagonalizable generators $n$ is called the
\emph{rank} of the algebra, and can be determined by the function
\com{Rank[\args{expr}]} in LieART. We denote all other generators as $e_\alpha$.
They satisfy $n$ eigenvalue equations with the generators of the Cartan
subalgebra $h_i$:
\begin{equation}\label{eq:DefinitionRootVector}
    \liebracket{h_i}{e_\alpha} = \alpha_i\,e_\alpha, \qquad  i=1,\ldots,n,
\end{equation}
which is a subset of \eqref{eq:DefinitionStructureConstants} and thus the
$\alpha_i$ are structure constants, which are real numbers due to the
hermiticity of the $h_i$'s. Since the $\alpha_i$ are the solutions to the
eigenvalue equation \eqref{eq:DefinitionRootVector} the vectors
$\alpha{=}(\alpha_1,\ldots,\alpha_n)$ are called the \emph{root vectors}, which
lie in an $n$-dimensional euclidian space, called the \emph{root space}.
\emph{Roots} are functionals mapping the Cartan subalgebra $H$ onto the real
numbers (the eigenvalues), for all generators $t_i$, which also includes the
$h_i$ where the eigenvalues are zero. Thus, a Lie algebra has as many roots as
generators. The roots are labeled by the root vectors, which we will use in its
place from now on.

A zero root with an $n$-fold degeneracy is associated with the Cartan
subalgebra. In the Cartan-Weyl basis the other generators come in conjugated
pairs $e_\alpha^\dagger{=}e_{{-}\alpha}$ and correspond to the ladder operators
of \SU2. So-called \emph{positive roots} correspond to the raising operator
$e_\alpha$ and negative roots to the lowering operators $e_{{-}\alpha}$. If
$\alpha$ is a root so is ${-}\alpha$.

Some of the positive roots can be written as sum of others. Those for which this
is not possible are called \emph{simple roots} and a Lie algebra has as many
simple roots as its rank. It is clear that specifying the simple roots fully
determines a Lie algebra and thus can be used to replace
\eqref{eq:DefinitionStructureConstants}, because all structure constants can be
derived therefrom.

\subsubsection{Classification of Lie Algebras}

Using the commutation relations and the Jacobi identity to analyze the
generators, constraints on the roots can be derived and eventually all possible
root systems found, which is identical to identifying all allowed Lie algebras.
It turns out that simple roots can only come in at most two lengths in one Lie
algebra and at four different angles between any pair of them. The simple roots
are in particular not orthogonal. The so-called \emph{Dynkin diagrams} are an
ingenious way to depict these relations: simple roots are represented by dots,
which are open, {\Large $\Circle$}, for the longer roots or for all roots if
they only come in one length, and filled, {\Large $\CIRCLE$}, for the shorter
roots. Angles between two simple roots are represented by lines connecting the
dots: no line for an angle of 90\textdegree, one line for 120\textdegree, two
lines for 135\textdegree\ and three for 150\textdegree. Figure
\ref{fig:DynkinDiagrams} shows the Dynkin diagrams for all simple Lie algebras.
Semi-simple Lie algebras have disjoint parts and can thus be reduced to two 
or more Dynkin diagrams of simple Lie algebras.

The simple Lie algebras fall into two types: four families of infinite series
algebras, \A{n}, \B{n}, \C{n} and \D{n}, also called the \emph{classical Lie
algebras} and five so-called \emph{exceptional algebras}, \E6, \E7, \E8, \F4 and
\G2, with their rank as subscript (see Table
\ref{tab:LieAlgebrasClassification}).
\begin{table}[t]
\begin{center}
\begin{tabular}{lllll}\toprule
\textbf{Type}   & \textbf{Cartan} & \textbf{Name} & \textbf{Rank}   & \textbf{Description}\\\midrule
classical       & \A{n}           & \SU{n{+}1}    & $n\geq1$        & Special unitary algebras of $n{+}1$ complex dimension\\
                & \B{n}           & \SO{2n{+}1}   & $n\geq3$        & Special orthogonal algebras of odd ($2n{+}1$) real dimension\\
                & \C{n}           & \Sp{2n}       & $n\geq2$        & Symplectic algebras of even ($2n$) complex dimension\\
                & \D{n}           & \SO{2n}       & $n\geq4$        & Special orthogonal algebras of even ($2n$) real dimension\\\midrule
exceptional     & \E6             & \E6           & 6               & Exceptional algebra of rank 6\\
                & \E7             & \E7           & 7               & Exceptional algebra of rank 7\\
                & \E8             & \E8           & 8               & Exceptional algebra of rank 8\\
                & \F4             & \F4           & 4               & Exceptional algebra of rank 4\\
                & \G2             & \G2           & 2               & Exceptional algebra of rank 2\\
\bottomrule
\end{tabular}
\caption{\label{tab:LieAlgebrasClassification} Classification of simple Lie algebras.}
\end{center}
\end{table}
The labels are according to the classification by Cartan. The classical Lie 
algebras are internally represented (i.e., in \com{FullForm}) in LieART by 
\com{Algebra[\args{algebraClass}][\args{n}]}, with \args{algebraClass} being 
either \com{A}, \com{B}, \com{C} or \com{D} and \args{n} being the rank. The 
exceptional algebras are defined in LieART as \com{E6}, \com{E7}, \com{E8}, 
\com{F4} and \com{G2} and short forms of the classical algebras are predefined 
up to rank 30 for the Cartan classification, i.e., \com{A1}$,\ldots,$\com{A30}, 
\com{B3}$,\ldots,$\com{B30}, \com{C2}$,\ldots,$\com{C30}, \com{D4}$,\ldots,$\com{D30}, 
and up to dimension 30 for the conventional name, i.e., \com{SU2}$,\ldots,$\com{SU30}, 
\com{SO7}$,\ldots,$\com{SO30} and \com{Sp4}, \com{Sp6}$,\ldots,$\com{Sp30}. In 
\com{StandardForm} the Cartan classification is explicitly displayed and in 
\com{TraditionalForm} the Lie algebra is written by its conventional name. 

Note the isomorphisms for low-dimension algebras:
\begin{subequations}
\begin{align}
	\SU2 &\sim \SO3 \sim \Sp2  & (\A1 &\sim \B1 \sim \C1   ), \label{eqn:SU2Isomorphism}\\
	\SO4 &\sim \SU2\otimes\SU2 & (\D2 &\sim \A1 \otimes \A1), \label{eqn:SO4Isomorphism}\\
	\SO5 &\sim \Sp4            & (\B2 &\sim \C2            ), \label{eqn:SO5Isomorphism}\\
	\SO6 &\sim \SU4            & (\D3 &\sim \A3).             \label{eqn:SO6Isomorphism}
\end{align}
\end{subequations}
When these low-dimension Lie algebras occur in a calculation, choose the \SU2 
form in \eqref{eqn:SU2Isomorphism} and \eqref{eqn:SO4Isomorphism}, the \Sp4 form 
in \eqref{eqn:SO5Isomorphism} and the \SU4 form in \eqref{eqn:SO6Isomorphism} 
when using LieART.

\subsubsection{Bases}
\label{ssec:Bases}

With respect to the Weyl reflection group, inherent in all compact Lie algebras, as
we will explain later, it is convenient to express the root space in an
orthogonal coordinate system, which is a subspace of Euclidian space. The
specific subspace varies with the Lie algebra. For \A{n} it is a subspace of
$\mathbb{R}^{n+1}$, where the coordinates sum to one. As the simple roots define the Lie
algebra, they are explicitly specified in LieART using orthogonal coordinates
and can be retrieved by \com{OrthogonalSimpleRoots[\args{algebra}]}. E.g., the
four simple roots of \A4 (\SU5) in orthogonal coordinates are: 
\begin{mathin}
OrthogonalSimpleRoots[A4]//Column
\end{mathin}
\begin{mathout}
\nohangingindent
\rootorthogonal{1, {-}1, 0, 0, 0}\newline
\rootorthogonal{0, 1, {-}1, 0, 0}\newline
\rootorthogonal{0, 0, 1, {-}1, 0}\newline
\rootorthogonal{0, 0, 0, 1, {-}1}
\end{mathout}

The so-called \emph{Cartan matrix} exhibits the non-orthogonality of the simple
roots. It is defined as
\begin{equation}\label{eq:CartanMatrix}
    A_{ij} = \frac{2 \scalarproduct{\alpha_i}{\alpha_j}}{\scalarproduct{\alpha_j}{\alpha_j}}\qquad i,j=1,\ldots,n
\end{equation}
where the scalar product $\scalarproduct{\cdot}{\cdot}$ is the ordinary scalar
product of $\mathbb{R}^{n+1}$ in the case of \A{n}. Most textbooks translate the
Dynkin diagrams to the corresponding Cartan matrix as a starting point. And in
fact, the rows of the Cartan matrix are the simple roots in the so-called
\emph{$\omega$-basis}, which is the bases of \emph{fundamental weights}, also
called the \emph{Dynkin basis}. (Weights will be introduced later in the context
of representations.) The Cartan matrix is implemented in LieART as the function
\com{CartanMatrix[\args{algebra}]} following the definition of
\eqref{eq:CartanMatrix}. The Cartan matrix for \A4 reads:
\begin{mathin}
CartanMatrix[A4]
\end{mathin}
\begin{mathout}
$\begin{pmatrix}
 2 & {-}1 & 0 & 0 \\
 {-}1 & 2 & {-}1 & 0 \\
 0 & {-}1 & 2 & {-}1 \\
 0 & 0 & {-}1 & 2 \\
\end{pmatrix}$
\end{mathout}
Besides the orthogonal basis, and the $\omega$-basis, the $\alpha$-basis is also
useful. As the name indicates it is the basis of simple roots and it explicitly
shows how, e.g., a root is composed out of simple roots. Neither the
$\omega$-basis nor the $\alpha$-basis is orthogonal. The Cartan matrix
mediates between the $\omega$- and $\alpha$-bases:
\begin{equation}
    \alpha_i
    = \sum_{j=1}^n A_{ij}\omega_j, \qquad \omega_i
    = \sum_{j=1}^n (A^{{-}1})_{ij}\alpha_j.
\end{equation}
where the $\omega_i$ are the fundamental weights, which we will define later.
These bases are dual to each other in the sense that
\begin{equation}\label{eq:AlphaOmegaDual}
    \frac{2\scalarproduct{\alpha_i}{\omega_j}}{\scalarproduct{\alpha_i}{\alpha_i}}
    \equiv \scalarproduct{\alpha_i^\vee}{\omega_j}
    = \delta_{ij},\qquad i,j=1,\ldots,n
\end{equation}
where $\alpha_i^\vee$ is the so-called \emph{coroot} of $\alpha_i$ defined as
\begin{equation}
    \alpha^\vee = \frac{2\alpha}{\scalarproduct{\alpha}{\alpha}}.
\end{equation}
The transformation to the orthogonal basis can be derived from
\eqref{eq:AlphaOmegaDual}: Expressing $\alpha_i$ and $\omega_j$ in orthogonal
coordinates as $\hat\alpha_i$ and $\hat\omega_j$\eqref{eq:AlphaOmegaDual} reads
\begin{equation}\label{eq:AlphaOmegaDualOrthogonal}
    \frac{2 \hat\alpha_i\cdot\hat\omega_j}{\hat\alpha_i\cdot\hat\alpha_i}
    \equiv \hat\alpha_i^\vee\cdot\hat\omega_j
    = \delta_{ij},\qquad i,j=1,\ldots,n
\end{equation}
using the ordinary scalar product of $\mathbb{R}^m$, where $m$ is the dimension
of the orthogonal subspace. Using the matrices $\matrixhat{A}$ and
$\matrixhat\Omega$ with the simple \emph{co}roots $\hat\alpha_i^\vee$ and the
fundamental weights $\hat\omega_j$ as rows, we can write
\eqref{eq:AlphaOmegaDualOrthogonal} as the matrix equation:
\begin{equation}\label{eq:AlphaOmegaDualOrthogonalMatrices}
    \matrixhat{A}\transpose{\matrixhat{\Omega}} = I_n
\end{equation}
where both $\matrixhat{A}$ and $\matrixhat\Omega$ are $n{\times}m$ matrices, 
where $\matrixhat\Omega$ is defined below in \ref{eq:FundamentalWeights}.
Please note that the dimension of the orthogonal space $m$ is not necessarily
the same as the rank of the algebra $n$. These exceptions are: \A{n} with
$m{=}n+1$, \E6 with $m{=}8$,  \E7 with $m{=}8$ and \G2 with $m{=}3$. For all others
$m{=}n$ holds. The matrix of the simple coroots in the orthogonal basis
$\matrixhat{A}$ is easily calculated from the simple roots given in LieART, but
the matrix of fundamental weights in the orthogonal basis $\matrixhat{\Omega}$
must be determined by \eqref{eq:AlphaOmegaDualOrthogonalMatrices}. In the cases
where $\matrixhat{A}$ is not a square matrix its inverse does not exist. Because
the rows of $\matrixhat{A}$, which are the simple coroots in the orthogonal
basis, are linear independent, $\matrixhat{A}\transpose{\matrixhat{A}}$ is
invertible and the so-called right-inverse $\matrixhat{A}^{\!+}$ can be found via
\begin{equation}
    \matrixhat{A}^{\!+}
    = \transpose{\matrixhat{A}}(\matrixhat{A}\transpose{\matrixhat{A}})^{{-}1}
\end{equation}
which satisfies: $\matrixhat{A}\matrixhat{A}^{\!+}{=}I_n$, i.e., 
by comparing with \eqref{eq:AlphaOmegaDualOrthogonalMatrices} the
matrix $\transpose{\matrixhat{\Omega}}$ can be identified with
$\matrixhat{A}^{\!+}\!$, in other words the fundamental weights as rows of
$\matrixhat{\Omega}$ in terms of simple coroots as rows of $\matrixhat{A}$ are
\begin{equation}\label{eq:FundamentalWeights}
    \matrixhat{\Omega} = \transpose{(\matrixhat{A}^{\!+}\!)}
    = (\matrixhat{A}\transpose{\matrixhat{A}})^{{-}1}\matrixhat{A}
\end{equation}
The Mathematica built-in function \com{PseudoInverse[\args{matrix}]} yields the
right-inverse for our case of a \args{matrix} with linear independent rows, i.e.,
the implementation of the second equality in \eqref{eq:FundamentalWeights} is not
needed. The matrix of the fundamental weights $\matrixhat{\Omega}$ is implemented as
\com{OmegaMatrix[\args{algebra}]}, e.g., for \A4:
\begin{mathin}
    OmegaMatrix[A4]
\end{mathin}
\begin{mathout}
\setlength\extrarowheight{2pt}
    $\begin{pmatrix}
        \frac{4}{5} & {-}\frac{1}{5} & {-}\frac{1}{5} & {-}\frac{1}{5} & {-}\frac{1}{5} \\
        \frac{3}{5} & \frac{3}{5} & {-}\frac{2}{5} & {-}\frac{2}{5} & {-}\frac{2}{5} \\
        \frac{2}{5} & \frac{2}{5} & \frac{2}{5} & {-}\frac{3}{5} & {-}\frac{3}{5} \\
        \frac{1}{5} & \frac{1}{5} & \frac{1}{5} & \frac{1}{5} & {-}\frac{4}{5}
    \end{pmatrix}$
\end{mathout}
and the function \com{OrthogonalFundamentalWeights[\args{algebra}]} adds the proper heads
to the rows of $\matrixhat{\Omega}$, to identify them as weights is the orthogonal basis.
We will discuss (fundamental) weights in Section \ref{ssec:Representations} in more detail.

The matrix of the fundamental weights in the orthogonal basis $\matrixhat\Omega$
mediates between the $\omega$-basis and the orthogonal basis:
\begin{equation}
    \omega_i = \sum_{j=1}^n \matrixhat\Omega_{ij}e_j, \qquad e_i
    = \sum_{j=1}^n (\matrixhat\Omega^{{-}1})_{ij}\omega_j.
\end{equation}

The LieART functions \com{AlphaBasis[\args{weightOrRoot}]},
\com{OmegaBasis[\args{weightOrRoot}]} and \linebreak
\com{OrthogonalBasis[\args{weightOrRoot}]} transform \args{weightOrRoot} from
any basis into the $\alpha$-basis, the $\omega$-basis and the
orthogonal basis, respectively. It is obvious how the simple roots in the
$\alpha$-basis look:
\begin{mathin}
AlphaBasis[OrthogonalSimpleRoots[A4]]//Column
\end{mathin}
\begin{mathout}
\nohangingindent
\rootorthogonal{1, 0, 0, 0}\newline
\rootorthogonal{0, 1, 0, 0}\newline
\rootorthogonal{0, 0, 1, 0}\newline
\rootorthogonal{0, 0, 0, 1}
\end{mathout}
and likewise the fundamental weights in the $\omega$-basis:
\begin{mathin}
OmegaBasis[OrthogonalFundamentalWeights[A4]]//Column
\end{mathin}
\begin{mathout}
\label{out:FundamentalWeightsOmegaBasis}
\nohangingindent
\weight{1, 0, 0, 0}\newline
\weight{0, 1, 0, 0}\newline
\weight{0, 0, 1, 0}\newline
\weight{0, 0, 0, 1}
\end{mathout}

(Roots and weights in the $\omega$-basis are displayed with framed boxes
following the notation of most textbooks.) 
A root in LieART is represented by three different heads: \com{RootOrthogonal[\args{algebraClass}][\args{label}]} for a
root in the orthogonal basis, \com{RootOmega[\args{algebraClass}][\args{label}]}
in the $\omega$-basis and in the $\alpha$-basis by \com{RootAlpha[\args{algebraClass}][\args{label}]}. The \args{algebraClass} can only be \com{A}, \com{B},
\com{C} or \com{D} to indicate a classical Lie algebra or \com{E6}, \com{E7},
\com{E8}, \com{F4} or \com{G2} for the exceptionals. The \args{label} stands for
the comma-separated coordinates. This form of the roots is displayed in
\com{InputForm} and \com{FullForm}. E.g., the first simple root of \A4 in all
three bases reads:
\begin{mathin}
\{\slot,OmegaBasis[\slot],AlphaBasis[\slot]\}\&@First[OrthogonalSimpleRoots[A4]]//InputForm
\end{mathin}
\begin{mathout}
\{RootOrthogonal[A][1,{-}1,0,0,0], RootOmega[A][2,{-}1,0,0], RootAlpha[A][1,0,0,0]\}
\end{mathout}

\subsubsection{Scalar Product}

The standard choice for the length factors $\scalarproduct{\alpha_j}{\alpha_j}$
in \eqref{eq:CartanMatrix} is 2 for the longer roots, if there are two root
lengths. The factors $2/\scalarproduct{\alpha_j}{\alpha_j}$ can only take three
different values which are: 1 for all roots of \A{n}, \D{n}, \E6, \E7, \E8 and
for the long roots of \B{n}, \C{n}, \F4 and \G2; 2 for the short roots of \B{n},
\C{n} and \F4 and 3 for the short root of \G2. Their implementation in LieART
is in the form of diagonal matrices with the inverse factors for the simple roots
corresponding to the row on the main diagonal, i.e.,
\begin{equation}
D=\text{diag}\left(\frac{1}{2}\scalarproduct{\alpha_1}{\alpha_1},\ldots, \frac{1}{2}\scalarproduct{\alpha_n}{\alpha_n}\right)
\end{equation}
as defined in \cite{klimyk_orbit_2006}. E.g., for \F4, to avoid a trivial
example, we have:
\begin{mathin}
DMatrix[F4]
\end{mathin}
\begin{mathout}
$\begin{pmatrix}
 1 & 0 & 0 & 0 \\
 0 & 1 & 0 & 0 \\
 0 & 0 & \frac{1}{2} & 0 \\
 0 & 0 & 0 & \frac{1}{2} \\
\end{pmatrix}$
\end{mathout}

In the $\omega$-basis the scalar product used in \eqref{eq:CartanMatrix} becomes:
\begin{equation}
    \scalarproduct{x}{y} = \sum_{i,j}^n x_i (A^{{-}1})_{ij}D_j y_j = \sum_{i,j}^n x_i G_{ij} y_j
\end{equation}
where the $x_i$ and $y_j$ are coordinates of $x$ and $y$ in the $\omega$-basis. The matrix
\begin{equation}
    G_{ij} = (A^{{-}1})_{ij} \frac{\scalarproduct{\alpha_j}{\alpha_j}}{2} = (A^{{-}1})_{ij} D_j
\end{equation}
is called \emph{quadratic-form matrix} or \emph{metric tensor} of the Lie
algebra. The scalar product is available in LieART as
\com{ScalarProduct[\args{weightOrRoot1},\args{weightOrRoot2}]}, where \args{weightOrRoot1} and
\args{weightOrRoot2} may be roots or weights in the orthogonal basis, the
$\alpha$-basis or the $\omega$-basis. The function recognizes the basis by the heads
of \args{weightOrRoot1} and \args{weightOrRoot2}. The LieART function for the metric tensor $G$ is
\com{MetricTensor[\args{algebra}]}, e.g., for \A4:
\begin{mathin}
MetricTensor[A4]
\end{mathin}
\begin{mathout}
\setlength\extrarowheight{2pt}
$\begin{pmatrix}
 \frac{4}{5} & \frac{3}{5} & \frac{2}{5} & \frac{1}{5} \\
 \frac{3}{5} & \frac{6}{5} & \frac{4}{5} & \frac{2}{5} \\
 \frac{2}{5} & \frac{4}{5} & \frac{6}{5} & \frac{3}{5} \\
 \frac{1}{5} & \frac{2}{5} & \frac{3}{5} & \frac{4}{5} \\
\end{pmatrix}$
\end{mathout}

\subsubsection{Representation}
\label{ssec:Representations}

A \emph{representation} is a linear map of the Lie algebra into the general
linear group, i.e., the matrix group, that preserves the Lie bracket relations.
It is a homomorphism that maps the generators $t_i$ onto invertible matrices
$T_i$, that satisfy the same ``commutation'' relations as the Lie algebra,
namely
\begin{equation}\label{eq:DefinitionRepresentation}
    \liebracket{T_i}{T_j} = f_{ijk} T_k,
\end{equation}
where the $\left[\cdot,\cdot\right]$ is now the commutator.

Points in the vector space that the matrices act on can be labeled by the set of
eigenvalues of the matrices representing the generators of the Cartan
subalgebra. Such a set of eigenvalues is called a \emph{weight vector}, and the
associated functional \emph{weight}, denoted by $\lambda$. They are defined in
root space which is called \emph{weight space} in this context. The weights and
weight vectors of a representation correspond to roots and root vectors of the
algebra. In fact, weights can be expressed as rational linear combinations of
roots, and, as pointed out in this section, eventually by simple roots. In
particular, the structure functions themselves form a representation of the
algebra: the \emph{adjoint representation}, which has the same dimension as the
algebra, namely the number of roots.

\subsection{Weyl Group Orbits}

\definition{
    \com{Reflect[\args{weightOrRoot},\args{simpleroots}]}       & reflects \args{weightOrRoot} at the hyperplanes orthogonal to the specified \args{simpleroots}.\\
    \com{Reflect[\args{weightOrRoot}]}                          & reflects \args{weightOrRoot} at the hyperplanes orthogonal to all simple roots of the Lie algebra of \args{weightOrRoot}.\\
    \com{ReflectionMatrices[\args{algebra}]}                    & gives the reflection matrices of the Weyl group of \args{algebra}.\\
    \com{Orbit[\args{weightOrRoot},\args{simpleroots}]}         & generates the Weyl group orbit of \args{weightOrRoot} using only the specified \args{simpleroots}.\\
    \com{Orbit[\args{weightOrRoot}]}                            & generates the full Weyl group orbit of \args{weightOrRoot} using all simple roots of the Lie algebra of \args{weightOrRoot}.\\
    \com{DimOrbit[\args{weightOrRoot},\args{simpleroots}]}      & gives the size of the orbit of \args{weightOrRoot} using only the \args{simpleroots}.\\
    \com{DimOrbit[\args{weightOrRoot}]}                         & gives the size of the orbit of \args{weightOrRoot} using all simple roots of the Lie algebra of \args{weightOrRoot}.\\
}{Weyl Group Orbits}

The finite group $W(L)$, called the Weyl group of the Lie algebra $L$, is a
reflection group inherent in the root systems of all simple Lie algebras. The
Coxeter groups are an abstraction of reflection groups and the so-called
\emph{Coxeter-Dynkin diagram} describing Coxeter groups are closely related to
the Dynkin diagrams presented here. In fact the Coxeter-Dynkin diagram
corresponding to the Dynkin diagram describes the Weyl group of the Lie algebra.

The transformations $r_i$ generating the Weyl group are reflections of a vector $x$
in root space at the hyperplanes orthogonal to the simple roots $\alpha_i$
of the Lie algebra defined by
\begin{equation}\label{eq:WeylReflection}
r_i x = x - \frac{2\scalarproduct{x}{\alpha_i}}{\scalarproduct{\alpha_i}{\alpha_i}}\alpha_i,\qquad i=1,\ldots,n,\qquad x\in \mathbb{R}^n.
\end{equation}

The LieART function \com{Reflect[\args{weightOrRoot},\args{simpleroots}]}
implements the reflections $r_i$ with \args{weightOrRoot} as $x$ and
\args{simpleroots} as a list of simple roots $\alpha_i$. The result is a list of
weights, because the reflection is preformed with several roots simultaneously.

If \args{weightOrRoots} are in the orthogonal basis and ought to be reflected
using all roots, the function pattern is \com{Reflect[\args{weightOrRoot}]},
without the simple roots as second argument. Instead of the definition with
scalar products following \eqref{eq:WeylReflection}, the implementation
multiplies the orthogonal coordinates with precomputed reflection matrices,
which have a simple form in the orthogonal basis. The function computing the
reflection matrices is \com{ReflectionMatrices[\args{algebra}]} and simply
applies the built-in Mathematica command \com{ReflectionMatrix} to all simple
roots and saves the result as \com{DownValues} of \linebreak \com{ReflectionMatrices[\args{algebra}]}.
E.g., the reflection matrices for \A4 (in the 5-dimensional orthogonal basis) are:
\begin{mathin}
Row[MatrixForm /@ ReflectionMatrices[A4]]
\end{mathin}
\begin{mathout}
$\begin{pmatrix}
 0 & 1 & 0 & 0 & 0 \\
 1 & 0 & 0 & 0 & 0 \\
 0 & 0 & 1 & 0 & 0 \\
 0 & 0 & 0 & 1 & 0 \\
 0 & 0 & 0 & 0 & 1 \\
\end{pmatrix}
\begin{pmatrix}
 1 & 0 & 0 & 0 & 0 \\
 0 & 0 & 1 & 0 & 0 \\
 0 & 1 & 0 & 0 & 0 \\
 0 & 0 & 0 & 1 & 0 \\
 0 & 0 & 0 & 0 & 1 \\
\end{pmatrix}
\begin{pmatrix}
 1 & 0 & 0 & 0 & 0 \\
 0 & 1 & 0 & 0 & 0 \\
 0 & 0 & 0 & 1 & 0 \\
 0 & 0 & 1 & 0 & 0 \\
 0 & 0 & 0 & 0 & 1 \\
\end{pmatrix}
\begin{pmatrix}
 1 & 0 & 0 & 0 & 0 \\
 0 & 1 & 0 & 0 & 0 \\
 0 & 0 & 1 & 0 & 0 \\
 0 & 0 & 0 & 0 & 1 \\
 0 & 0 & 0 & 1 & 0 \\
\end{pmatrix}$
\end{mathout}
The Weyl group of \A{n} is particularly simple in the orthogonal basis: It is
the symmetric group $S_{n+1}$. The reflection matrices for \A4 above represent
the generators of $S_5$, i.e., the coordinate permutations (12), (23), (34) and
(45), respectively.

Acting on a vector $x$ in root space by all elements of the Weyl group gives a
set of points, of which some may coincide. The subset of distinct points is
called the \emph{orbit} of $x$ and denoted as $O(x)$. The LieART function
\com{Orbit[\args{weightOrRoot},\args{simpleroots}]} gives the orbit of
\args{weightOrRoot} using the \args{simpleroots}. If the second argument is
omitted, all simple roots of the algebra associated with \args{weightOrRoot} are
used. The function applies \com{Reflect} in a nested fashion and removes
duplicate points in every step. The orbit of an \A{n} root or weight is
constructed in a special way for performance reasons: The \args{weightOrRoot} is
transformed to the orthogonal basis and the other points of its orbit are
constructed by permuting its coordinates using the built-in Mathematica function
\com{Permutations}. For example, the orbit of the first simple root of \A4 is
\begin{mathin}
Orbit[First[OrthogonalSimpleRoots[A4]]]
\end{mathin}
\begin{mathout}
\label{out:A4RootOrbit}
\{%
\rootorthogonal{{-}1, 0, 0, 0, 1},%
\rootorthogonal{{-}1, 0, 0, 1, 0},%
\rootorthogonal{{-}1, 0, 1, 0, 0},%
\rootorthogonal{{-}1, 1, 0, 0, 0},%
\rootorthogonal{0, {-}1, 0, 0, 1},\newline
\rootorthogonal{0, {-}1, 0, 1, 0},%
\rootorthogonal{0, {-}1, 1, 0, 0},%
\rootorthogonal{0, 0, {-}1, 0, 1},%
\rootorthogonal{0, 0, {-}1, 1, 0},%
\rootorthogonal{0, 0, 0, {-}1, 1},\newline
\rootorthogonal{0, 0, 0, 1, {-}1},%
\rootorthogonal{0, 0, 1, {-}1, 0},%
\rootorthogonal{0, 0, 1, 0, {-}1},%
\rootorthogonal{0, 1, {-}1, 0, 0},%
\rootorthogonal{0, 1, 0, {-}1, 0},\newline
\rootorthogonal{0, 1, 0, 0, {-}1},%
\rootorthogonal{1, {-}1, 0, 0, 0},%
\rootorthogonal{1, 0, {-}1, 0, 0},%
\rootorthogonal{1, 0, 0, {-}1, 0},%
\rootorthogonal{1, 0, 0, 0, {-}1}%
\}
\end{mathout}
which is in fact the \A4 root system without the zero roots.

With the same set of Weyl group generators, defined by the roots used, every
vector is uniquely associated with only one orbit. In turn every element of an
orbit allows us to generate the entire orbit by reflecting at the hyperplanes
defined by the roots. The hyperplanes divide the space into so-called \emph{Weyl
chambers}. An orbit has no more than one distinct element in every chamber and the Weyl
group permutes the chambers. The so-called \emph{dominant chamber} has elements
with only positive coordinates in the $\omega$-basis, which serves as a
definite element for the orbits associated with them. The test function
\com{DominantQ[\args{weightOrRoot}]} gives \com{True} if \args{weightOrRoot} is
in the dominant chamber and \com{False} otherwise. The dominant root of
\outref{out:A4RootOrbit} in the $\omega$-basis is
\begin{mathin}
OmegaBasis[Select[\%, DominantQ]]
\end{mathin}
\begin{mathout}
$\left\{\rootomega{1, 0, 0 ,1}\right\}$
\end{mathout}
If an orbit is created by LieART it
is saved as a \com{DownValue} of \com{Orbit} associated with its dominant root
or weight. Whenever an orbit of a non-dominant weight or root is needed, LieART
first seeks the \com{DownValue}s of \com{Orbit} for the weight or root, to see
if the orbit has already been generated. Reusing computed orbits saves CPU time
especially for Lie algebras other than \A{n} and the described procedure avoids
saving the same orbit multiple times as \com{DownValue} involving different
roots or weights.

The size of the orbit, i.e., its numbers of elements, denoted by
$\left|O(x)\right|$, is implemented as the function
\com{DimOrbit[\args{weightOrRoot},\args{simpleroots}]} or
\com{DimOrbit[\args{weightOrRoot}]} if all simple roots of the associated Lie
algebras should be used. The size of the orbit in \outref{out:A4RootOrbit} is
\begin{mathin}
DimOrbit[First[OrthogonalSimpleRoots[A4]]]
\end{mathin}
\begin{mathout}
20
\end{mathout}

\subsection{Roots}

\definition{
    \com{RootSystem[\args{algebra}]}            & root system of \args{algebra}\\
    \com{ZeroRoots[\args{algebra}]}             & zero roots associated with the Cartan subalgebra of \args{algebra}\\
    \com{Height[\args{root}]}                   & height of a \args{root} within the root system\\
    \com{HighestRoot[\args{algebra}]}           & highest root of the root system of \args{algebra}\\
    \com{PositiveRootQ[\args{root}]}            & gives \com{True} if \args{root} is a positive root and \com{False} otherwise\\
    \com{NumberOfPositiveRoots[\args{algebra}]} & number of positive roots of \args{algebra}\\
    \com{PositiveRoots[\args{algebra}]}         & gives only the positive roots of \args{algebra}\\
}{Roots}

The roots of a Lie algebra can be built from the simple roots. There are two
traditional approaches: (a)~building the roots from linear combinations of
simple roots. Since not all linear combinations of simple roots are roots, the
difficulty lies in filtering out combinations that are roots.  (b)~Starting from
a \emph{highest root} the roots can be constructed by subtracting simple roots.
LieART uses yet another approach: It builds the orbits of the simple roots by
applying the Weyl group of the Lie algebra and adds the n-fold degenerated zero
roots corresponding to the Cartan subalgebra. The simple roots of the same
length belong to the same orbit, e.g., for \A{n} there is only one orbit besides
the zero orbit (see \outref{out:A4RootOrbit}). Nevertheless, the orbits of all
simple roots are generated and then united. The fact that non-zero roots are
non-degenerate allows us to remove duplicate roots obtained by the described
procedure.

The function \com{RootSystem[\args{algebra}]} constructs the root system by the
procedure described above. As a non-trivial example we demonstrate the
procedure on \G2, which has two non-trivial orbits and the zero orbit: The two simple roots of \G2
\begin{mathin}
OmegaBasis[OrthogonalSimpleRoots[G2]]
\end{mathin}
\begin{mathout}
$\left\{\rootomega{2, {-}1},\rootomega{{-}3, 2}\right\}$
\end{mathout}
have different lengths:
\begin{mathin}
DMatrix[G2]
\end{mathin}
\begin{mathout}
$\begin{pmatrix}
    \frac{1}{3} & 0 \\
    0 & 1 \\
 \end{pmatrix}$
\end{mathout}
Generating the Weyl group orbits of each of the simple roots
\begin{mathin}
Orbit /@ OmegaBasis[OrthogonalSimpleRoots[G2]]
\end{mathin}
\begin{mathout}
$\begin{pmatrix}
\rootomega{-2, 1} & \rootomega{-1, 0} & \rootomega{-1, 1} & \rootomega{1, -1} & \rootomega{1, 0}  & \rootomega{2, -1}\\
\rootomega{-3, 1} & \rootomega{-3, 2} & \rootomega{0, -1} & \rootomega{0, 1}  & \rootomega{3, -2} & \rootomega{3, -1}\\
\end{pmatrix}$
\end{mathout}
and adding the twofold degenerated zero roots constructed by \com{ZeroRoots[\args{algebra}]}
\begin{mathin}
ZeroRoots[G2]
\end{mathin}
\begin{mathout}
$\left\{\rootomega{0, 0},\rootomega{0, 0}\right\}$
\end{mathout}
yields the full \G2 root system, displayed in spindle shape, as defined below in section~\ref{ssec:WeightSystem}
\begin{mathin}
 RootSystem[G2, SpindleShape -> True]
\end{mathin}
\begin{mathout}
\begin{minipage}{0.8in}
\begin{center}
\rootomega{0, 1}\linebreak
\rootomega{3, {-}1}\linebreak
\rootomega{1, 0}\linebreak
\rootomega{{-}1, 1}\linebreak
\rootomega{{-}3, 2}%
\rootomega{2, {-}1}\linebreak
\rootomega{0, 0}%
\rootomega{0, 0}\linebreak
\rootomega{{-}2, 1}%
\rootomega{3, {-}2}\linebreak
\rootomega{1, {-}1}\linebreak
\rootomega{{-}1, 0}\linebreak
\rootomega{{-}3, 1}\linebreak
\rootomega{0, {-}1}
\end{center}
\end{minipage}
\end{mathout}
where a row stands for the same height of the roots. The \emph{height} of a root
is defined as the sum of coefficients in its linear combination of simple roots,
i.e., the sum of coordinates in the $\alpha$-basis. It is implemented by
\com{Height[\args{root}]}. The \emph{highest root} has the largest height,
implemented in LieART as \com{HighestRoot[\args{algebra}]}, which simply returns
the first root of the root system, since the latter is sorted by the height of
the roots decreasingly. E.g. for \G2:
\begin{mathin}
HighestRoot[G2]
\end{mathin}
\begin{mathout}
\rootomega{0, 1}
\end{mathout}

The \emph{positive roots} are the roots that are only positive linear
combinations of simple roots, i.e., the coordinates in the $\alpha$-basis are
all positive, with at least one being non-zero, to exclude the zero roots. The
function \com{PositiveRootQ[\args{root}]} tests if \args{root} is positive. The
\args{root} may be in any basis and will be transformed into the $\alpha$-basis,
where its coordinates are tested accordingly. The number of positive roots are
explicitly stated as \com{NumberOfPositiveRoots[\args{algebra}]} in LieART. It
serves as a limiter to the nested reflections for the generation of Weyl group
orbits. There is a theorem stating that the maximum number of reflections
building an element of the Weyl group is equal to the number of positive roots of
the corresponding Lie algebra.

Since the root system is sorted by height, the positive roots come first.
\com{PositiveRoots[\args{algebra}]} extracts only those with the use of
\com{NumberOfPositiveRoots[\args{algebra}]}. E.g., for \G2:
\begin{mathin}
PositiveRoots[G2]
\end{mathin}
\begin{mathout}
$\left\{\rootomega{0, 1},\rootomega{3, -1},\rootomega{1, 0},\rootomega{-1, 1},\rootomega{-3, 2},\rootomega{2, -1}\right\}$
\end{mathout}

\subsection{Representations}

\definition{
    \com{WeightOrthogonal[\args{algebraClass}][\args{label}]}  & weight in the orthogonal basis defined by its algebra \args{algebraClass} and Dynkin \args{label}\\
    \com{WeightAlpha[\args{algebraClass}][\args{label}]}  & weight in the $\alpha$-basis defined by its algebra \args{algebraClass} and Dynkin \args{label}\\
    \com{Weight[\args{algebraClass}][\args{label}]}  & weight in the $\omega$-basis defined by its algebra \args{algebraClass} and Dynkin \args{label}\\
    \com{Irrep[\args{algebraClass}][\args{label}]}  & irrep described by its algebra \args{algebraClass} and Dynkin \args{label}\\
    \com{WeightLevel[\args{weight},\args{irrep}]}          & Level of the \args{weight} within the \args{irrep}\\
    \com{Height[\args{irrep}]}                      & height of \args{irrep}\\
    \com{SingleDominantWeightSystem[\args{irrep}]}          & dominant weights of \args{irrep} without their multiplicities\\
    \com{WeightMultiplicity[\args{weight},\args{irrep}]}   & computes the multiplicity of \args{weight} within \args{irrep}\\
    \com{DominantWeightSystem[\args{irrep}]}                & dominant weights of \args{irrep} with their multiplicities\\
    \com{WeightSystem[\args{irrep}]}                        & full weight system of \args{irrep}\\
    \com{Irrep[\args{algebra}][\args{dimname}]}     & irrep entered by its \args{algebra} and \args{dimname}\\
    \com{ProductIrrep[\args{irreps}]}               & head of product \args{irreps}\\
    \com{Delta[\args{algebra}]}                     & half the sum of positive roots of \args{algebra} ($\delta{=}\dynkincomma{1,1,\ldots}$)\\
    \com{WeylDimensionFormula[\args{algebra}]}      & explicit Weyl dimension formula for \args{algebra}\\
    \com{Dim[\args{irrep}]}                         & dimension of \args{irrep}\\
    \com{DimName[\args{irrep}]}                     & dimensional name of \args{irrep}\\
    \com{Index[\args{irrep}]}                       & index of \args{irrep}\\
    \com{CongruencyClass[\args{irrep}]}             & congruency class number of \args{irrep}\\
}{Basic Properties of Irreps}

As explained in Section \ref{ssec:Representations} a \emph{representation} is a set of
matrices that satisfies the same commutation relations as the algebra. Each of
the matrices can be labeled by the \emph{weight vector} with the eigenvalues of
the matrices corresponding to the generators of the Cartan subalgebra, and we
will refer to the weight vector simply as \emph{weight}. The weight vector has
the dimension of the Cartan subalgebra, i.e., the rank of the algebra, and not
the dimension of the space the matrices act on. The latter depends on the
particular representation.

The weights $\lambda$ can be written as linear combination of simple roots and a
crucial theorem states that the so-called \emph{Dynkin labels} $a_i$ defined as
\begin{equation}
    a_i = \frac{2\scalarproduct{\lambda}{\alpha_i}}{\scalarproduct{\alpha_i}{\alpha_i}},\qquad i=1,\ldots,n
\end{equation}
are integers for all simple roots $\alpha_i$. (Please note that this is also
true if $\lambda$ is replaced by any simple root, since this constitutes an
element of the Cartan matrix as defined in \eqref{eq:CartanMatrix}.) The Dynkin
labels are in particular used to label weights (and roots). The smallest
non-zero weights with $a_i\geq 0$ are called the \emph{fundamental weights}
$\omega_i$. They define the $\omega$-basis or Dynkin basis already introduced.
They are implemented in LieART as
\com{OrthogonalFundamentalWeights[\args{algebra}]} in the orthogonal basis and
we have given an example for \A4 in \outref{out:FundamentalWeightsOmegaBasis}.
The Dynkin labels $a_i$ of a weight $\lambda$ are the coefficients of its linear
combination of fundamental weights, i.e., the $a_i$ are the coordinates in the
$\omega$-basis, which can be displayed as a row vector with comma separated
entries or as a framed box following the convention of some textbooks:
\begin{equation}
    \lambda = \sum_{i=1}^n a_i\omega_i
            = \dynkincomma{a_1, a_2, \ldots, a_n}
            = \weight{a_1, a_2, \ldots, a_n}.
\end{equation}

A weight in LieART is represented by three different heads, depending on its
basis, in analogy with the roots: \com{WeightOrthogonal[\args{algebraClass}][\args{label}]}
for a weight in the orthogonal basis, in the $\alpha$-basis
\com{WeightAlpha[\args{algebraClass}][\args{label}]}  and
simply \com{Weight[\args{algebraClass}][\args{label}]} in the $\omega$-basis,
where we omit the explicit ``\com{Omega}'' for brevity, because the
$\omega$-basis is the natural basis for weights. (The same can be said for the
$\alpha$-basis for roots, favoring the shorter head \com{Root} instead of
\com{RootAlpha} in the $\alpha$-basis. Unfortunately this would clash with the
built-in Mathematica function \com{Root[\args{f},\args{k}]} representing the
\args{k}th root of a polynomial equation defined by $f[x]=0$.) The
\args{algebraClass} can only be \com{A}, \com{B}, \com{C} or \com{D} to indicate
a classical Lie algebra or \com{E6}, \com{E7}, \com{E8}, \com{F4} or \com{G2}
for the exceptionals. The \args{label} stands for the comma-separated
coordinates. This form of the weight is displayed in \com{InputForm} and
\com{FullForm}. E.g., the first fundamental weight of \A4 in all three bases
reads:
\begin{mathin}
\{\slot,AlphaBasis[\slot],OmegaBasis[\slot]\}\&\newline
@First[OrthogonalFundamentalWeights[A4]]//InputForm
\end{mathin}
\begin{mathout}
\{WeightOrthogonal[A][4/5,{-}1/5,{-}1/5,{-}1/5,{-}1/5], WeightAlpha[A][4/5,3/5,2/5,1/5], Weight[A][1,0,0,0]\}
\end{mathout}

Since weights are linear combinations of roots, many properties of roots
translate to weights. The Weyl group also applies to weights and the weight space is
also divided into Weyl chambers. A weight with only positive coordinates lies in
the dominant Weyl chamber and is called a \emph{dominant weight}. In analogy
with the highest root, every irreducible representation (irrep) has a
non-degenerate \emph{highest weight}, denoted as $\Lambda$, which is also a
dominant weight, but not necessarily the only dominant weight of the irrep. The
weight system of the irrep can be computed from the highest weight $\Lambda$ by
subtracting simple roots. Thus, a highest weight $\Lambda$ uniquely defines the
irrep, and since a particular Lie algebra has infinitely many irreps, it serves
as a label for the irrep itself using the same denotation, $\Lambda$.

In LieART an irrep is represented by
\com{Irrep[\args{algebraClass}][\args{label}]}, where \args{algebraClass}
defines the Lie algebra class in the same manner as for weights and roots, and
\args{label} is the comma-separated label of the highest weight of the irrep.
E.g., the 10 dimensional irrep of \A4 has the highest weight
\dynkincomma{0,1,0,0} and thus the irrep can be entered as
\com{Irrep[A][0,1,0,0]}.

The so-called \emph{Dynkin label} of an irrep is similar to the notation of a
weight, but since the highest weight has only positive label entries the commas
between them can be ommitted, as long as this is unambiguous. The Dynkin label
in LieART is displayed in \com{StandardForm}, e.g., the \irrep{10} of \A4:
\begin{mathin}
Irrep[A][0,1,0,0]//StandardForm
\end{mathin}
\begin{mathout}
\dynkin{0,1,0,0}
\end{mathout}
If at least one of the entries in the Dynkin labels has more than a single
digit, all entries are separated by commas to avoid ambiguities, which is the
standard textbook convention:
\begin{mathin}
Irrep[A][0, 10, 3, 1] // StandardForm
\end{mathin}
\begin{mathout}
\dynkincomma{0,10,3,1}
\end{mathout}

\subsubsection{Weight System}
\label{ssec:WeightSystem}

The conventional approach to computing all weights of an irrep is to subtract
simple roots from the highest weight $\Lambda$ that defines the irrep. The Dynkin
label of the highest weight $\Lambda=\dynkincomma{a_1, a_2, \ldots, a_n}$
reveals how many times each simple root can be subtracted: the $i$th root can be
subtracted $a_i$ times. The \emph{level} of a weight is the number of simple
root that need to be subtracted from the highest weight to obtain it. A weight
may be obtained by different subtraction routes, but it always involves the same
number of simple roots, thus its level is unique. As explained earlier, the
$\alpha$-basis exhibits the coefficients of the linear combination of simple
roots, which are rational numbers in general. The difference between these
coefficients of the weight and the highest weight show how many times each
simple roots has been subtracted from the latter. The sum over these
differences, for each simple root, gives the level of the weight:
\begin{equation}
    L(\lambda, \Lambda) = \sum_{i=1}^n \left(\bar{\Lambda}_i-\bar{\lambda}_i\right)
\end{equation}
where $\lambda=\dynkincomma{\bar{\lambda}_1,\ldots,\bar{\lambda}_n}$ and
$\Lambda=\dynkincomma{\bar{\Lambda}_1,\ldots,\bar{\Lambda}_n}$ is the weight and
highest weight in the $\alpha$-basis, respectively. The LieART function
\com{WeightLevel[\args{weight},\args{irrep}]} implements this procedure. The
highest level of an irrep is called its \emph{height}. Please note that the
highest weight has the lowest level, which is zero. The weight with the highest
level has the coefficients of the highest weight in the alpha basis, with
negative sign and rearranged (if the irrep is complex), i.e., the sum over them
is the negative of the sum of the highest weight. Thus, the height of an irrep
with highest weight $\Lambda$ is
\begin{equation}
    \height(\Lambda) = 2 \sum_{i=1}^n \bar{\Lambda}_i,
\end{equation}
which is available in LieART as \com{Height[\args{irrep}]}.

The algorithm to compute the weight system used in LieART is an implementation 
of the scheme developed in \cite{Moody:1982}. It deviates from the traditional 
procedure described above for performance reasons. Some weights of an irrep may 
be degenerated and the procedure with subtracting simple roots only yields an 
upper limit of this degeneracy, which is the number of subtraction routes that 
lead to a weight. To compute the \emph{multiplicity} $m_\lambda$ of a weight 
$\lambda$ of the irrep with highest weight $\Lambda$, the so-called 
\emph{Freudenthal recursion formula} is usually used:
\begin{equation}\label{eq:FreudenthalFormula}
    2\sum_{\alpha\in\Delta^{\!+}\!}\,\sum_{k\geq0}\scalarproduct{\lambda{+}k\alpha}{\alpha}m_{\lambda+k\alpha}
    = \left[\scalarproduct{\Lambda{+}\delta}{\Lambda{+}\delta}-\scalarproduct{\lambda{+}\delta}{\lambda{+}\delta}\right]m_\lambda
\end{equation}
where $\Delta^{\!+}$ denotes the positive root system,
$\delta{=}\dynkincomma{1,1,\ldots,1}$ is half the sum of all positive roots in
the $\omega$-basis and $m_{\lambda+k\alpha}$ is the already computed
multiplicity of a weight $\lambda+k\alpha$ that is higher than $\lambda$. The
sum over $k$ is finite because the weight $\lambda+k\alpha$ must be a member of
the weight system of the irrep under consideration.

The recursive nature of the Freudenthal formula makes the computation of weight
multiplicities the most CPU time consuming procedure in the
determination of the weight system of an irrep. The algorithm developed in
\cite{Moody:1982} exploits the Weyl group in both the weight and the
root system. The weight system of an irrep is a collection of Weyl group orbits,
represented by their unique dominant weight. The multiplicity of the dominant weight is
the same for all weights of the associated orbit. Thus, a weight system can be
constructed by (a) determining the dominant weights of the irrep, (b) computing
their multiplicity and (c) generating the orbits of the dominants weights with the
same multiplicity by application of the Weyl group of the associated algebra.

In LieART the function \com{SingleDominantWeightSystem[\args{irrep}]} determines
the dominant weights of \args{irrep} by successively subtracting positive roots
starting from the highest weight and keeping only the dominant weight of the
result in every step. This process terminates, because there are smallest
dominant weights, i.e., the fundamental weights, constituting a lower boundary.
E.g., the \irrep{40} of \A4 has two distinct dominant weights:
\begin{mathin}
SingleDominantWeightSystem[Irrep[A][0, 0, 1, 1]]
\end{mathin}
\begin{mathout}\label{out:SingleDominantWeights}
\{\weight{0, 0, 1, 1},\weight{0, 1, 0, 0}\}
\end{mathout}

Thus, an improved version of the Freudenthal formula considers only dominant
weights $\lambda$. The second exploitable property is the existence of a
stabilizer of the weight, a subgroup of its Weyl group $W$ that fixes the
weight:
\begin{equation}\label{eqn:DefinitionStabilizer}
\text{Stab}_W(\lambda) = W_T := \left\{w\in W\,|\: w\lambda=\lambda\right\}.
\end{equation}
The stabilizer $W_T$ reduces the number of independent scalar products and
previous computed multiplicities, because for $w\in W_T$,
$\scalarproduct{\lambda{+}k\alpha}{\alpha}{=}\scalarproduct{w(\lambda{+}
k\alpha_i)}{w\alpha_i}{=}\scalarproduct{(\lambda{+}kw\alpha_i)}{w\alpha_i}$ and
$m_{\lambda+k\alpha_i}{=}m_{w(\lambda+k\alpha_i)}{=}m_{\lambda+kw\alpha_i}$.
Since the elements of the Weyl group $W$ are reflections at simple roots, the
stabilizer group is defined by the reflection at simple roots that map $\lambda$
onto itself. Because of \eqref{eq:WeylReflection} this is the case, when
$\scalarproduct{\lambda}{\alpha_i}=0$. If $\lambda$ is expressed in the
$\omega$-basis as $\lambda{=}\sum n_i\omega_i$ the scalar product with the $i$th
simple root is zero, when $n_i=0$. Let $T$ be a set of these indices, i.e.,
$T{=}\left\{i\,|\: n_i{=}0\right\}$, and let $\Delta_T$ be the root system based
on the simple roots labeled by $T$.

The group $\hat{W}_T$, which is the inclusion of $W_T$ and the negative
identity element $w{=}{-}1$ as $\hat{W}_T{=}\left<W_T,{-}1\right>$, decomposes the root
system into orbits $o_1,\ldots,o_r$, defined by $\hat{W}_T\alpha_i$. Each
orbit has a unique representative $\xi_i$ in the positive roots
($\xi_i{\in}\Delta^{\!+}$). The $\xi_i$'s are those positive roots, that have
non-zero coefficients in the $\omega$-basis at the positions where $\lambda$
has zeros, i.e., $\xi_i{=}\sum m_i\omega_i$ with $m_i{\geq}0$ for $i{\in}T$.

The computation of the multiplicity $m_\lambda$ of the dominant weight $\lambda$
is then accomplished by the \emph{modified Freudenthal formula}:
\begin{equation}
    \sum_{i=1}^n \left|o_i\right|\sum_{k=1}^\infty\scalarproduct{\lambda{+}k\xi_i}{\xi_i}m_{\lambda+k\xi_i}
    = \left[\scalarproduct{\Lambda{+}\delta}{\Lambda{+}\delta}-\scalarproduct{\lambda{+}\delta}{\lambda{+}\delta}\right]m_\lambda
\end{equation}
where $\left|o_i\right|$ are the sizes of the orbits. It is important to note
that these sizes are $\left|o_i\right|{=}\left|W_T\xi_i\right|$ if
$\xi_i{\in}\Delta_T$ and $\left|o_i\right|{=}2\left|W_T\xi_T\right|$ if
$\xi_i{\notin}\Delta_T$, because in the former case the negative roots are
included in $W_T\xi_i$, i.e., the $-1$, while only positive roots are in the
orbit $W_T\xi_i$ if $\xi_i{\notin}\Delta_T$, requiring a factor of 2 for the
same reason as on the left-hand side of \eqref{eq:FreudenthalFormula}. It is
$\xi_i{\in}\Delta_T$ if $\scalarproduct{\lambda}{\xi_i}{=}0$.

The higher weight $\lambda+k\xi_i$ is not necessarily a dominant weight, but can
always be reflected to the dominant chamber to obtain the corresponding
multiplicity that is already computed.

The computation of weight multiplicities is implemented in LieART as
\com{WeightMultiplicity[\args{weight},\args{irrep}]} following the above
algorithm, using several helper functions: \com{T[\args{weight}]} gives the
set $T$, the positions of zeros of the coefficients of \args{weight} in the
$\omega$-basis, \com{Xis[\args{algebra},\args{t}]} determines the $\xi$'s based
on the set $T$ which should be supplied via the argument \args{t},
\com{Alphas[\args{algebra},\args{t}]} gives $\alpha_i$ with $i\in T$ to
construct the orbit $W_T\xi_i$. \com{XisAndMul[\args{algebra},\args{t}]} yields
a list of the $\xi_i$'s together with their associated orbit size
$\left|o_i\right|$. Since the possible subsets $T$ of zeros in the weight
coefficients for a specific algebra are limited, we follow the suggestion of
\cite{Moody:1982} and save this in list form as \com{XisAndMul} for reuse
upon first evaluation in the course of a calculation. Saved values of
\com{XisAndMul} can be retrieved by \com{Definition[XisAndMul]}.

Take for example the dominant weight \weight{0, 1, 0, 0} of the \irrep{40} of
\A4 from \outref{out:SingleDominantWeights}. The set of indices $T$ is
\begin{mathin}
    T[Weight[A][0,1,0,0]]
\end{mathin}
\begin{mathout}
    $\begin{pmatrix} 1 \\ 3 \\ 4 \end{pmatrix}$
\end{mathout}
or in \com{InputForm}: \com{\{\{1\},\{3\},\{4\}\}}. (The structure as ``list of
lists'' is due to the use of the Mathematica built-in functions \com{Position}
and \com{Extract}.) The $\xi_i$'s and the sizes of their associated orbits
$|o_i|$ are
\begin{mathin}
    XisAndMul[A4, T[Weight[A][0,1,0,0]]]
\end{mathin}
\begin{mathout}
    $\begin{pmatrix}
         \weight{1, 0, 0, 1} & 12 \\
         \weight{0, {-}1, 1, 1} & 6 \\
         \weight{2, {-}1, 0, 0} & 2 \\
     \end{pmatrix}$
\end{mathout}
and the weight multiplicity of the dominant weight in the \irrep{40} of \A4 is:
\begin{mathin}
    WeightMultiplicity[Weight[A][0,1,0,0], Irrep[A][0,0,1,1]]
\end{mathin}
\begin{mathout}
    2
\end{mathout}

The LieART function \com{DominantWeightSystem[\args{irrep}]} gives a list of the
dominant weights of \args{irrep} along with their multiplicities:
\begin{mathin}
DominantWeightSystem[Irrep[A][0,0,1,1]]
\end{mathin}
\begin{mathout}
        $\begin{pmatrix}
         \weight{0, 0, 1, 1} & 1 \\
         \weight{0, 1, 0, 0} & 2 \\
     \end{pmatrix}$
\end{mathout}
Thus, the weight system of the \irrep{40} consists of the Weyl group orbit of
\weight{0, 0, 1, 1} and \weight{0, 1, 0, 0}, where every weight of the latter
has a multiplicity of two. The function \com{WeightSystem[\args{irrep}]}
generates these orbits based on \com{DominantWeightSystem} and explicitly
duplicates weights according to the multiplicity. If no further processing is
intended the option \com{SpindleShape->True} turns the output into a spindle
shape display, with weights of the same level in one row. E.g., the weight
system of the \irrep{40} of \A4 in spindle shape is
\begin{mathin}
    WeightSystem[Irrep[A][0,0,1,1]]
\end{mathin}
\begin{mathout}
    \begin{minipage}{4.4in}
        \begin{center}
            \weight{0, 0, 1, 1}\linebreak
            \weight{0, 0, 2, {-}1}\weight{0, 1, {-}1, 2}\linebreak
            \weight{0, 1, 0, 0}\weight{0, 1, 0, 0}\weight{1, {-}1, 0, 2}\linebreak
            \weight{{-}1, 0, 0, 2}\weight{0, 1, 1, {-}2}\weight{0, 2, {-}2, 1}\weight{1, {-}1, 1, 0}\weight{1, {-}1, 1, 0}\linebreak
            \weight{{-}1, 0, 1, 0}\weight{{-}1, 0, 1, 0}\weight{0, 2, {-}1, {-}1}\weight{1, {-}1, 2, {-}2}\weight{1, 0, {-}1, 1}\weight{1, 0, {-}1, 1}\linebreak
            \weight{{-}1, 0, 2, {-}2}\weight{{-}1, 1, {-}1, 1}\weight{{-}1, 1, {-}1, 1}\weight{1, 0, 0, {-}1}\weight{1, 0, 0, {-}1}\weight{2, {-}2, 0, 1}\linebreak
            \weight{{-}1, 1, 0, {-}1}\weight{{-}1, 1, 0, {-}1}\weight{0, {-}1, 0, 1}\weight{0, {-}1, 0, 1}\weight{1, 1, {-}2, 0}\weight{2, {-}2, 1, {-}1}\linebreak
            \weight{{-}2, 0, 0, 1}\weight{{-}1, 2, {-}2, 0}\weight{0, {-}1, 1, {-}1}\weight{0, {-}1, 1, {-}1}\weight{2, {-}1, {-}1, 0}\linebreak
            \weight{{-}2, 0, 1, {-}1}\weight{0, 0, {-}1, 0}\weight{0, 0, {-}1, 0}\linebreak
            \weight{{-}2, 1, {-}1, 0}\weight{1, {-}2, 0, 0}\linebreak
            \weight{{-}1, {-}1, 0, 0}\linebreak
        \end{center}
    \end{minipage}
\end{mathout}

\subsubsection{Properties of Irreducible Representations}

\paragraph{Dimension}
The dimension of an irrep, i.e., the number of its weights, can be calculated
without explicitly generating the weight system. The \emph{Weyl dimension
formula}, which is a special case of  Weyl's character formula, gives the
dimension of an irrep in terms of its highest weight $\Lambda$, positive roots
$\alpha{\in}\Delta^{\!+}$ and $\delta{=}\dynkincomma{1,1,\ldots,1}$:
\begin{equation}
\dim(\Lambda) = \prod_{\alpha\in\Delta^{\!+}\!}\frac{\scalarproduct{\alpha}{\Lambda + \delta}}{\scalarproduct{\alpha}{\delta}}.
\end{equation}
It is implemented in LieART as \com{Dim[\args{irrep}]}. E.g., the dimension of
the \irrep{40} of \A4 can be obtained by
\begin{mathin}
    Dim[Irrep[A][0,0,1,1]]
\end{mathin}
\begin{mathout}
    40
\end{mathout}
The Dynkin label of \args{irrep} does not need to be numerical. By using
variables the simple structure of the formula becomes explicit, e.g., for a
general irrep of \A4:
\begin{mathin}
    Dim[Irrep[A][a,b,c,d]]
\end{mathin}
\begin{mathout}
    $\displaystyle\frac{1}{288}(a+1)(b+1)(c+1)(d+1)(a+b+2)(b+c+2)(c+d+2)(a+b+c+3)(b+c+d+3)(a+b+c+d+4)$
\end{mathout}
Internally, the Weyl dimension formula is computed by
\com{WeylDimensionFormula[\args{algebra}]} as a pure function with the digits of
the Dynkin label as parameters. E.g. for \A4:
\begin{mathin}
    WeylDimensionFormula[A4]//InputForm
\end{mathin}
\begin{mathout}
    Function[\{a1,a2,a3,a4\},((1{+}a1){*}(1{+}a2){*}(1{+}a3){*}(1{+}a4){*}(2{+}a1{+}a2){*}(2{+}a2{+}a3)\newline
    {*}(2{+}a3{+}a4){*}(3{+}a1{+}a2{+}a3){*}(3{+}a2{+}a3{+}a4){*}(4{+}a1{+}a2{+}a3{+}a4))/288]
\end{mathout}
The function \com{Dim[\args{irrep}]} acts only as a wrapper applying the pure
function to the specified Dynkin label in the argument of \com{Dim}.

\paragraph{Index}
Another important property of an irrep $\Lambda$ is its \emph{index}, denoted as
$l(\Lambda)$, which is an eigenvalue of the \emph{Casimir invariant} normalized
to be an integer:
\begin{equation}
    l(\Lambda) = \frac{\text{dim}(\Lambda)}{\text{ord}(L)}\scalarproduct{\Lambda}{\Lambda+2\delta},
\end{equation}
where $\text{ord}(L)$ is the order of the Lie algebra $L$, which is equivalent to
the number of roots or the dimension of the adjoint irrep. The index is related
to the length of the weights and has applications in renormalization group
equations and elsewhere. The corresponding LieART function is
\com{Index[\args{irrep}]}. E.g., the index of the \irrep{40} of \A4 is:
\begin{mathin}
    Index[Irrep[A][0,0,1,1]]
\end{mathin}
\begin{mathout}
    22
\end{mathout}
The label of \args{irrep} does not need to be numerical, similar to \com{Dim}.

\paragraph{Congruency Class}
The \emph{congruency class} expands the concept of $n$-ality of \SU{N}, which in
turn is a generalization of \SU3 triality, to all other simple Lie
algebras. LieART uses congruency classes to characterize irreps, especially for
the distinction of irreps of the same dimension and with the same index. We
follow the definitions of \cite{McKay:99021,lemire_congruence_1980}, labeling
the congruency class by the \emph{congruency number}, which is a single number
for \A{n}, \B{n}, \C{n}, \E6, \E7, \E8, \F4 and \G2 and a two component vector
for \D{n}. For an irreducible representation \dynkin{a_1a_2\ldots a_n} the
congruency number (vector) $c$ is:
\begin{subequations}
\begin{align}
         \A{n} &:\quad  c = \sum_{k=1}^n k a_k\ (\text{mod}\ n+1)\\
         \B{n} &:\quad  c = a_n\ (\text{mod}\ 2)\\
         \C{n} &:\quad  c = \sum_{k=0}^{\floor{\frac{n-1}{2}}} a_{2k+1}\ (\text{mod}\ 2)\\
         \D{n} &:\quad  c = \begin{pmatrix}
                                  a_{n-1}+a_n\ (\text{mod}\ 2) \\
                                  2\sum_{k=0}\limits^{\floor{\frac{n-3}{2}}} a_{2k+1} + (n-2)a_{n-1}+n a_n\ (\text{mod}\ 4)
                           \end{pmatrix}&&\\
           \E6 &:\quad  c = a_1 - a_2 + a_4 - a_5 \ (\text{mod}\ 3)\\
           \E7 &:\quad  c = a_4 + a_6 + a_7\ (\text{mod}\ 2)\\
    \E8,\F4,\G2&:\quad  c = 0
\end{align}
\end{subequations}
Please note that the congruency class definitions of
\cite{McKay:99021,lemire_congruence_1980}, which we use, differ from
\cite{Slansky} for the \D{n}'s : For \D4, i.e.\ \SO8, the second component of
\cite{Slansky} is half of the definition above. The congruency class for \D5 is
only a single number in \cite{Slansky}, which is the same as the second
component of the \D5 congruency class vector of our definition from
\cite{McKay:99021,lemire_congruence_1980} (with no factor of 2).

These congruency numbers (vectors) $c$ are implemented as
\com{CongruencyClass[\args{irrep}]} in LieART. E.g., the congruency numbers of
the three eight dimensional irreps of \SO8 are all distiguished by their
congruency number vector:
\begin{mathin}
    CongruencyClass[\{Irrep[D][1,0,0,0], Irrep[D][0,0,1,0], Irrep[D][0,0,0,1]\}]
\end{mathin}
\begin{mathout}
    \{(02),(12),(10)\}
\end{mathout}
The head of a congruency vector is \com{CongruencyVector} and it is displayed as
$(c_1c_2)$, i.e. without commas separating the two components similar to the
Dynkin label of irreps.

\subsubsection{Representation Names}
\label{ssec:irrepNames}

The Dynkin label of an irreducible representation together with its Lie algebra
uniquely specifies it, e.g., \dynkin{0,0,1,1} of \A4. An irrep in LieART is also
represented (\com{FullForm}) by the Dynkin label and a Mathematica head that
indicates the algebra class. E.g., the irrep \dynkin{0,0,1,1} of \A4 is
represented by \com{Irrep[A][0,0,1,1]} in LieART.

However, it is common practice to name representations by their dimension, the
\emph{dimensional name}, which is often times shorter. The dimension of a
representation is not unique, i.e., there are different irreps with the same
dimensions, which might be accidental or because of a relation between them. If
it is accidental, irreps with the same dimension have primes ($\text{\bf
dim}^\prime$) in their dimensional name, e.g., the \irrep[1]{175} of \A4 is
unrelated to the \irrep{175}. Irreps can be related by conjugation, when they
are complex. One of the irreps is written with an overbar ($\irrepbar{dim}$).
E.g., the \irrepbar{10} of \A4 is the conjugate of the \irrep{10}. Due to the
high symmetry of \SO8 irrep, more than two related irreps of the same dimension
exist. In the case of \SO8 subscripts specify the irreps completely.

The introduced properties of representations, the dimension, the index and the
congruency class serve us well to discriminate between irreps with the same
dimension. LieART has an algorithm implemented that determines the dimensional
name of an irrep, following the naming conventions of \cite{Slansky}:
\begin{enumerate}
    \item
        To determine the dimensional name of a specific irrep, LieART collects
        other irreps of the same dimensionality by brute-force scanning through a
        generated set of irreps.
    \item
        Irreps that are related by conjugation or the symmetry of \SO8 not only
        have the same dimension, but also the same index. \emph{Un}related irreps of
        equal dimension have different indices and can be organized and labeled by their
        indices. They are sorted by ascending index and labeled with primes ($\text{\bf
        dim}^\prime$) accordingly, starting with no prime. E.g., the names of the two
        unrelated 70 dimensional irreps of \A4 are (the congruency class of \A4 is
        called ``Quintality''):
        \begin{center}
            \begin{tabular}{lllll}
                \toprule
                \textbf{Dynkin}     & \textbf{Dimension} & \textbf{Index} & \textbf{Quintality} & \textbf{Name}  \\
                \midrule
                \dynkin{2, 0, 0, 1} & 70                 & 49             & 1                   & \irrep{70}    \\
                \dynkin{0, 0, 0, 4} & 70                 & 84             & 1                   & \irrep[1]{70} \\
                \bottomrule
            \end{tabular}
        \end{center}
    \item
        Related irreps of the same dimensionality have the same index, but
        mostly (see below) different congruency class numbers. For Lie algebras other
        than \SO8, only the conjugate of complex irreps are related. The convention here
        is that the irrep with \emph{higher} congruency class number of the conjugated
        pair is labeled with an overbar ($\irrepbar{dim}$). Since e.g. the \irrep[1]{70}
        is a complex irrep it has a related conjugated irrep, the \irrepbar[1]{70},
        i.e., overbars and primes may both appear in the labeling of an irrep. The above
        table for the determination of the primes involves only the lower congruency
        class number irreps of same dimensional and same index irreps. Consider the
        \irrep[1]{70} and its conjugate, the \irrepbar[1]{70}:
        \begin{center}
            \begin{tabular}{lllll}
                \toprule
                \textbf{Dynkin}     & \textbf{Dimension}& \textbf{Index} & \textbf{Quintality}  & \textbf{Name}  \\
                \midrule
                \dynkin{0, 0, 0, 4} & 70                & 84             & 1                    & \irrep[1]{70} \\
                \dynkin{4, 0, 0, 0} & 70                & 84             & 4                    & \irrepbar[1]{70} \\
                \bottomrule
            \end{tabular}
        \end{center}
        If the congruency class number of a complex irrep is zero, it conjugate
        also has a congruency class number of zero. In this case, where all three, the
        dimension, the index and the congruency class number are the same, the structure
        of the Dynkin labels are consulted: With the Dynkin label interpreted as digits
        of an integer number, the \emph{smaller} ``number'' is labeled with the overbar.
        E.g., the 126 dimensional irreps of \A4 are
        \begin{center}
            \begin{tabular}{lllll}
                \toprule
                \textbf{Dynkin}     & \textbf{Dimension}& \textbf{Index} & \textbf{Quintality}  & \textbf{Name}     \\
                \midrule
                \dynkin{2, 0, 1, 0} & 126               & 105            & 0                    & \irrep{126}       \\
                \dynkin{0, 1, 0, 2} & 126               & 105            & 0                    & \irrepbar{126}    \\
                \dynkin{5, 0, 0, 0} & 126               & 210            & 0                    & \irrep[1]{126}    \\
                \dynkin{0, 0, 0, 5} & 126               & 210            & 0                    & \irrepbar[1]{126} \\
                \bottomrule
            \end{tabular}
        \end{center}
        Observe that this rule only applies for zero congruency class number:
        The \irrepbar[1]{70} has a ``larger'' number \dynkin{4, 0, 0, 0} than the
        \irrep[1]{70} with \dynkin{0, 0, 0, 4}.
    \item
        For \SO8 irreps the convention for the labeling with primes are the same
        as for all other Lie algebras. Due to the three-fold symmetry most irreps come
        in sets of three with the same dimension and index. If only one digit of the
        Dynkin label is non-zero it is called the spinor, vector or conjugate irrep,
        depending on the dot in the Dynkin diagram it corresponds to. Usually they can
        be distinguished by the congruency class number, which is a two component vector
        for \SO8: (01) for a vector irrep, (10) for a spinor and (11) for the conjugate.
        The irrep is then labeled by the subscripts ``v'', ``s'' and ``c'', resp. E.g.,
        the three 8 dimensional irreps of \SO8 are
        \begin{center}
            \begin{tabular}{lllll}
                \toprule
                \textbf{Dynkin}     & \textbf{Dimension}& \textbf{Index} & \textbf{Congruency vector}  & \textbf{Name}  \\
                \midrule
                \dynkin{1, 0, 0, 0} & 8                 & 1             & (01)                    & \irrepsub{8}{v} \\
                \dynkin{0, 0, 0, 1} & 8                 & 1             & (10)                    & \irrepsub{8}{s} \\
                \dynkin{0, 0, 1, 0} & 8                 & 1             & (11)                    & \irrepsub{8}{c} \\
                \bottomrule
            \end{tabular}
        \end{center}
        Some irreps with more than one non-zero digit of the Dynkin label with
        the same congruency vectors as above are labeled the same way if they are
        unique. However, if there is more than one irrep with the same dimension, index
        and also congruency vector, there is more than one digit of the Dynkin label
        non-zero. The subscript label is then a mixture like ``sv'', and the ordering is
        determined by the Dynkin digit beginning with the largest. E.g. the 224
        dimensional irreps of \SO8:
        \begin{center}
            \begin{tabular}{lllll}
                \toprule
                \textbf{Dynkin}     & \textbf{Dimension}& \textbf{Index} & \textbf{Congruency vector}  & \textbf{Name}  \\
                \midrule
                \dynkin{2, 0, 1, 0} & 224 & 100 & (12) & \irrepsub{224}{vc} \\
                \dynkin{2, 0, 0, 1} & 224 & 100 & (10) & \irrepsub{224}{vs} \\
                \dynkin{1, 0, 2, 0} & 224 & 100 & (02) & \irrepsub{224}{cv} \\
                \dynkin{1, 0, 0, 2} & 224 & 100 & (02) & \irrepsub{224}{sv} \\
                \dynkin{0, 0, 2, 1} & 224 & 100 & (10) & \irrepsub{224}{cs} \\
                \dynkin{0, 0, 1, 2} & 224 & 100 & (12) & \irrepsub{224}{sc} \\
                \bottomrule
            \end{tabular}
        \end{center}
        There are also cases where the congruency vector is zero in both
        components for all irreps of the same dimension and index. In this case
        subtracting the same integer from every Dynkin digit to obtain irreps with
        non-zero congruency class vector has proven to be a reliable way to label the
        irreps. E.g., the 35 dimensional irreps can be related to the 8 dimensional ones
        and, e.g., the primed 840 dimensional irreps to the 56 dimensional ones:
        \begin{center}
            \begin{tabular}{lllll}
                \toprule
                \textbf{Dynkin}     & \textbf{Dimension}& \textbf{Index} & \textbf{Congruency vector}  & \textbf{Name}  \\
                \midrule
                \dynkin{1, 0, 0, 0} & 8   & 1   & (02) & \irrepsub{8}{v}      \\
                \dynkin{0, 0, 1, 0} & 8   & 1   & (12) & \irrepsub{8}{c}      \\
                \dynkin{0, 0, 0, 1} & 8   & 1   & (10) & \irrepsub{8}{s}      \\
                \dynkin{2, 0, 0, 0} & 35  & 10  & (00) & \irrepsub{35}{v}     \\
                \dynkin{0, 0, 2, 0} & 35  & 10  & (00) & \irrepsub{35}{c}     \\
                \dynkin{0, 0, 0, 2} & 35  & 10  & (00) & \irrepsub{35}{s}     \\  \midrule
                \dynkin{1, 0, 1, 0} & 56  & 15  & (10) & \irrepsub{56}{s}     \\
                \dynkin{1, 0, 0, 1} & 56  & 15  & (12) & \irrepsub{56}{c}     \\
                \dynkin{0, 0, 1, 1} & 56  & 15  & (02) & \irrepsub{56}{v}     \\
                \dynkin{2, 0, 2, 0} & 840 & 540 & (00) & \irrepsub[1]{840}{s} \\
                \dynkin{2, 0, 0, 2} & 840 & 540 & (00) & \irrepsub[1]{840}{c} \\
                \dynkin{0, 0, 2, 2} & 840 & 540 & (00) & \irrepsub[1]{840}{v} \\
                \bottomrule
            \end{tabular}
        \end{center}
\end{enumerate}

In LieART the function \com{DimName[\args{irrep}]} determines the dimensional
name according to the algorithm described above, which is automatically
displayed if an irrep is displayed in \com{TraditionalForm}. Several internal
helper functions are called by \com{DimName}. The function
\com{GetIrrepByDim[\args{algebra},\args{dim},\args{maxDynkinDigit}]} provides
irreps with the same dimension, which are then gathered into sublists by
\com{DimName}.
The function \com{SortSameDimAndIndex} sorts the irreps of same dimension and
index by their congruency class, and automatically by the Dynkin label viewed as
``numbers'', if the congruency class numbers are the same. The positions of the
lists of same-index irreps determines the number of primes to apply and the
position of the irrep within the same-index list, whether it should be labeled
by an overbar. In case of an \SO8 irrep \com{DimName} branches to the function
\com{SO8Label[\args{irrep}]}, which uses \com{SimpleSO8Label} to give a
subscript of ``v'', ``s'' and ``c'' in the case where the congruency vector in
unique. If the congruency vector is not unique, but non-zero,
\com{ConcatSO8Label} concatenates the mixed subscripts like ``sv'' in the
correct ordering. If the congruency vector is zero in both components the irrep
is related to irreps with non-zero congruency vector by
\com{ReducedDynkinLabel}.

\paragraph{Limitations}
The determination of the primes has one limitation, which requires explanation:
The function \com{GetIrrepByDim[\args{algebra},\args{dim},\args{maxDynkinDigit}]}
determines irreps of the same dimension. In a brute-force fashion it generates
``all'' irreps and extracts only those that have the dimension \args{dim}. Since
there are infinite many irreps of any Lie algebra, it must be constrained. This
is done by imposing a maximum Dynkin digit to use for the generation of Dynkin
labels. Since the numbers of possible Dynkin labels grow rapidly with the
maximum Dynkin digit allowed, the limit should be very low. To compare the irrep
in question with others it should be at least its maximum Dynkin digit, e.g., for
\dynkin{2,0,3,1} it is ``3''. The related irreps only have a permutation of the
Dynkin label, thus they are included in the generated list of irreps up to a
Dynkin digit of ``3'' in the example. However, for the determination of the
primes for the unrelated irreps it may not suffice to generate irreps only up to
the maximum Dynkin digit of the irrep in question: The number of primes are
determined by the position in a list of same-dimensional irreps sorted by the
index. If there is an irrep with a higher maximal Dynkin digit, e.g.,
``4'' in our example, but at the same time has a lower index than the irrep it
question, this procedure would give the irrep in question too few primes. This
situation rarely happens, especially in \A{n}'s, but e.g.\ for \G2 it
happens as early as the 77 dimensional irrep:
        \begin{center}
            \begin{tabular}{lllll}
                \toprule
                \textbf{Dynkin}     & \textbf{Dimension}& \textbf{Index} & \textbf{Congruency number}  & \textbf{Name}  \\
                \midrule
                \dynkin{3, 0} & 77 & 44 & 0 & \irrep{77} \\
                \dynkin{0, 2} & 77 & 55 & 0 & \irrep[1]{77} \\
                \bottomrule
            \end{tabular}
        \end{center}
When determining the name of \dynkin{0, 2} of \G2 the Dynkin labels would
only be generated up to a maximum Dynkin digit of ``2'', the \dynkin{3, 0} would
not appear and thus the \dynkin{0, 2} would be labeled without any prime. The
determination of the name of \dynkin{3, 0} would ``see'' the \dynkin{0, 2}, but
would determine no prime for \dynkin{3, 0} because of its lower index compared
to \dynkin{0, 2}. For these two irreps the problem can be solved by generating
irreps up to the maximum Dynkin digit \emph{plus one} for the irrep in question,
i.e., up to ``3'' for \dynkin{0, 2}. Because the Dynkin label of a \G2 irrep is
small, this is easily manageable. In fact, we have implemented the addition of
three to the maximum Dynkin digit for \G2, because for some higher dimensions
the problem will reappear. However, for Lie algebras with long Dynkin labels,
the number of generated Dynkin labels becomes large and its construction slows
LieART down and consumes a large amount of memory. We have found a balance
between accuracy and efficiency, which pushes this problem to very high
dimensional irreps, by defining the following number to add to the maximum
Dynkin digit of the irrep in question: 1 for \A{n},\B{n},\C{n} and \D{n} with
$n\leq 4$, 0 for \A{n},\B{n},\C{n} and \D{n} with $n\geq 5$, 1 for \E6 and \F4,
0 for \E7 and \E8 and 3 for \G2. Please note that this limitation is only
connected to the labeling of irreps with primes. Computations in LieART are
always performed using the Dynkin label as in the \com{FullForm}. If in doubt
one can always use the Dynkin label displayed in \com{StandardForm},
\com{InputForm} and \com{FullForm} which serves as the unique identifier of an
irrep.

Besides its Dynkin label, the alternative definition of \com{Irrep} as 
\com{Irrep[\args{algebra}][\args{dimname}]} can be used to specify an irrep by 
its dimensional name as \args{dimname} and its algebra as \args{algebra}. The 
\args{algebra} must be fully specified, such as \com{A4}, \com{SU5}, \com{E6}, 
not only the algebra class such as \com{A}. The \args{dimname} is an integer for 
the dimension with a \com{Bar[\args{dim}]} wrapped around it for a conjugated 
irrep or an \com{IrrepPrime[\args{dim},\args{n}]} for an irrep with \args{n} 
primes. If only one prime is needed the second argument \args{n} may be omitted. 
The \com{Bar} and \com{IrrepPrime} can be combined in any sequence. E.g., the 
\irrepbar[1]{175} can be entered by
\begin{mathin}
    Irrep[A4][IrrepPrime[Bar[175]]]//InputForm
\end{mathin}
\begin{mathout}
    Irrep[A][0, 0, 2, 1]
\end{mathout}
Alternatives are \texttt{Irrep[SU5][IrrepPrime[Bar[175]]]}, 
\texttt{Irrep[SU5][Bar[IrrepPrime[175]]]} and 
\texttt{Irrep[A4][IrrepPrime[Bar[175],1]]}. Internally the function 
\com{GetIrrepByDimName[\args{algebra}, \args{dimname}]} determines the 
corresponding Dynkin label. It uses the function \com{GetIrrepByDim} mentioned 
above to find all irreps with the same dimension and then extract the irrep with 
the identical dimensional name. If the user specifies an irrep that does not 
exist, e.g. an \irrepbar{11} of \A4, the comparison must stop at some point. It 
has been chosen that \com{GetIrrepByDim} generates only irreps with a maximum 
Dynkin digit as set by the global variable \com{\$MaxDynkinDigit}. The default 
is \com{\$MaxDynkinDigit}=3. The consequence is that the determination of the 
correct Dynkin label of the entered irrep may abort, because the irrep does not 
exist or that it involves a Dynkin digit higher that \com{\$MaxDynkinDigit}=3. 
The latter is the case for the \irrep[1]{70} with a Dynkin label of 
\dynkin{0,0,0,4}. LieART prints an error message indicating the two possible 
scenarios:
\begin{mathin}
    Irrep[A4][IrrepPrime[70]]//InputForm
\end{mathin}
\begin{mathout}
\color{darkbrown}Irrep::noirrep: Either an irrep with the dimension name \irrep[1]{70} does not exist in SU(5) or it has at least one Dynkin digit higher than 3. Try with \$MaxDynkinDigit set to a higher value than 3. >>
\end{mathout}
Increasing \com{\$MaxDynkinDigit} to 4 resolves the issue:
\begin{mathin}
    \nohangingindent
    \$MaxDynkinDigit=4;\newline
    Irrep[A4][IrrepPrime[70]]//InputForm
\end{mathin}
\begin{mathout}
    Irrep[A][0,0,0,4]
\end{mathout}

\subsection{Tensor Product Decomposition}

\definition{
    \com{DecomposeProduct[\args{irreps}]} & decomposes the tensor product of \args{irreps}\\
    \com{DominantWeightsAndMul[\args{weights}]} & filters and tallies dominant weights of \args{weights} by multiplicities\\
    \com{SortOutIrrep[\args{dominantWeightsAndMul}]} & sorts out the irrep of largest height from the collection of dominant weights \args{dominantWeightsAndMul}\\
    \com{WeightSystemWithMul[\args{irrep}]} & full weight system of \args{irrep} with multiplicities\\
    \com{TrivialStabilizerWeights[\args{weights}]} & drops weights that lie on a chamber wall\\
    \com{ReflectToDominantWeightWithMul[\args{weightAndMul}]} & reflects \args{weightAndMul} to the dominant chamber and multiplies the parity of the reflection to the multiplicity
}{Tensor product decomposition.}

Tensor products of irreps can be decomposed into a direct sum of irreps. The 
product of two irreps $R_1$ and $R_2$ can be decomposed as
\begin{equation}\label{eq:TensorProduct}
    R_1\otimes R_2 = \sum_i m_i R_i
\end{equation}
with the following dimension and index sum rules:
\begin{align}
    \dim(R_1\otimes R_2)&= \dim(R_1)\cdot\dim(R_2)=\sum_i m_i \dim(R_i)\\
    l(R_1\otimes R_2)&= l(R_1)\dim(R_2)+\dim(R_1)l(R_2)=\sum_i m_i\, l(R_i).
\end{align}

\subsubsection{Generic Algorithm}
\label{ssec:TensorProductGeneric}

A straight-forward method to compute the right-side of \eqref{eq:TensorProduct} 
is the following: Add all weights of $R_2$ to each weight of $R_1$. The 
resulting $\dim(R_1)\cdot\dim(R_2)$ weights belong to the different irreps 
$R_i$, which must be sorted out. Instead of all weights, one can consider just 
the dominant weights in the product, as each of the dominant weights represents 
an orbit in the irreps $R_i$. As an irrep is a collection of orbits, some of the 
dominant weights in the product represent the highest weight of an irrep in the 
decomposition. There is a unique dominant weight that represents the irrep of 
largest height in the decomposition. The sorting procedure should start with 
this dominant weight viewed as the highest weight of an irrep and then construct 
the dominant weight system of the corresponding irrep. The dominant weight 
system of the irrep with largest height should then be subtracted from the 
combined dominant weights of the product to filter it out. The same procedure is 
applied recursively to the remaining set of dominant weights until it is empty, 
i.e., all irreps have been filtered out. 

LieART provides the function \com{DecomposeProduct[\args{irreps}]} for the 
decomposition of the tensor product of arbitrary many \args{irreps} of any 
classical or exceptional Lie algebra as argument. As a demonstration of generic 
algorithm we consider the decomposition of the \SU3 tensor product 
$\irrep{8}{\otimes}\irrep{8}$, which is 
\begin{mathin}
    DecomposeProduct[Irrep[SU3][8], Irrep[SU3][8]]
\end{mathin}
\begin{mathout}
    $\irrep{1}+2(\irrep{8})+\irrep{10}+\irrepbar{10}+\irrep{27}$
\end{mathout}
In the straight forward approach one adds all weights of the second \irrep{8} to 
each weight of the first \irrep{8}, using the built-in Mathematica function 
\com{Outer}. One filters out only the dominant weights and tallies multiple 
occurrences thereof using the LieART function 
\com{DominantWeightsAndMul[\args{weights}]}, which also sorts the weights 
according to their height, when viewed as a highest weight of an irrep. For the 
\SU3 tensor product $\irrep{8}{\otimes}\irrep{8}$ the dominant weights along 
with their multiplicities are \begin{mathin}
    DominantWeightsAndMul[Flatten[Outer[Plus, WeightSystem[Irrep[SU3][8]], WeightSystem[Irrep[SU3][8]]]]]
\end{mathin}
\begin{mathout}\label{out:8otimes8}
    $\begin{pmatrix}
        \weight{2, 2} & 1\\
        \weight{0, 3} & 2\\
        \weight{3, 0} & 2\\
        \weight{1, 1} & 6\\
        \weight{0, 0} & 10
    \end{pmatrix}$
\end{mathout}
The dominant weight with the largest height \weight{2, 2} must be the highest 
weight of an irrep. The dominant weight system of the \dynkin{2,2} (the 
\irrep{27}) of \SU3 is
\begin{mathin}
    DominantWeightSystem[Irrep[A][2, 2]]
\end{mathin}
\begin{mathout}\label{out:27}
    $\begin{pmatrix}
        \weight{2, 2} & 1\\
        \weight{0, 3} & 1\\
        \weight{3, 0} & 1\\
        \weight{1, 1} & 2\\
        \weight{0, 0} & 3
    \end{pmatrix}$
\end{mathout}
It contains all dominant weights appearing in the tensor product, but with 
mostly smaller multiplicities. The irrep \dynkin{2,2} can be filtered out by 
subtracting the multiplicities in \outref{out:27} from \outref{out:8otimes8}. 
The LieART function \com{SortOutIrrep[\args{dominantWeightsAndMul}]} performs 
the task of computing the dominant weight system of the irrep corresponding to 
the largest height weight and subtracting it from the tensor product. It returns 
the dominant weights of the tensor product with the ones of the irrep removed 
and passes the latter using the \com{Sow} and \com{Reap} mechanism of 
Mathematica: \begin{mathin}
Reap[SortOutIrrep[\%\%]]
\end{mathin}
\begin{mathout}\label{out:27filteredout}
    $\{\begin{pmatrix}
        \weight{0, 3} & 1\\
        \weight{3, 0} & 1\\
        \weight{1, 1} & 4\\
        \weight{0, 0} & 7
    \end{pmatrix},\,(\irrep{27})\}$
\end{mathout}
The function \com{SortOutIrrep} is applied recursively until the list of 
dominant weights with multiplicities of the tensor product is empty. E.g., 
applying \com{SortOutIrrep} to the dominants weights of 
\outref{out:27filteredout} filters out the \dynkin{0,3} (the \irrepbar{10}) of 
\SU3:
\begin{mathin}
    Reap[SortOutIrrep[First[\%]]]
\end{mathin}
\begin{mathout}
    $\{\begin{pmatrix}
        \weight{3, 0} & 1\\
        \weight{1, 1} & 3\\
        \weight{0, 0} & 6
    \end{pmatrix},\,(\irrepbar{10})\}$
\end{mathout}
The irreps filtered out by \com{SortOutIrrep} are collected by the LieART function
\com{GetIrreps} and can be displayed as the result of the decomposition. However,
LieART computes the tensor product decomposition by an implementation of Klimyk's 
formula, which is far more efficient than the straight-forward procedure described 
above, but the sorting-out algorithm is still used for subalgebra decomposition in 
Section\ \ref{ssec:SubalgebraDecomposition}.

\subsubsection{Algorithm Based on Klimyk's Formula}

Adding all weights of $R_2$ to each weight of $R_1$ is costly for large irreps. 
LieART's algorithm to decompose tensor-product implements 
Klimyk's formula~\cite{klimyk_translation,Humphreys:1980dw}, which improves the runtime 
of tensor product decompositions considerably: Let $\lambda_1$ and $\lambda_2$ 
be weights and $\Lambda_1$ and $\Lambda_2$ the highest weights of $R_1$ and 
$R_2$, respectively. Instead of adding all weights $\lambda_1$ to each weight 
$\lambda_2$, the weights $\lambda_1$ are added only to the highest weight 
$\Lambda_2$ of $R_2$ together with half the sum of positive simple roots, 
$\delta{=}\dynkincomma{1,\ldots,1}$, building the set of weights 
\begin{equation}\label{eqn:mu}
	\mu = \lambda_1 + \Lambda_2 + \delta.
\end{equation}
Each $\mu$ is reflected to the dominant chamber, yielding a highest weight 
denoted as $\{\mu\}$, with $\sgn(\mu)$ as the parity of the reflection. Of these 
dominant weights all that lie on a chamber wall are dropped. The irreps in the 
decomposition are $R(\{\mu\}-\delta)$. Klimyk's formula reads 
\begin{equation}
	R_1(\Lambda_1)\otimes R_2(\Lambda_2) = \sum\limits_{\lambda_1} m_{1\lambda_1} t(\lambda_1+ \Lambda_2 + \delta) R(\{\lambda_1+ \Lambda_2 + \delta\}-\delta),
\end{equation}
where $m_{1\lambda_1}$ denotes the multiplicity of $\lambda_1$ in $R_1$. We 
define $t(\mu)$ to be $\sgn(\mu)$ if $\mu$ has a trivial-stabilizer subgroup $\text{Stab}(\mu) = \{1\}$ 
(see \eqref{eqn:DefinitionStabilizer}), and zero if the stabilizer subgroup 
is non-trivial, i.e. the weight lies on a chamber wall:
\begin{equation}
	t(\mu) = \left\{
		\begin{array}{rcl}
			\sgn(\mu) & : & \text{Stab}(\mu) = \{1\}\\
			          0 & : & \text{else}
		\end{array}
		\right..
\end{equation}
We will demonstrate this algorithm with LieART in the following paragraphs.

As a demonstration of the algorithm implemented we consider the decomposition of 
the \SU3 tensor product $\irrep{6}{\otimes}\irrep{3}$, which is 
\begin{mathin}
    DecomposeProduct[Irrep[SU3][6], Irrep[SU3][3]]
\end{mathin}
\begin{mathout}\label{out:6times3}
    $\irrep{8}+\irrep{10}$
\end{mathout}

LieART normally reorders the arguments of \com{DecomposeProduct} by their 
dimension, which usually simplifies the application of Klimyk's formula. But for 
didactical reasons we assume in the following that the irreps are not reordered. 
LieART generates the weight system with weight multiplicities of the \irrep{6}: 
\begin{mathin}
    WeightSystemWithMul[Irrep[SU3][6]]
\end{mathin}
\begin{mathout}
    $\begin{pmatrix}
        \weight{ 2, 0} & 1\\
        \weight{-2, 2} & 1\\
        \weight{ 0,-2} & 1\\
        \weight{ 0, 1} & 1\\
        \weight{ 1,-1} & 1\\
        \weight{-1, 0} & 1\\
    \end{pmatrix}$
\end{mathout}
We add all weights of the \irrep{6} to the highest weight of the \irrep{3}, 
i.e.\ $\com{HighestWeight[Irrep[SU3][3]]}=\weight{ 1, 0}$, and 
$\com{Delta[SU3]=\weight{1,1}}$, according to \eqref{eqn:mu}. The highest weight 
$\Lambda_2$ and $\delta$ can be added directly, but to add the sum to all 
weights in the weight system with multiplicities LieART provides the command 
\com{Add}: 
\begin{mathin}
    mu = Add[WeightSystemWithMul[Irrep[SU3][6]], HighestWeight[Irrep[SU3][3]] + Delta[SU3]]
\end{mathin}
\begin{mathout}
	$\begin{pmatrix}
        \weight{ 4, 1} & 1\\
        \weight{ 0, 3} & 1\\
        \weight{ 2,-1} & 1\\
        \weight{ 2, 2} & 1\\
        \weight{ 3, 0} & 1\\
        \weight{ 1, 1} & 1\\
    \end{pmatrix}$
\end{mathout}
LieART reflects these weights to the dominant chamber yielding the corresponding 
dominant weights and the parity of the reflection, i.e.\ $\sgn(\mu)=(-1)^l$, 
where $l$ is the number of reflections needed to reach the dominant chamber: 
\begin{mathin}
	ReflectToDominantWeightWithMul /@ mu
\end{mathin}
\begin{mathout}
	$\begin{pmatrix}
        \weight{ 4, 1} &  1\\
        \weight{ 0, 3} &  1\\
        \weight{ 1, 1} & -1\\
        \weight{ 2, 2} &  1\\
        \weight{ 3, 0} &  1\\
        \weight{ 1, 1} &  1\\
    \end{pmatrix}$
\end{mathout}
The weights \weight{ 4, 1}, \weight{ 0, 3}, \weight{ 2, 2}, \weight{ 3, 0} and 
\weight{ 1, 1} were already in the dominant chamber, while \weight{ 2,-1} needed 
one reflection to become the dominant weight $\weight{ 1, 1}$ with a parity of $-1$.

Weights on walls of the dominant chamber do not contribute in Klimyk's formula 
and must be dropped. Weights not lying on a chamber wall have a 
trivial-stabilizer subgroup. To keep only these weights LieART applies the 
command \com{TrivialStabilizerWeights}, which drops all weights containing at 
least one zero anywhere in their Dynkin label, which corresponds to lying on a 
chamber wall:
\begin{mathin}
	TrivialStabilizerWeights[ReflectToDominantWeightWithMul /@ mu]
\end{mathin}
\begin{mathout}
	$\begin{pmatrix}
        \weight{ 4, 1} &  1\\
        \weight{ 1, 1} & -1\\
        \weight{ 2, 2} &  1\\
        \weight{ 1, 1} &  1\\
    \end{pmatrix}$
\end{mathout}
LieART subtracts $\delta$ yielding highest weights and constructs the irreps in the 
decomposition $R(\{\lambda_1{+}\Lambda_2{+}\delta\}{-}\delta)$:
\begin{mathin}
	ToIrrep/@Add[TrivialStabilizerWeights[ReflectToDominantWeightWithMul/@mu], -delta]
\end{mathin}
\begin{mathout}
	$\begin{pmatrix}
        \irrep{10} &  1\\
        \irrep{1}  & -1\\
        \irrep{8}  &  1\\
        \irrep{1}  &  1\\
    \end{pmatrix}$
\end{mathout}
While the irreps in the left column correspond to $R(\{\lambda_1+ \Lambda_2 + 
\delta\}-\delta)$ in Klimyks formula, the multiplicities in the right column 
correspond to $m_{1\lambda_1} \sgn(\lambda_1 + \Lambda_2 + \delta)$. Summing 
up the decomposition accordingly we see that the \irrep{1} drops out and we 
obtain the result $\irrep{8}+\irrep{10}$ as already stated in 
\outref{out:6times3}.

\subsubsection{\SU{N} Decomposition via Young Tableaux}

A correspondence of \SU{N} irreps and Young tableaux is very useful for the 
calculation of tensor products and subalgebra decomposition by hand. We have 
found that the algorithm for the \SU{N} tensor product decomposition via Young 
tableaux also performs better on the computer, with respect to CPU time and 
memory consumption, than the procedure described in the previous section. Thus, 
LieART uses the Young tableaux algorithm for the tensor-product decomposition of 
\SU{N} irreps and the procedure of adding weights and filtering out irreps for 
all other classical and exceptional Lie algebras.

A \emph{Young tableau} is a left-aligned set of boxes, with successive rows 
having an equal or smaller number of boxes. Young tableaux correspond to the 
symmetry of the tensors of \SU{N} irreps, by first writing each index of the 
tensor into one box of a Young tableau and the prescription that they ought to 
be first symmetrized in the rows and then antisymmetrized in the columns. Please 
see \outref{out:YoungTableau720SU5} in Section~\ref{sec:QuickStart} for a 
non-trivial example for a Young tableau displayed by LieART.

To demonstrate the algorithm for the tensor product decomposition via Young 
tableaux we use the same \SU3 tensor product as in Section~\ref{ssec:TensorProductGeneric}, 
$\irrep{8}{\otimes}\irrep{8}$.

The construction principle (by hand) is to put the Young tableau with the most 
boxes to the left and bump all boxes of the right Young tableau row by row to 
the left one following certain rules. To understand these rules label the boxes 
of each row of the right Young tableau alphabetically: 
\begin{equation}
    \irrep{8} \otimes \irrep{8} = \young(\hfil\hfil,\hfil)\otimes\young(aa,b)
\end{equation}
Bump the boxes of the first row of the right tableau (labeled with $a$'s) to the 
ends of the left tableau to form valid Young tableaux and the additional 
condition that no $a$'s are in the same column, as this would change symmetric 
indices to antisymmetric ones. (see crossed out tableau, which is also not a 
valid \SU3 tableau, because it has four (or $N{+}1$ in general for \SU{N}) boxes 
in a column): 
 \begin{equation}
    \young(\hfil\hfil,\hfil)\otimes\young(aa,b) =
    \left(\,
        \young(\hfil\hfil aa,\hfil) \oplus
        \young(\hfil\hfil a,\hfil a) \oplus
        \young(\hfil\hfil a,\hfil,a) \oplus
        \young(\hfil\hfil,\hfil a,a) \oplus
        \cancel{\young(\hfil\hfil,\hfil,a,a)}\;
    \right)\otimes\,\young(b)
\end{equation}
As the next step bump the boxes of the next row to all tableaux obtained by the 
first step obeying the same rules, but keep only tableaux with an 
\emph{admissible} sequence of letters, when reading the letters from right to 
left, row by row. A sequence of letters is admissible if at any point at least as 
many $a$'s have occurred as $b$'s and likewise for $b$'s and $c$'s, etc. The 
sequences $abcd$, $aaabc$, $abab$ are admissible, but $abba$, $acb$ are not. In 
our example only the sequence $baa$ is not admissible, which appears in the four 
crossed-out tableaux:
\begin{equation}
\begin{aligned}
    \young(\hfil\hfil,\hfil)\otimes\young(aa,b) =\;
    &
        \cancel{\young(\hfil\hfil aab,\hfil)} \oplus
        \young(\hfil\hfil aa,\hfil b) \oplus
        \young(\hfil\hfil aa,\hfil,b) \oplus
    \\&
        \cancel{\young(\hfil\hfil ab,\hfil a)} \oplus
        \young(\hfil\hfil a,\hfil ab) \oplus
        \young(\hfil\hfil a,\hfil a,b) \oplus
    \\&
        \cancel{\young(\hfil\hfil ab,\hfil,a)} \oplus
        \young(\hfil\hfil a,\hfil b,a) \oplus
    \\ &
        \cancel{\young(\hfil\hfil b,\hfil a,a)} \oplus
        \young(\hfil\hfil,\hfil a,ab)
\end{aligned}
\end{equation}
Removing the labeling with $a$'s and $b$'s, we obtain the \SU3 tensor product 
decomposition of $\irrep{8}{\otimes}\irrep{8}$ in terms of Young tableaux:
\newcommand\oplusvar{\:\oplus\:}
\newcommand\otimesvar{\:\otimes\:}
\begin{equation}
    \begin{array}{c@{\otimesvar}c@{\:=\:}c@{\oplusvar}c@{\oplusvar}c@{\oplusvar}c@{\oplusvar}c@{\oplusvar}c}
        \yng(2,1) & \yng(2,1) & \yng(4,2) & \yng(4,1,1) & \yng(3,3)     & \yng(3,2,1) & \yng(3,2,1) & \yng(2,2,2)
    \end{array}
\end{equation}
Finally we knock out triples (full columns with three boxes) to find:
\begin{equation}
    \begin{array}{c@{\otimesvar}c@{\:=\:}c@{\oplusvar}c@{\oplusvar}c@{\oplusvar}c@{\oplusvar}c@{\oplusvar}c}
        \yng(2,1) & \yng(2,1) & \yng(4,2) & \yng(3) & \yng(3,3)     & \yng(2,1) & \yng(2,1) & \:\bullet\\[20pt]
        \irrep{8} & \irrep{8} & \irrep{27}& \irrep{10}  & \irrepbar{10} & \irrep{8}   & \irrep{8}   & \irrep{1}
    \end{array}
\end{equation}

In LieART the Young tableau algorithm is automatically applied to tensor 
products of \SU{N} irreps using \com{DecomposeProduct[\args{irreps}]}. After 
sorting the irrep with fewer boxes to the right (which we will call 
\args{irrep2} opposed to the first one named \args{irrep1}), the function 
processes through \args{irrep2}'s rows to bump boxes to \args{irrep1}. The 
function \com{BoxesToBump[\args{irrep2},\,\args{row}]} gives the number of boxes 
in the current \args{row} to bump to the tableau of \args{irrep1}. The function 
\com{AllowedRows[\args{irrep1},\,\args{nboxes}]} determines the rows of 
\args{irrep1} that are allowed to bump boxes to yielding a valid young tableau 
with an admissible sequence. The latter is checked by the helper function 
\com{AllowedCombination}. The function 
\com{AddToTableau[\args{irrep1},\,\args{rowcombinations}]} performs the bumping 
of the boxes of one row of \args{irrep2} in all allowed combinations 
(\args{rowcombinations}) to \args{irrep1}. The result of the bumping is directly 
expressed in terms of a changed Dynkin label.

\subsection{Subalgebra Decomposition}
\label{ssec:SubalgebraDecomposition}

\definition{
    \com{DecomposeIrrep[\args{irrep},\,\args{subalgebra}]} & decomposes \args{irrep} to the specified \args{subalgebra}.\\
    \com{DecomposeIrrep[\args{productIrrep},\,\args{subalgebra},\,\args{pos}]} & decomposes \args{productIrrep} at position \args{pos}.\\
    \com{ProjectionMatrix[\args{origin},\args{target}]} & defines the projection matrix for the algebra-subalgebra pair specified by \args{origin} and \args{target}\\
    \com{Project[\args{projectionMatrix},\args{weights}]} & applies the \args{projectionMatrix} to the \args{weights}\\
    \com{GroupProjectedWeights[\args{projectedWeights},\args{target}]} & groups the projected weights according to the subalgebra specified in \args{target}\\
    \com{NonSemiSimpleSubalgebra[\args{origin},\args{simpleRootToDrop}]} & computes the projection matrix of a maximal non-semi-simple subalgebra by dropping one dot of the Dynkin diagram \args{simpleRootToDrop} and turning it into a \U1 charge\\
    \com{SemiSimpleSubalgebra[\args{origin},\args{simpleRootToDrop}]} & computes the projection matrix of a maximal semi-simple subalgebra by dropping one dot from the extended Dynkin diagram.\\
    \com{ExtendedWeightScheme[\args{algebra},\args{simpleRootToDrop}]} & adds the Dynkin label associated with the extended simple root ${-}\gamma$ to each weight of the lowest orbit of \args{algebra} and drops the simple root \args{simpleRootToDrop}\\
    \com{SpecialSubalgebra[\args{origin},\args{targetirreps}]} & computes the projection matrix of a maximal special subalgebra by specifying the branching rule of the generating irrep.
}{Subalgebra decomposition of irreps and product algebra irreps.}

The LieART function \com{DecomposeIrrep[\args{irrep},\,\args{subalgebra}]}
decomposes an irrep of a simple Lie algebra into a maximal subalgebra specified
by \args{subalgebra}, which can be simple, semi-simple or non-semi-simple. To
decompose an irrep of a semi-simple or non-semi-simple irrep, a third argument
\args{pos} allows one to specify which part of \args{productIrrep} should be
decomposed into the \args{subalgebra}.

The implementation of \com{DecomposeIrrep} in LieART uses so-called projection 
matrices. These matrices project the weights of an irrep into the specified 
subalgebra. The resulting weights are further processed in the same manner as 
in Section~\ref{ssec:SubalgebraDecomposition}: Only the dominant weights of the 
decomposed weights are kept, because they uniquely define the orbits of the 
subalgebra and thus its irreps. In the next step the irreps comprised in the 
collection of dominant weights are sorted out using the same LieART functions as 
for the generic tensor product decomposition, discussed in section 
\ref{ssec:TensorProductGeneric}.
It is clear that the major task is the determination of the projection matrices.
They are different for each algebra-maximal-subalgebra pair and are not unique.
I.e., a projection is not unique and thus the matrices
are not unique in general. Our matrices are correct
and consistent, but one may find different projection matrices in the
literature, which are also correct. (An extensive collection of projection matrices can be found in
\cite{larouche_branching_2009} for the Lie algebra \A{n} and in \cite{larouche_branching_2011} for the Lie algebras \B{n}, \C{n} and \D{n}.)
Once a projection matrix is known it can be used for the decomposition of all irreps of the algebra-maximal-subalgebra pair. E.g., the projection matrix for the branching
$\SU5\to\SU3{\otimes}\SU2{\otimes}\U1$ is
\begin{mathin}
    ProjectionMatrix[SU5,ProductAlgebra[SU3,SU2,U1]]
\end{mathin}
\begin{mathout}\label{out:ProjectionMatrix}
    $\begin{pmatrix}
        1 & 0 & 0 & 0 \\
        0 & 1 & 0 & 0 \\
        0 & 0 & 0 & 1 \\
        2 & 4 & 6 & 3
    \end{pmatrix}$
\end{mathout}
The determination of the projection matrices is closely connected to the problem of finding maximal subalgebras and we defer the description of its implementation in
LieART to the next section. Taking the projection matrix \outref{out:ProjectionMatrix} as given we demonstrate the algorithm of \com{DecomposeIrrep}
to find the branching rule for the \irrep{10} of \SU5 to $\SU3{\otimes}\SU2{\otimes}\U1$, which is
\begin{mathin}
    DecomposeIrrep[Irrep[SU5][10],ProductAlgebra[SU3,SU2,U1]]
\end{mathin}
\begin{mathout}
    $(\irrep{1},\irrep{1})(-6)+(\irrepbar{3},\irrep{1})(4)+(\irrep{3},\irrep{2})(-1)$
\end{mathout}
The LieART function \com{Project[\args{projectionMatrix},\args{weights}]} applies the \args{projectionMatrix} to each of the \args{weights} and a subsequent
\com{GroupProjectedWeights[\args{projectedWeights},\args{target}]} groups the Dynkin label of each of the \args{projectedWeights} according to the subalgebra specified by
\args{target}. In the case of our example each weight of the \irrep{10} of \SU5 decomposes to $\SU3{\otimes}\SU2{\otimes}\U1$ as:
\begin{mathin}
IrrepRule @@@ Transpose[\{WeightSystem[Irrep[SU5][10]],\newline
Row/@GroupProjectedWeights[Project[ProjectionMatrix[SU5, ProductAlgebra[SU3,SU2,U1]],WeightSystem[Irrep[SU5][10]]], ProductAlgebra[SU3,SU2,U1]]\}]
\end{mathin}
\begin{mathout}\label{out:Projected10SU5}
\nohangingindent
    $\weight{0, 1, 0, 0}   \rightarrow \weight{0, 1} \weight{0} \weight{4}$\newline
    $\weight{1, {-}1, 1, 0}  \rightarrow \weight{1, {-}1}\weight{0} \weight{4}$\newline
    $\weight{{-}1, 0, 1, 0}  \rightarrow \weight{{-}1, 0}\weight{0} \weight{4}$\newline
    $\weight{1, 0, {-}1, 1}  \rightarrow \weight{1, 0} \weight{1} \weight{{-}1}$\newline
    $\weight{{-}1, 1, {-}1, 1} \rightarrow \weight{{-}1, 1}\weight{1} \weight{{-}1}$\newline
    $\weight{1, 0, 0, {-}1}  \rightarrow \weight{1, 0} \weight{{-}1}\weight{{-}1}$\newline
    $\weight{{-}1, 1, 0, {-}1} \rightarrow \weight{{-}1, 1}\weight{{-}1}\weight{{-}1}$\newline
    $\weight{0, {-}1, 0, 1}  \rightarrow \weight{0, {-}1}\weight{1} \weight{{-}1}$\newline
    $\weight{0, {-}1, 1, {-}1} \rightarrow \weight{0, {-}1}\weight{{-}1}\weight{{-}1}$\newline
    $\weight{0, 0, {-}1, 0}  \rightarrow \weight{0, 0} \weight{0} \weight{{-}6}$
\end{mathout}
The algorithm of \com{DecomposeIrrep} differs slightly and keeps only the dominant weights after projection and groups only them yielding
\begin{equation}
    \left(\begin{array}{l}
        \weight{0, 1} \weight{0} \weight{4}\\
        \weight{1, 0} \weight{1} \weight{{-}1}\\
        \weight{0, 0} \weight{0} \weight{{-}6}
    \end{array}\right)
\end{equation}
for our example. A combination of the functions \com{GetAllProductIrrep} and  \com{GetProductIrrep} filter out the product irreps, $(\irrepbar{3},\irrep{1})(4)$, $(\irrep{1},\irrep{1})(-6)$ and $(\irrep{3},\irrep{2})(-1)$ in our case, by applying
the function \com{GetIrrep} known from section \ref{ssec:TensorProductGeneric} to the weights.

\subsubsection{Branching Rules and Maximal Subalgebras}

To determine the projection matrices we start with the algorithm to find maximal subalgebras. Subalgebras fall into two classes: \emph{regular} and \emph{special} subalgebras, with the
first one being further categorized into non-semisimple and semisimple subalgebras. In the following we describe the derivation of the three types of maximal subalgebras: regular non-semisimple, regular semisimple and special subalgebras,
originally developed by Dynkin \cite{Dynkin:1957um,Dynkin:1957dm} and demonstrate how it is utilized by LieART to determine the projection matrices.

\paragraph{Non-Semisimple Subalgebras}

A non-semisimple subalgebra is a semisimple subalgebra times a \U1 factor, e.g. $\SU3{\otimes}\SU2{\otimes}\U1$. A subalgebra of this type is obtained by removing a dot from the Dynkin diagram. The resulting two or more disconnected
Dynkin diagrams symbolize the semisimple subalgebra and the removed dot, i.e., simple root, becomes the \U1 generator. E.g., the non-semisimple subalgebra $\SU3{\otimes}\SU2{\otimes}\U1$ can be
obtained from \SU5 by removing the third dot from its Dynkin diagram:
\begin{center}
\includegraphics{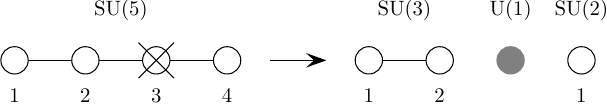}
\end{center}
Since the Dynkin label of a weight represents it composition of simple roots (explicitly in the $\alpha$-basis), dropping a simple root (dot) from the Dynkin diagram corresponds to
dropping the associated digit from the Dynkin label. The \U1 charge is the coefficient of the dropped simple root in the weight's linear combination of simple roots, i.e., the associated digit
of the Dynkin label in the $\alpha$-basis, which is often normalized to give integer values. Accordingly, in \outref{out:Projected10SU5} the third Dynkin digit of the weight of the \irrep{10} has been removed after the projection
and by expressing the weight system in the $\alpha$-basis
\begin{mathin}
    AlphaBasis[WeightSystem[Irrep[SU5][10]]]//Column
\end{mathin}
\begin{mathout}\nohangingindent
    (3/5, 6/5, 4/5, 2/5)\newline
    (3/5, 1/5, 4/5, 2/5)\newline
    (-2/5, 1/5, 4/5, 2/5)\newline
    (3/5, 1/5, -1/5, 2/5)\newline
    (-2/5, 1/5, -1/5, 2/5)\newline
    (3/5, 1/5, -1/5, -3/5)\newline
    (-2/5, 1/5, -1/5, -3/5)\newline
    (-2/5, -4/5, -1/5, 2/5)\newline
    (-2/5, -4/5, -1/5, -3/5)\newline
    (-2/5, -4/5, -6/5, -3/5)
\end{mathout}
we see that the \U1 charge at the end are the third coordinate of the weight in the $\alpha$-basis multiplied by 5 to give integer values.
To summarize, the third Dynkin digit has been moved to the end
(i.e., third and forth digits have been interchanged)
and only it has been projected into the $\alpha$ basis and rescaled.

Writing the weights of the \irrep{10} as \emph{columns} of a matrix $\matrixhat{W}$ and the weights with the third digit expressed in non-normalized $\alpha$-basis coordinates moved to the end as rows of a matrix $\matrixhat{W}'$,
the projection matrix $\matrixhat{P}$ can be determined from
\begin{equation}
    \matrixhat{P}\matrixhat{W} = \matrixhat{W}'
\end{equation}
with the right-inverse $\matrixhat{W}^{\!+}$ of $\matrixhat{W}$ (see section \ref{ssec:Bases}), since $\matrixhat{W}$ is in general not a rectangular matrix:
\begin{equation}
    \matrixhat{P} = \matrixhat{W}'\matrixhat{W}^{\!+}.
\end{equation}

As mentioned above the projection matrix found by this procedure can now be used to decompose any \SU5 irrep into $\SU3{\otimes}\SU2{\otimes}\U1$. The \irrep{10} is actually not the smallest irrep needed for the determination
of the projection matrix. The \irrep{10} as well as all other irreps can be built from tensor products of the \irrep{5}, which we call the \emph{generating irrep} of \SU5.
In the orthogonal algebras only tensor products of the so-called \emph{spinor representations} can construct all other irreps of the algebra. Thus, they must be used for the determination of the projection matrices.
The generating irreps of representative Lie algebras are listed in Table \ref{tab:GeneratingIrreps}.
\begin{table}[h]
    \begin{center}
        \begin{tabular}{lll}
            \toprule
            \textbf{Algebra} & \textbf{Irrep}    & \textbf{Irrep}\\
                             & \textbf{(Dynkin)} & \textbf{(Name)} \\
            \midrule
            \A4 (\SU5) & \dynkin{1,0,0,0}          & \irrep{5}       \\
            \B4 (\SO9) & \dynkin{0,0,0,1}          & \irrep{16}      \\
            \C4 (\Sp8) & \dynkin{1,0,0,0}          & \irrep{8}       \\
            \D4 (\SO8) & \dynkin{0,0,0,1}          & \irrepsub{8}{s} \\
            \E6 & \dynkin{1,0,0,0,0,0}      & \irrep{27}      \\
            \E7 & \dynkin{0,0,0,0,0,1,0}    & \irrep{56}      \\
            \E8 & \dynkin{0,0,0,0,0,0,1,0}  & \irrep{248}     \\
            \F4 & \dynkin{0,0,0,1}          & \irrep{26}      \\
            \G2 & \dynkin{1,0}              & \irrep{7}       \\
            \bottomrule
        \end{tabular}
        \caption{\label{tab:GeneratingIrreps} Generating Irreps of representative Lie algebras}
    \end{center}
    \vspace{-15pt}
\end{table}

In fact LieART excludes the zero-weights from the generating irreps, if any, i.e., only the lowest non-trivial orbit is needed for the determination of the projection matrices.

The calculation of a projection requires the knowledge of the simple root to drop from the Dynkin diagram for a specified algebra-subalgebra pair.
LieART provides an extra package file called \texttt{BranchingRules.m}, listing this information for the implemented branching rules along with special embeddings to be discussed later.
The file will be extended to encompass more branching rules in future versions of LieART, but may also be extended by the user. The definition for the more general branching rule $\SU{n}\to\SU{N{-}k}{\otimes}\SU{k}{\otimes}\U1$,
including the demonstrated case $\SU5\to\SU3{\otimes}\SU2{\otimes}\U1$, reads:

{\ttfamily\hangingindent
ProjectionMatrix[\pattern{origin}:Algebra[A][\pattern{n}\_],\newline
\hspace*{4.3ex}ProductAlgebra[Algebra[A][\pattern{m}\_],Algebra[A][\pattern{k}\_],Algebra[U][1]]] :=\newline
\hspace*{4.3ex}NonSemiSimpleSubalgebra[\pattern{origin},-\pattern{k}-1] /; \pattern{m}==(\pattern{n}-\pattern{k}-1)
}

\begin{figure}[t]
    \begin{center}
        \includegraphics[scale=0.95]{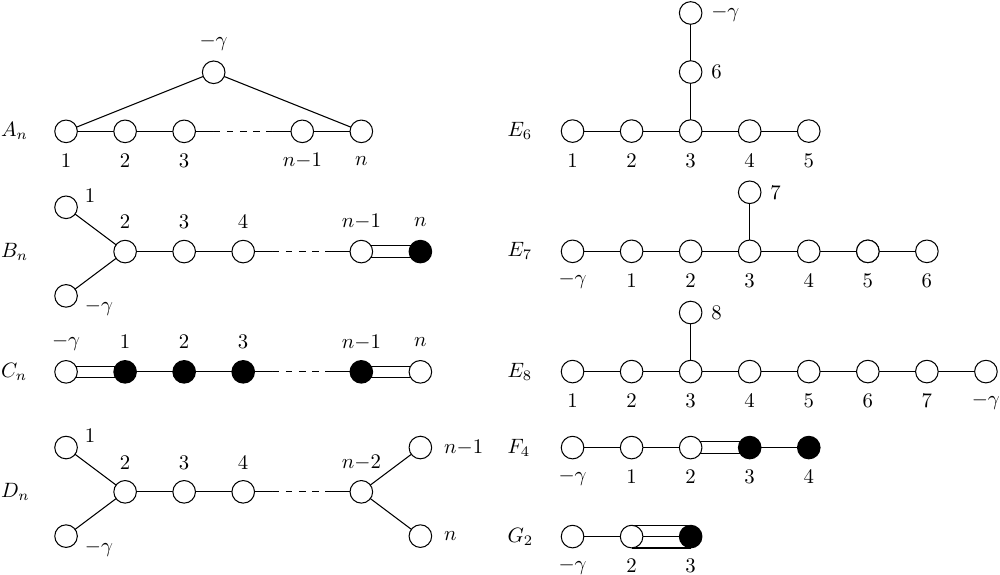}
        \caption{\label{fig:ExtendedDynkinDiagrams} Extended Dynkin Diagrams of classical and exceptional simple Lie algebras.}
    \end{center}
\end{figure}

\paragraph{Semisimple Subalgebras}

To obtain a semisimple subalgebra without a \U1 generator, a root from the
so-called \emph{extended Dynkin diagram} is removed. The extended Dynkin diagram
is constructed by adding the most negative root to the set of simple roots. (The
negative of the highest root $\gamma$ gives the most negative root $-\gamma$ to
form the extended Dynkin diagram.) The resulting set of roots is linearly
dependent, but removing one root restores the linear independence yielding a
valid system of simple root of a subalgebra, which in general is semisimple. The
highest roots $\gamma$ and the according extended root $-\gamma$ for
representative Lie algebras are listed in Table \ref{tab:MostNegativeRoots}.
\begin{table}[h]
    \begin{center}
        \begin{tabular}{lll}
            \toprule
            \textbf{Algebra} & \textbf{Highest Root}    & \textbf{Extended Root}\\
                             & \boldmath$(\gamma)$               & \boldmath$({-}\gamma)$  \\
            \midrule
            \A4 (\SU5) & \rootomega{1, 0, 0, 1}              & \rootomega{{-}1, 0, 0, {-}1}            \\
            \B4 (\SO9) & \rootomega{0, 1, 0, 0}              & \rootomega{0, {-}1, 0, 0}             \\
            \C4 (\Sp8) & \rootomega{2, 0, 0, 0}              & \rootomega{{-}2, 0, 0, 0}             \\
            \D4 (\SO8) & \rootomega{0, 1, 0, 0}              & \rootomega{0, {-}1, 0, 0}             \\
            \E6        & \rootomega{0, 0, 0, 0, 0, 1}        & \rootomega{0, 0, 0, 0, 0, {-}1}       \\
            \E7        & \rootomega{1, 0, 0, 0, 0, 0, 0}     & \rootomega{{-}1, 0, 0, 0, 0, 0, 0}    \\
            \E8        & \rootomega{0, 0, 0, 0, 0, 0, 1, 0}  & \rootomega{0, 0, 0, 0, 0, 0, {-}1, 0} \\
            \F4        & \rootomega{1, 0, 0, 0}              & \rootomega{{-}1, 0, 0, 0}             \\
            \G2        & \rootomega{0, 1}                    & \rootomega{0, {-}1}                   \\
            \bottomrule
        \end{tabular}
        \caption{\label{tab:MostNegativeRoots} Highest roots $\gamma$ and most negative roots $-\gamma$ of representative Lie algebras.}
    \end{center}
    \vspace{-15pt}
\end{table}
The non-zero entries in the Dynkin label of ${-}\gamma$ prescribe to which 
existing dot in the Dynkin diagram it should connect, since the Dynkin label in 
the $\omega$-basis encode the angle between two simple roots. A ``1'' is an 
angle of 120\textdegree, symbolized by a single connected line in the Dynkin 
diagram. A ``2'' is an angle of 135\textdegree, expressed by a double line in 
the Dynkin diagram. The minus sign gives negative angles or reverses the order 
of roots. The extended Dynkin diagrams for all classical and exceptional Lie 
Algebras are shown in Figure~\ref{fig:ExtendedDynkinDiagrams}. Please note that the 
double line connecting the extended root $-\gamma$ for \C{n} is according to 
the ``${-}2$'' in its Dynkin label.

To demonstrate the determination of the projection matrix from using the
generating irrep, we cannot use an irrep of \SU{N}, because dropping a root
from the extended Dynkin diagram of \SU{N} returns \SU{N}. Thus, \SU{N} has no
\emph{regular} \emph{maximal} semisimple subalgebra. (Please note that some
\SU{N}'s have \emph{special} maximal semisimple subalgebras, e.g.,
$\SU4\to\SU2{\otimes}\SU2$.) Instead we consider the subalgebra branching of
$\SO7{\to}\SU2{\otimes}\SU2{\otimes}\SU2$
($\B3{\to}\A1{\otimes}\A1{\otimes}\A1$). The maximal subalgebra
$\SU2{\otimes}\SU2{\otimes}\SU2$ is obtained from the extended Dynkin diagram of
\SO7 (\B3) by removing the second dot:
\begin{center}
\includegraphics{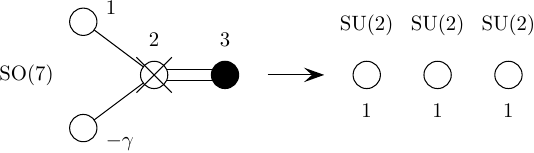}
\end{center}
To derive the projection matrix, we investigate the decomposition of the \SO7 generating irrep (the \irrep{8}) into three $\SU2$s.
Extending the Dynkin diagram with ${-}\gamma$ has the effect that each weight $w$ gets extended by one entry with the coefficient of
the weight relative to $-\gamma$, obtained by their scalar product: $\scalarproduct{w}{{-}\gamma}$. The so-called \emph{extended weight scheme}
of the lowest non-trivial orbit of a generating irrep is determined by the LieART function \com{ExtendedWeightScheme[\args{algebra},\args{simpleRootToDrop}]},
which directly removes the Dynkin digits associated to the simple root to drop, specified by \args{simpleRootToDrop}.
For the lowest non-trivial orbit of the generating irrep of \SO7 these two steps are:
\begin{equation}\label{eq:DropExtendedWeightScheme}
    \begin{matrix}
        \weight{0, 0, 1}\\\weight{0, 1, {-}1}\\\weight{1, {-}1, 1}\\\weight{{-}1, 0, 1}\\\weight{1, 0, {-}1}\\\weight{{-}1, 1, {-}1}\\\weight{0, {-}1, 1}\\\weight{0, 0, {-}1}
    \end{matrix}
    \xrightarrow{\text{insert ${{-}}\gamma$}}
    \begin{matrix}
        \weight{0, {-}1, 0, 1}\\\weight{0, {-}1, 1, {-}1}\\\weight{1, 0, {-}1, 1}\\\weight{{-}1, 0, 0, 1}\\\weight{1, 0, 0, {-}1}\\\weight{{-}1, 0, 1, {-}1}\\\weight{0, 1, {-}1, 1}\\\weight{0, 1, 0, {-}1}
        \end{matrix}
    \xrightarrow{\text{drop 2}}
    \begin{matrix}
        \weight{0, {-}1, 1}\\\weight{0, {-}1, {-}1}\\\weight{1, 0, 1}\\\weight{{-}1, 0, 1}\\\weight{1, 0, {-}1}\\\weight{{-}1, 0, {-}1}\\\weight{0, 1, 1}\\\weight{0, 1, {-}1}
    \end{matrix}
\end{equation}
With the weight of the \SO7 generating irrep as columns of the matrix
$\matrixhat{W}$ and the weights in the $3\,\SU2$ decomposition (right-hand side
of \eqref{eq:DropExtendedWeightScheme}) as columns of $\matrixhat{W}'$ the
projection matrix $\matrixhat{P}$ is computed as described for non-semisimple
regular subalgebras as $\matrixhat{P}{=}\matrixhat{W}'\matrixhat{W}^{\!+}$ with
the right-inverse $\matrixhat{W}^{\!+}\!$ of $\matrixhat{W}$.

The definition for the branching rule $\SO7{\to}\SU2{\otimes}\SU2{\otimes}\SU2$ in the file \texttt{BranchingRules.m} reads:

{\ttfamily\hangingindent
ProjectionMatrix[\pattern{origin}:Algebra[B][3],\newline
\hspace*{4.3ex}ProductAlgebra[Algebra[A][1],Algebra[A][1],Algebra[A][1]]]:=\newline
\hspace*{4.3ex}SemiSimpleSubalgebra[\pattern{origin},2]
}

\paragraph{Special Subalgebras}

Special maximal subalgebras cannot be derived from the root system. The embedding of a special subalgebra does not follow a general pattern and must be derived for every algebra-subalgebra pair individually.
Generating irreps are also used to derive the subalgebra embedding, which may be simple or semisimple and can involve more than one irrep of the subalgebra. LieART is not equipped with an algorithm
to determine the maximal special subalgebras, but provides an interface to declare the embeddings (\texttt{BranchingRules.m}), which can be taken from the literature \cite{Slansky,McKay:99021}.

As an example we consider \SO7 (\B3) again, which has \G2 as special maximal subalgebra. The generating spinor irrep of \SO7, the \irrep{8}, decomposes to the \G2 singlet plus the \irrep{7}.
The weights of the \irrep{8} of \SO7 and the weights of both the \irrep{1} and \irrep{7} of \G2 are brought into lexicographical order to define the projection matrix:
\begin{equation}\label{eq:SO7ToG2}
    \begin{array}{l@{\:\to\:}l}
        \weight{1, 0, {-}1}  & \weight{2, {-}1} \\
        \weight{1, {-}1, 1}  & \weight{1, 0}  \\
        \weight{0, 1, {-}1}  & \weight{1, {-}1} \\
        \weight{0, 0, 1}   & \weight{0, 0}  \\
        \weight{0, 0, {-}1}  & \weight{0, 0}  \\
        \weight{0, {-}1, 1}  & \weight{{-}1, 1} \\
        \weight{{-}1, 1, {-}1} & \weight{{-}1, 0} \\
        \weight{{-}1, 0, 1}  & \weight{{-}2, 1} \\
    \end{array}
\end{equation}
Arranging the weights on the left-hand side of \eqref{eq:SO7ToG2} as columns of $\matrixhat{W}$ and the weights of the right-hand side as columns of $\matrixhat{W}'$, the projection matrix $\matrixhat{P}$ is
again computed via $\matrixhat{P}{=}\matrixhat{W}'\matrixhat{W}^{\!+}$ with the right-inverse $\matrixhat{W}^{\!+}\!$ of $\matrixhat{W}$.

These procedures are performed by the LieART function
\com{SpecialSubalgebra[\args{origin},\args{targetirreps}]}. The definition for the branching rule $\SO7{\to}\G2$ in the file \texttt{BranchingRules.m} reads:

{\ttfamily\hangingindent
ProjectionMatrix[\pattern{origin}:Algebra[B][3],ProductAlgebra[G2]]:=\newline
\hspace*{4.3ex}SpecialSubalgebra[\pattern{origin},\newline
\hspace*{4.3ex}\{ProductIrrep[Irrep[G2][0,0]],ProductIrrep[Irrep[G2][1,0]]\}]
}

Please note that irreps of the subalgebra must be gathered in a list (\{\ldots\}), even if it is a single irrep. The projection matrix for $\SO7{\to}\G2$ is
\begin{mathin}
ProjectionMatrix[B3,ProductAlgebra[G2]]
\end{mathin}
\begin{mathout}
    $\begin{pmatrix}
    2 & 1 & 0 \\
    {-}1 & {-}1 & 0 \\
    \end{pmatrix}$
\end{mathout}

\pagebreak

\section{Benchmarks}
\label{sec:Benchmarks}

In this section we give runtime benchmarks for the tensor decomposition and 
subalgebra decomposition. Because exceptional algebras have complicated Weyl 
reflection groups of high orders, computations involving them are much more CPU 
and memory demanding than with classical algebras of equal rank. In the 
following we give runtime benchmarks for the tensor-product decomposition and 
subalgebra decomposition of irreps of exceptional algebras. 

We use the Mathematica command 
\texttt{Timing[]}, which gives the CPU time spent in the Mathematica kernel in 
seconds. It does not include the time needed for the display of results in the 
front end. As pointed out in Section\ \ref{ssec:irrepNames}, in 
\texttt{TraditionalForm} irreps are displayed by their dimensional names, while 
computations are performed by their Dynkin labels. The determination of the 
dimensional name can be very time consuming depending on the algebra and the 
irrep. This time is allotted to the display of computation results and thus not 
measured with \texttt{Timing[]}. For most of the examples below you will find 
significant differences between the displayed time and the wall-clock time due to 
this effect, but this is intended as we want to give an estimate of the 
actual time needed for decompositions and not the display of results. Please 
note that each of the following timings are taken with a newly launched Kernel 
to avoid speedup due to caching of intermediate results from previous 
computations.

The following timings were taken with Mathematica 8.0.1.\ on an 
Apple\textsuperscript{\textregistered} iMac\textsuperscript{\textregistered} 
with an Intel\textsuperscript{\textregistered} 
Core\textsuperscript{\texttrademark} i5 750 (2.66\,GHz) processor and 4\,GB RAM.

We compute ${\irrep{27}}^n$ for $n=2,\ldots,8$ in \E6:
\begin{mathin}
Timing[Irrep[E6][27]\textasciicircum2]
\end{mathin}
\begin{mathout}
$\{0.136494,\irrepbar{27}+\irrepbar{351}+\irrepbar[1]{351}\}$
\end{mathout}

\begin{mathin}
Timing[Irrep[E6][27]\textasciicircum3]
\end{mathin}
\begin{mathout}
$\{0.143171,\irrep{1}+2(\irrep{78})+3(\irrep{650})+\irrep{2925}+\irrep{3003}+2(\irrep{5824})\}$
\end{mathout}

\begin{mathin}
Timing[Irrep[E6][27]\textasciicircum4]
\end{mathin}
\begin{mathout}
$\{0.43606,6(\irrep{27})+6(\irrep{351})+3(\irrep[1]{351})+8(\irrep{1728})+6(\irrep{7371})+6(\irrep{7722})+\irrep{17550}+\irrep[1]{19305}+2(\irrep{34398})+3(\irrep{51975})+3(\irrep{54054})\}$
\end{mathout}

\begin{mathin}
Timing[Irrep[E6][27]\textasciicircum5]
\end{mathin}
\begin{mathout}
$\{0.715819,15(\irrepbar{27})+26(\irrepbar{351})+20(\irrepbar[1]{351})+24(\irrepbar{1728})+30(\irrepbar{7371})+15(\irrepbar{7722})+20(\irrepbar{17550})+20(\irrepbar{19305})+\irrepbar{34398}+\irrepbar{46332}+10(\irrepbar{51975})+10(\irrepbar{61425})+\irrepbar{100386}+20(\irrepbar{112320})+4(\irrepbar{314496})+4(\irrepbar[1]{359424})+5(\irrepbar{386100})+5(\irrepbar{412776})+6(\irrepbar{494208})\}$
\end{mathout}

\begin{mathin}
Timing[Irrep[E6][27]\textasciicircum6]
\end{mathin}
\begin{mathout}
$\{1.53349,15(\irrep{1})+65(\irrep{78})+130(\irrep{650})+45(\irrep{2430})+110(\irrep{2925})+50(\irrep{3003})+15(\irrepbar{3003})+136(\irrep{5824})+80(\irrepbar{5824})+144(\irrep{34749})+\irrep{43758}+90(\irrep{70070})+90(\irrep{78975})+45(\irrepbar{78975})+45(\irrep{85293})+40(\irrep{105600})+40(\irrep{146432})+80(\irrep{252252})+16(\irrepbar{252252})+15(\irrep{371800})+\irrep{442442}+5(\irrep{600600})+30(\irrepbar{600600})+5(\irrep{812175})+45(\irrep{852930})+45(\irrep{972972})+5(\irrep{1337050})+5(\irrep{1559376})+5(\irrep{1896180})+9(\irrep{2453814})+10(\irrep{2977975})+9(\irrep{3007368})+10(\irrep{3309696})+16(\irrep{4752384})\}$
\end{mathout}

\begin{mathin}
Timing[Irrep[E6][27]\textasciicircum7]
\end{mathin}
\begin{mathout}
$\{12.9962,210(\irrep{27})+385(\irrep{351})+225(\irrep[1]{351})+630(\irrep{1728})+735(\irrep{7371})+595(\irrep{7722})+525(\irrep{17550})+300(\irrep{19305})+105(\irrep[1]{19305})+336(\irrep{34398})+315(\irrep{46332})+735(\irrep{51975})+441(\irrep{54054})+105(\irrep{61425})+560(\irrep{112320})+504(\irrep{314496})+504(\irrep{359424})+210(\irrep{386100})+71(\irrep{393822})+21(\irrep{412776})+21(\irrep{459459})+106(\irrep{494208})+210(\irrep{579150})+105(\irrep{638820})+6(\irrep{741312})+70(\irrep{853281})+420(\irrep{967680})+210(\irrep{1123200})+140(\irrep{1253070})+210(\irrep{1640925})+\irrep{1706562}+21(\irrep{1837836})+210(\irrep{2088450})+90(\irrep{4200768})+14(\irrep{4582656})+15(\irrep{5553900})+105(\irrep{5776056})+84(\irrep{6110208})+14(\irrep{6243237})+105(\irrep{6747300})+15(\irrep{7528950})+126(\irrep{7601958})+6(\irrep{8401536})+14(\irrep{10378368})+14(\irrep{14805504})+14(\irrep{16540524})+15(\irrep{17453475})+21(\irrep{17918901})+21(\irrep{19297278})+20(\irrep{19768320})+35(\irrep{30115800})+35(\irrep[1]{34906950})\}$
\end{mathout}
\pagebreak

\begin{mathin}
Timing[Irrep[E6][27]\textasciicircum8]
\end{mathin}
\begin{mathout}
$\{93.8079,820(\irrepbar{27})+1960(\irrepbar{351})+1435(\irrepbar[1]{351})+2800(\irrepbar{1728})+4165(\irrepbar{7371})+2520(\irrepbar{7722})+3780(\irrepbar{17550})+3220(\irrepbar{19305})+105(\irrepbar[1]{19305})+1316(\irrepbar{34398})+1736(\irrepbar{46332})+3675(\irrepbar{51975})+1092(\irrepbar{54054})+1960(\irrepbar{61425})+196(\irrepbar{100386})+5040(\irrepbar{112320})+4144(\irrepbar{314496})+2400(\irrepbar{359424})+1120(\irrepbar[1]{359424})+2660(\irrepbar{386100})+1260(\irrepbar{393822})+1652(\irrepbar{412776})+1260(\irrepbar{459459})+2856(\irrepbar{494208})+421(\irrepbar{579150})+420(\irrepbar{638820})+112(\irrepbar{741312})+2240(\irrepbar{967680})+2240(\irrepbar{1123200})+420(\irrepbar{1253070})+1345(\irrepbar{1640925})+238(\irrepbar{2088450})+1344(\irrepbar{2559843})+420(\irrepbar{3281850})+210(\irrepbar{3675672})+112(\irrepbar{4088448})+2688(\irrepbar{4200768})+840(\irrepbar{4582656})+20(\irrepbar{5501925})+420(\irrepbar{5553900})+1260(\irrepbar{5776056})+\irrepbar{5895396}+455(\irrepbar{6243237})+21(\irrepbar{6675669})+140(\irrepbar{6747300})+140(\irrepbar{7528950})+637(\irrepbar{7601958})+1260(\irrepbar{7757100})+28(\irrepbar[1]{7757100})+630(\irrepbar{9189180})+560(\irrepbar{10378368})+448(\irrepbar{12648636})+560(\irrepbar{13478400})+672(\irrepbar{14017536})+140(\irrepbar{17918901})+28(\irrepbar{19297278})+14(\irrepbar{22007700})+140(\irrepbar{23629320})+28(\irrepbar{26702676})+272(\irrepbar{30115800})+140(\irrepbar{30718116})+7(\irrepbar{32424678})+35(\irrepbar{36100350})+301(\irrepbar{37459422})+14(\irrepbar{41442192})+252(\irrepbar{46542600})+280(\irrepbar{48243195})+35(\irrepbar{49017150})+64(\irrepbar{54991872})+448(\irrepbar{66830400})+20(\irrepbar{74826180})+64(\irrepbar{75119616})+21(\irrepbar{77026950})+28(\irrepbar[1]{89791416})+42(\irrepbar{93459366})+35(\irrepbar{103169430})+56(\irrepbar{123803316})+56(\irrepbar{136547775})+70(\irrepbar{138881925})+70(\irrepbar{184864680})+64(\irrepbar{192067200})+90(\irrepbar{219490128})\}$
\end{mathout}

and ${\irrep{78}}^n$ for $n=2,\ldots,7$ in \E6:
\begin{mathin}
Timing[Irrep[E6][78]\textasciicircum2]
\end{mathin}
\begin{mathout}
$\{0.182129,\irrep{1}+\irrep{78}+\irrep{650}+\irrep{2430}+\irrep{2925}\}$
\end{mathout}

\begin{mathin}
Timing[Irrep[E6][78]\textasciicircum3]
\end{mathin}
\begin{mathout}
$\{0.20391,\irrep{1}+5(\irrep{78})+4(\irrep{650})+3(\irrep{2430})+4(\irrep{2925})+2(\irrep{5824})+2(\irrepbar{5824})+3(\irrep{34749})+\irrep{43758}+\irrep{70070}+2(\irrep{105600})\}$
\end{mathout}

\begin{mathin}
Timing[Irrep[E6][78]\textasciicircum4]
\end{mathin}
\begin{mathout}
$\{2.33734,5(\irrep{1})+17(\irrep{78})+24(\irrep{650})+18(\irrep{2430})+26(\irrep{2925})+2(\irrep{3003})+2(\irrepbar{3003})+16(\irrep{5824})+16(\irrepbar{5824})+27(\irrep{34749})+6(\irrep{43758})+15(\irrep{70070})+6(\irrep{78975})+6(\irrepbar{78975})+3(\irrep{85293})+16(\irrep{105600})+8(\irrep{252252})+8(\irrepbar{252252})+\irrep{537966}+\irrep{600600}+\irrepbar{600600}+6(\irrep{812175})+6(\irrep{852930})+2(\irrep{1337050})+3(\irrep{1911195})+3(\irrep{2453814})\}$
\end{mathout}

\begin{mathin}
Timing[Irrep[E6][78]\textasciicircum5]
\end{mathin}
\begin{mathout}
$\{5.25617,17(\irrep{1})+90(\irrep{78})+150(\irrep{650})+110(\irrep{2430})+175(\irrep{2925})+24(\irrep{3003})+24(\irrepbar{3003})+140(\irrep{5824})+140(\irrepbar{5824})+255(\irrep{34749})+50(\irrep{43758})+170(\irrep{70070})+90(\irrep{78975})+90(\irrepbar{78975})+51(\irrep{85293})+160(\irrep{105600})+16(\irrep{146432})+16(\irrepbar{146432})+120(\irrep{252252})+120(\irrepbar{252252})+10(\irrep{537966})+40(\irrep{600600})+40(\irrepbar{600600})+95(\irrep{812175})+120(\irrep{852930})+24(\irrep{972972})+24(\irrepbar{972972})+30(\irrep{1337050})+\irrep{1559376}+\irrepbar{1559376}+40(\irrep{1911195})+75(\irrep{2453814})+30(\irrep{2977975})+30(\irrepbar{2977975})+15(\irrep{3490695})+20(\irrep{4752384})+20(\irrepbar{4752384})+\irrep{4969107}+20(\irrep{5054400})+20(\irrepbar{5054400})+10(\irrep{11655930})+11(\irrep{12514788})+4(\irrep{19160064})+4(\irrepbar{19160064})+4(\irrep{22843392})+20(\irrep{23795200})+5(\irrep{29422393})+5(\irrep{34906950})+6(\irrep{42134742})\}$
\end{mathout}

\begin{mathin}
Timing[Irrep[E6][78]\textasciicircum6]
\end{mathin}
\begin{mathout}
$\{14.811,90(\irrep{1})+542(\irrep{78})+1100(\irrep{650})+840(\irrep{2430})+1390(\irrep{2925})+270(\irrep{3003})+270(\irrepbar{3003})+1264(\irrep{5824})+1264(\irrepbar{5824})+2496(\irrep{34749})+465(\irrep{43758})+1905(\irrep{70070})+1170(\irrep{78975})+1170(\irrepbar{78975})+720(\irrep{85293})+1680(\irrep{105600})+320(\irrep{146432})+320(\irrepbar{146432})+1615(\irrep{252252})+1615(\irrepbar{252252})+40(\irrep{371800})+40(\irrepbar{371800})+115(\irrep{537966})+745(\irrep{600600})+745(\irrepbar{600600})+1370(\irrep{812175})+1890(\irrep{852930})+585(\irrep{972972})+585(\irrepbar{972972})+495(\irrep{1337050})+65(\irrep{1559376})+65(\irrepbar{1559376})+585(\irrep{1911195})+1350(\irrep{2453814})+760(\irrep{2977975})+760(\irrepbar{2977975})+45(\irrep{3007368})+45(\irrepbar{3007368})+66(\irrep{3162159})+66(\irrepbar{3162159})+90(\irrep{3309696})+90(\irrepbar{3309696})+450(\irrep{3490695})+15(\irrep{4548180})+560(\irrep{4752384})+560(\irrepbar{4752384})+15(\irrep{4969107})+480(\irrep{5054400})+480(\irrepbar{5054400})+80(\irrep{7779200})+80(\irrepbar{7779200})+245(\irrep{11655930})+515(\irrep{12514788})+240(\irrep{19160064})+240(\irrepbar{19160064})+80(\irrep{22843392})+640(\irrep{23795200})+115(\irrep{29422393})+144(\irrep{32752512})+144(\irrepbar{32752512})+210(\irrep{34906950})+\irrep{36685506}+225(\irrep{42134742})+5(\irrep{44767800})+5(\irrepbar{44767800})+41(\irrep{47783736})+41(\irrepbar{47783736})+90(\irrep{47849373})+90(\irrepbar{47849373})+90(\irrep{53557504})+90(\irrepbar{53557504})+15(\irrep{59073300})+15(\irrepbar{59073300})+46(\irrep{64205141})+40(\irrep{64414350})+40(\irrepbar{64414350})+45(\irrep{66023100})+80(\irrep{115287744})+80(\irrepbar{115287744})+15(\irrep{119189070})+5(\irrep{200449886})+5(\irrep{203365305})+5(\irrep{221077350})+30(\irrep{226459233})+9(\irrep{252808452})+9(\irrepbar{252808452})+10(\irrep{303388800})+10(\irrepbar{303388800})+50(\irrep{348985350})+45(\irrep{350895402})+9(\irrep{366235506})+10(\irrep{476952476})+16(\irrep{734557824})\}$
\end{mathout}

\begin{mathin}
Timing[Irrep[E6][78]\textasciicircum7]
\end{mathin}
\begin{mathout}
$\{334.518,542(\irrep{1})+3962(\irrep{78})+9156(\irrep{650})+7413(\irrep{2430})+12481(\irrep{2925})+3024(\irrep{3003})+3024(\irrepbar{3003})+12474(\irrep{5824})+12474(\irrepbar{5824})+26481(\irrep{34749})+5055(\irrep{43758})+22204(\irrep{70070})+14910(\irrep{78975})+14910(\irrepbar{78975})+9786(\irrep{85293})+19376(\irrep{105600})+5236(\irrep{146432})+5236(\irrepbar{146432})+21420(\irrep{252252})+21420(\irrepbar{252252})+1155(\irrep{371800})+1155(\irrepbar{371800})+1505(\irrep{537966})+12145(\irrep{600600})+12145(\irrepbar{600600})+19621(\irrep{812175})+28161(\irrep{852930})+10899(\irrep{972972})+10899(\irrepbar{972972})+8085(\irrep{1337050})+1701(\irrep{1559376})+1701(\irrepbar{1559376})+175(\irrep{1896180})+175(\irrepbar{1896180})+8910(\irrep{1911195})+22365(\irrep{2453814})+14770(\irrep{2977975})+14770(\irrepbar{2977975})+1575(\irrep{3007368})+1575(\irrepbar{3007368})+2016(\irrep{3162159})+2016(\irrepbar{3162159})+2870(\irrep{3309696})+2870(\irrepbar{3309696})+9729(\irrep{3490695})+686(\irrep{4548180})+11760(\irrep{4752384})+11760(\irrepbar{4752384})+231(\irrep{4969107})+9450(\irrep{5054400})+9450(\irrepbar{5054400})+2990(\irrep{7779200})+2990(\irrepbar{7779200})+5075(\irrep{11655930})+13216(\irrep{12514788})+420(\irrep{14152320})+420(\irrepbar{14152320})+6804(\irrep{19160064})+6804(\irrepbar{19160064})+1624(\irrep{22843392})+15114(\irrep{23795200})+2730(\irrep{29422393})+5544(\irrep{32752512})+5544(\irrepbar{32752512})+6111(\irrep{34906950})+21(\irrep{36685506})+5901(\irrep{42134742})+196(\irrep{42398720})+196(\irrepbar{42398720})+455(\irrep{44767800})+455(\irrepbar{44767800})+490(\irrep{45741696})+490(\irrepbar{45741696})+2156(\irrep{47783736})+2156(\irrepbar{47783736})+3870(\irrep{47849373})+3870(\irrepbar{47849373})+3430(\irrep{53557504})+3430(\irrepbar{53557504})+300(\irrep[1]{54991872})+300(\irrepbar[1]{54991872})+1260(\irrep{59073300})+1260(\irrepbar{59073300})+2541(\irrep{64205141})+1400(\irrep{64414350})+1400(\irrepbar{64414350})+2079(\irrep{66023100})+560(\irrep{85974525})+560(\irrepbar{85974525})+315(\irrep{89791416})+315(\irrepbar{89791416})+3360(\irrep{115287744})+3360(\irrepbar{115287744})+525(\irrep{119189070})+106(\irrep{152423700})+175(\irrep{200449886})+140(\irrep{203365305})+70(\irrep{212838912})+70(\irrepbar{212838912})+386(\irrep{221077350})+\irrep{225961450}+1575(\irrep{226459233})+560(\irrep{236487680})+560(\irrepbar{236487680})+819(\irrep{252808452})+819(\irrepbar{252808452})+840(\irrep{303388800})+840(\irrepbar{303388800})+3465(\irrep{348985350})+2100(\irrep{350895402})+294(\irrep{366235506})+210(\irrep{392837445})+210(\irrepbar{392837445})+505(\irrep{466237200})+505(\irrepbar{466237200})+525(\irrep{476952476})+14(\irrep{532097280})+14(\irrepbar{532097280})+504(\irrep{537567030})+504(\irrepbar{537567030})+70(\irrep{598998400})+70(\irrepbar{598998400})+210(\irrep{625532544})+210(\irrepbar{625532544})+15(\irrep{649806300})+15(\irrepbar{649806300})+126(\irrep{688740975})+966(\irrep{734557824})+105(\irrep{797489550})+140(\irrep{929510400})+140(\irrepbar{929510400})+21(\irrep{944929700})+420(\irrep{1051315200})+420(\irrepbar{1051315200})+216(\irrep{1177830720})+216(\irrepbar{1177830720})+6(\irrep{1445558400})+216(\irrep{1478062080})+210(\irrep{1525620096})+210(\irrepbar{1525620096})+84(\irrep{1544524800})+84(\irrepbar{1544524800})+14(\irrep[1]{1643241600})+14(\irrepbar[1]{1643241600})+14(\irrep{3116305920})+20(\irrep{3203785728})+20(\irrepbar{3203785728})+14(\irrep{3256917300})+84(\irrep{3548188800})+119(\irrep{3863940795})+15(\irrep{4035297123})+105(\irrep{4129204716})+21(\irrep{4790483775})+35(\irrep{4942962024})+35(\irrepbar{4942962024})+21(\irrep{4973434830})+141(\irrep{4991693850})+35(\irrep{8713554850})\}$
\end{mathout}

and ${\irrep{248}}^n$ for $n=2,\ldots,7$ in \E8:
\begin{mathin}
Timing[Irrep[E8][248]\textasciicircum2]
\end{mathin}
\begin{mathout}
$\{0.850115,\irrep{1}+\irrep{248}+\irrep{3875}+\irrep{27000}+\irrep{30380}\}$
\end{mathout}

\begin{mathin}
Timing[Irrep[E8][248]\textasciicircum3]
\end{mathin}
\begin{mathout}
$\{0.883473,\irrep{1}+5(\irrep{248})+3(\irrep{3875})+3(\irrep{27000})+4(\irrep{30380})+2(\irrep{147250})+3(\irrep{779247})+\irrep{1763125}+\irrep{2450240}+2(\irrep{4096000})\}$
\end{mathout}

\begin{mathin}
Timing[Irrep[E8][248]\textasciicircum4]
\end{mathin}
\begin{mathout}
$\{31.9855,5(\irrep{1})+16(\irrep{248})+17(\irrep{3875})+18(\irrep{27000})+23(\irrep{30380})+13(\irrep{147250})+21(\irrep{779247})+6(\irrep{1763125})+12(\irrep{2450240})+16(\irrep{4096000})+3(\irrep{4881384})+6(\irrep{6696000})+8(\irrep{26411008})+6(\irrep{70680000})+6(\irrep{76271625})+\irrep{79143000}+\irrep{146325270}+2(\irrep{203205000})+3(\irrep{281545875})+3(\irrep{344452500})\}$
\end{mathout}

\begin{mathin}
Timing[Irrep[E8][248]\textasciicircum5]
\end{mathin}
\begin{mathout}
$\{49.9295,16(\irrep{1})+79(\irrep{248})+90(\irrep{3875})+100(\irrep{27000})+136(\irrep{30380})+100(\irrep{147250})+170(\irrep{779247})+50(\irrep{1763125})+109(\irrep{2450240})+140(\irrep{4096000})+36(\irrep{4881384})+70(\irrep{6696000})+100(\irrep{26411008})+75(\irrep{70680000})+90(\irrep{76271625})+10(\irrep{79143000})+36(\irrep{146325270})+30(\irrep{203205000})+40(\irrep{281545875})+24(\irrep{301694976})+60(\irrep{344452500})+15(\irrep{820260000})+30(\irrep{1094951000})+20(\irrep{2172667860})+20(\irrep{2275896000})+\irrep{2642777280}+10(\irrep{3929713760})+10(\irrep{4825673125})+\irrep{6899079264}+20(\irrep{8634368000})+4(\irrep{12692520960})+5(\irrep{17535336000})+4(\irrep{20288765952})+5(\irrep{21039669000})+6(\irrep{23592339045})\}$
\end{mathout}
\pagebreak

\begin{mathin}
Timing[Irrep[E8][248]\textasciicircum6]
\end{mathin}
\begin{mathout}
$\{90.112,79(\irrep{1})+421(\irrep{248})+575(\irrep{3875})+675(\irrep{27000})+924(\irrep{30380})+775(\irrep{147250})+1386(\irrep{779247})+415(\irrep{1763125})+1011(\irrep{2450240})+1240(\irrep{4096000})+405(\irrep{4881384})+765(\irrep{6696000})+1144(\irrep{26411008})+895(\irrep{70680000})+1125(\irrep{76271625})+115(\irrep{79143000})+554(\irrep{146325270})+410(\irrep{203205000})+510(\irrep{281545875})+456(\irrep{301694976})+855(\irrep{344452500})+315(\irrep{820260000})+605(\irrep{1094951000})+405(\irrep{2172667860})+470(\irrep{2275896000})+15(\irrep{2642777280})+15(\irrep{2903770000})+195(\irrep{3929713760})+45(\irrep{4076399250})+325(\irrep{4825673125})+125(\irrep{6899079264})+480(\irrep{8634368000})+80(\irrep[1]{8634368000})+80(\irrep{12692520960})+115(\irrep{17535336000})+216(\irrep{20288765952})+165(\irrep{21039669000})+180(\irrep{23592339045})+144(\irrep{45329752170})+45(\irrep{63513702720})+45(\irrep{66393847000})+\irrep{69176971200}+90(\irrep{83080364250})+90(\irrep{85424220000})+40(\irrep{110977024000})+40(\irrep{124436480000})+15(\irrep{152883490500})+15(\irrep{220778105625})+80(\irrep{234550030000})+\irrep{267413986840}+30(\irrep{355647996000})+5(\irrep{417933862500})+5(\irrep{492957660000})+45(\irrep{508731738750})+45(\irrep{574197082368})+5(\irrep{627099023250})+9(\irrep{841900509450})+5(\irrep{919045960000})+10(\irrep{1041872676000})+9(\irrep{1283242632840})+10(\irrep{1349926375875})+16(\irrep{1813461073920})\}$
\end{mathout}

\begin{mathin}
Timing[Irrep[E8][248]\textasciicircum7]
\end{mathin}
\begin{mathout}
$\{7636.29,421(\irrep{1})+2674(\irrep{248})+4081(\irrep{3875})+5061(\irrep{27000})+7007(\irrep{30380})+6580(\irrep{147250})+12306(\irrep{779247})+3850(\irrep{1763125})+9779(\irrep{2450240})+11830(\irrep{4096000})+4452(\irrep{4881384})+8226(\irrep{6696000})+12830(\irrep{26411008})+10465(\irrep{70680000})+13566(\irrep{76271625})+1330(\irrep{79143000})+7651(\irrep{146325270})+5370(\irrep{203205000})+6405(\irrep{281545875})+7014(\irrep{301694976})+11445(\irrep{344452500})+5250(\irrep{820260000})+9744(\irrep{1094951000})+6714(\irrep{2172667860})+8204(\irrep{2275896000})+231(\irrep{2642777280})+455(\irrep{2903770000})+3290(\irrep{3929713760})+1281(\irrep{4076399250})+6475(\irrep{4825673125})+3171(\irrep{6899079264})+8840(\irrep{8634368000})+2296(\irrep[1]{8634368000})+1414(\irrep{12692520960})+2240(\irrep{17535336000})+5124(\irrep{20288765952})+3696(\irrep{21039669000})+3717(\irrep{23592339045})+4410(\irrep{45329752170})+1449(\irrep{63513702720})+1674(\irrep{66393847000})+21(\irrep{69176971200})+2765(\irrep{83080364250})+3150(\irrep{85424220000})+1190(\irrep{110977024000})+1820(\irrep{124436480000})+420(\irrep{152883490500})+1155(\irrep{220778105625})+300(\irrep{223850628000})+2835(\irrep{234550030000})+546(\irrep{267413986840})+1155(\irrep{355647996000})+105(\irrep{417532087000})+140(\irrep{417933862500})+175(\irrep{492957660000})+1575(\irrep{508731738750})+315(\irrep{560213725500})+2170(\irrep{574197082368})+280(\irrep{627099023250})+294(\irrep{841900509450})+875(\irrep{919045960000})+420(\irrep{1041872676000})+560(\irrep{1198018560000})+735(\irrep{1283242632840})+756(\irrep{1349926375875})+\irrep{1473701482500}+756(\irrep{1813461073920})+105(\irrep{3067797300750})+504(\irrep{3191795712000})+504(\irrep{3233052753920})+105(\irrep{3431612952000})+\irrep{3754721200320}+210(\irrep{3950782290000})+70(\irrep{4007202600000})+210(\irrep{4189713446646})+21(\irrep{4490627295000})+70(\irrep{4779643627500})+\irrep{5006235840320}+210(\irrep{6458110083072})+21(\irrep{7723951192125})+420(\irrep{8145764352000})+140(\irrep{8715491428800})+6(\irrep{10701806469120})+210(\irrep{12737135385000})+210(\irrep{13532264250750})+84(\irrep{19994148864000})+84(\irrep{26125438976000})+105(\irrep{26461348084080})+14(\irrep{28123973939490})+105(\irrep{29369472656250})+14(\irrep{30014459904000})+15(\irrep{33372802062000})+6(\irrep{33699815424000})+126(\irrep{33943999320000})+14(\irrep{46678711926784})+21(\irrep{53540697687750})+14(\irrep{56860936405000})+21(\irrep{57306919524192})+20(\irrep{57591234560000})+15(\irrep{58549130859375})+35(\irrep{85471274280000})+35(\irrep{107701303073000})\}$
\end{mathout}
\pagebreak

As an example for subalgebra decomposition of a large irrep we decompose
the $\irrep{600600}$ of \E6 to $\SU3\times\SU3\times\SU3:$
\begin{mathin}
Timing[DecomposeIrrep[Irrep[E6][600600], ProductAlgebra[SU3, SU3, SU3]]]
\end{mathin}
\begin{mathout}
$\{124.17,(\irrep{1},\irrep{1},\irrep{1})+10(\irrepbar{3},\irrep{3},\irrep{3})+10(\irrep{3},\irrepbar{3},\irrepbar{3})+5(\irrep{8},\irrep{1},\irrep{1})+5(\irrep{1},\irrep{8},\irrep{1})+5(\irrep{1},\irrep{1},\irrep{8})+10(\irrep{6},\irrep{3},\irrep{3})+10(\irrepbar{3},\irrep{3},\irrepbar{6})+10(\irrepbar{3},\irrepbar{6},\irrep{3})+10(\irrep{3},\irrep{6},\irrepbar{3})+10(\irrep{3},\irrepbar{3},\irrep{6})+10(\irrepbar{6},\irrepbar{3},\irrepbar{3})+(\irrep{10},\irrep{1},\irrep{1})+(\irrepbar{10},\irrep{1},\irrep{1})+(\irrep{1},\irrep{10},\irrep{1})+(\irrep{1},\irrepbar{10},\irrep{1})+(\irrep{1},\irrep{1},\irrep{10})+(\irrep{1},\irrep{1},\irrepbar{10})+8(\irrep{6},\irrep{3},\irrepbar{6})+8(\irrep{6},\irrepbar{6},\irrep{3})+8(\irrepbar{3},\irrepbar{6},\irrepbar{6})+8(\irrep{3},\irrep{6},\irrep{6})+8(\irrepbar{6},\irrep{6},\irrepbar{3})+8(\irrepbar{6},\irrepbar{3},\irrep{6})+11(\irrep{8},\irrep{8},\irrep{1})+11(\irrep{8},\irrep{1},\irrep{8})+11(\irrep{1},\irrep{8},\irrep{8})+7(\irrep{6},\irrepbar{6},\irrepbar{6})+7(\irrepbar{6},\irrep{6},\irrep{6})+5(\irrep{10},\irrep{8},\irrep{1})+4(\irrep{10},\irrep{1},\irrep{8})+5(\irrep{8},\irrep{10},\irrep{1})+4(\irrep{8},\irrepbar{10},\irrep{1})+4(\irrep{8},\irrep{1},\irrep{10})+5(\irrep{8},\irrep{1},\irrepbar{10})+4(\irrepbar{10},\irrep{8},\irrep{1})+5(\irrepbar{10},\irrep{1},\irrep{8})+4(\irrep{1},\irrep{10},\irrep{8})+5(\irrep{1},\irrep{8},\irrep{10})+4(\irrep{1},\irrep{8},\irrepbar{10})+5(\irrep{1},\irrepbar{10},\irrep{8})+10(\irrepbar{15},\irrep{3},\irrep{3})+10(\irrepbar{3},\irrep{15},\irrep{3})+10(\irrepbar{3},\irrep{3},\irrep{15})+10(\irrep{15},\irrepbar{3},\irrepbar{3})+10(\irrep{3},\irrepbar{15},\irrepbar{3})+10(\irrep{3},\irrepbar{3},\irrepbar{15})+(\irrep{10},\irrep{10},\irrep{1})+2(\irrep{10},\irrepbar{10},\irrep{1})+2(\irrep{10},\irrep{1},\irrep{10})+2(\irrep{10},\irrep{1},\irrepbar{10})+2(\irrepbar{10},\irrep{10},\irrep{1})+2(\irrepbar{10},\irrepbar{10},\irrep{1})+2(\irrepbar{10},\irrep{1},\irrep{10})+(\irrepbar{10},\irrep{1},\irrepbar{10})+2(\irrep{1},\irrep{10},\irrep{10})+2(\irrep{1},\irrep{10},\irrepbar{10})+(\irrep{1},\irrepbar{10},\irrep{10})+2(\irrep{1},\irrepbar{10},\irrepbar{10})+2(\irrepbar[1]{15},\irrep{3},\irrep{3})+2(\irrepbar{3},\irrep[1]{15},\irrep{3})+2(\irrepbar{3},\irrep{3},\irrep[1]{15})+2(\irrep[1]{15},\irrepbar{3},\irrepbar{3})+2(\irrep{3},\irrepbar[1]{15},\irrepbar{3})+2(\irrep{3},\irrepbar{3},\irrepbar[1]{15})+25(\irrep{8},\irrep{8},\irrep{8})+9(\irrep{6},\irrep{15},\irrep{3})+9(\irrep{6},\irrep{3},\irrep{15})+9(\irrepbar{15},\irrep{3},\irrepbar{6})+9(\irrepbar{15},\irrepbar{6},\irrep{3})+9(\irrepbar{3},\irrep{15},\irrepbar{6})+9(\irrepbar{3},\irrepbar{6},\irrep{15})+9(\irrep{15},\irrep{6},\irrepbar{3})+9(\irrep{15},\irrepbar{3},\irrep{6})+9(\irrep{3},\irrep{6},\irrepbar{15})+9(\irrep{3},\irrepbar{15},\irrep{6})+9(\irrepbar{6},\irrepbar{15},\irrepbar{3})+9(\irrepbar{6},\irrepbar{3},\irrepbar{15})+(\irrep{6},\irrep[1]{15},\irrep{3})+(\irrep{6},\irrep{3},\irrep[1]{15})+(\irrepbar[1]{15},\irrep{3},\irrepbar{6})+(\irrepbar[1]{15},\irrepbar{6},\irrep{3})+(\irrepbar{3},\irrep[1]{15},\irrepbar{6})+(\irrepbar{3},\irrepbar{6},\irrep[1]{15})+(\irrep[1]{15},\irrep{6},\irrepbar{3})+(\irrep[1]{15},\irrepbar{3},\irrep{6})+(\irrep{3},\irrep{6},\irrepbar[1]{15})+(\irrep{3},\irrepbar[1]{15},\irrep{6})+(\irrepbar{6},\irrepbar[1]{15},\irrepbar{3})+(\irrepbar{6},\irrepbar{3},\irrepbar[1]{15})+8(\irrep{10},\irrep{8},\irrep{8})+8(\irrep{8},\irrep{10},\irrep{8})+8(\irrep{8},\irrep{8},\irrep{10})+8(\irrep{8},\irrep{8},\irrepbar{10})+8(\irrep{8},\irrepbar{10},\irrep{8})+8(\irrepbar{10},\irrep{8},\irrep{8})+6(\irrep{6},\irrep{15},\irrepbar{6})+6(\irrep{6},\irrepbar{6},\irrep{15})+6(\irrepbar{15},\irrepbar{6},\irrepbar{6})+6(\irrep{15},\irrep{6},\irrep{6})+6(\irrepbar{6},\irrep{6},\irrepbar{15})+6(\irrepbar{6},\irrepbar{15},\irrep{6})+(\irrep{6},\irrep[1]{15},\irrepbar{6})+(\irrep{6},\irrepbar{6},\irrep[1]{15})+(\irrepbar[1]{15},\irrepbar{6},\irrepbar{6})+(\irrep[1]{15},\irrep{6},\irrep{6})+(\irrepbar{6},\irrep{6},\irrepbar[1]{15})+(\irrepbar{6},\irrepbar[1]{15},\irrep{6})
+2(\irrep{10},\irrep{10},\irrep{8})+2(\irrep{10},\irrep{8},\irrep{10})+2(\irrep{10},\irrep{8},\irrepbar{10})+2(\irrep{10},\irrepbar{10},\irrep{8})+2(\irrep{8},\irrep{10},\irrep{10})+2(\irrep{8},\irrep{10},\irrepbar{10})+2(\irrep{8},\irrepbar{10},\irrep{10})+2(\irrep{8},\irrepbar{10},\irrepbar{10})+2(\irrepbar{10},\irrep{10},\irrep{8})+2(\irrepbar{10},\irrep{8},\irrep{10})+2(\irrepbar{10},\irrep{8},\irrepbar{10})+2(\irrepbar{10},\irrepbar{10},\irrep{8})+2(\irrep{27},\irrep{1},\irrep{1})+2(\irrep{1},\irrep{27},\irrep{1})+2(\irrep{1},\irrep{1},\irrep{27})+4(\irrepbar{24},\irrep{3},\irrep{3})+4(\irrepbar{3},\irrep{24},\irrep{3})+4(\irrepbar{3},\irrep{3},\irrep{24})+4(\irrep{24},\irrepbar{3},\irrepbar{3})+4(\irrep{3},\irrepbar{24},\irrepbar{3})+4(\irrep{3},\irrepbar{3},\irrepbar{24})+(\irrepbar{21},\irrepbar{6},\irrep{3})+(\irrep{6},\irrep{3},\irrep{21})+(\irrepbar{3},\irrep{21},\irrepbar{6})+(\irrep{3},\irrep{6},\irrepbar{21})+(\irrep{21},\irrepbar{3},\irrep{6})+(\irrepbar{6},\irrepbar{21},\irrepbar{3})+8(\irrepbar{15},\irrep{15},\irrep{3})+8(\irrepbar{15},\irrep{3},\irrep{15})+8(\irrepbar{3},\irrep{15},\irrep{15})+8(\irrep{15},\irrepbar{15},\irrepbar{3})+8(\irrep{15},\irrepbar{3},\irrepbar{15})+8(\irrep{3},\irrepbar{15},\irrepbar{15})+3(\irrepbar{24},\irrep{3},\irrepbar{6})+4(\irrepbar{24},\irrepbar{6},\irrep{3})+3(\irrep{6},\irrep{24},\irrep{3})+4(\irrep{6},\irrep{3},\irrep{24})+2(\irrepbar{15},\irrep[1]{15},\irrep{3})+(\irrepbar{15},\irrep{3},\irrep[1]{15})+(\irrepbar[1]{15},\irrep{15},\irrep{3})+2(\irrepbar[1]{15},\irrep{3},\irrep{15})+(\irrepbar{3},\irrep[1]{15},\irrep{15})+2(\irrepbar{3},\irrep{15},\irrep[1]{15})+4(\irrepbar{3},\irrep{24},\irrepbar{6})+3(\irrepbar{3},\irrepbar{6},\irrep{24})+2(\irrep[1]{15},\irrepbar{15},\irrepbar{3})+(\irrep[1]{15},\irrepbar{3},\irrepbar{15})+(\irrep{15},\irrepbar[1]{15},\irrepbar{3})+2(\irrep{15},\irrepbar{3},\irrepbar[1]{15})+3(\irrep{24},\irrep{6},\irrepbar{3})+4(\irrep{24},\irrepbar{3},\irrep{6})+3(\irrep{3},\irrepbar{24},\irrep{6})+4(\irrep{3},\irrep{6},\irrepbar{24})+(\irrep{3},\irrepbar{15},\irrepbar[1]{15})+2(\irrep{3},\irrepbar[1]{15},\irrepbar{15})+4(\irrepbar{6},\irrepbar{24},\irrepbar{3})+3(\irrepbar{6},\irrepbar{3},\irrepbar{24})+6(\irrep{6},\irrep{15},\irrep{15})+6(\irrepbar{15},\irrep{15},\irrepbar{6})+6(\irrepbar{15},\irrepbar{6},\irrep{15})+6(\irrep{15},\irrep{6},\irrepbar{15})+6(\irrep{15},\irrepbar{15},\irrep{6})+6(\irrepbar{6},\irrepbar{15},\irrepbar{15})+4(\irrep{27},\irrep{8},\irrep{1})+4(\irrep{27},\irrep{1},\irrep{8})+4(\irrep{8},\irrep{27},\irrep{1})+4(\irrep{8},\irrep{1},\irrep{27})+4(\irrep{1},\irrep{27},\irrep{8})+4(\irrep{1},\irrep{8},\irrep{27})+2(\irrepbar{24},\irrepbar{6},\irrepbar{6})+(\irrep{6},\irrep{15},\irrep[1]{15})+2(\irrep{6},\irrep{24},\irrepbar{6})+2(\irrep{6},\irrepbar{6},\irrep{24})+(\irrepbar{15},\irrep[1]{15},\irrepbar{6})+(\irrepbar[1]{15},\irrepbar{6},\irrep{15})+(\irrep[1]{15},\irrepbar{15},\irrep{6})+(\irrep{15},\irrep{6},\irrepbar[1]{15})+2(\irrep{24},\irrep{6},\irrep{6})+2(\irrepbar{6},\irrepbar{24},\irrep{6})+2(\irrepbar{6},\irrep{6},\irrepbar{24})+(\irrepbar{6},\irrepbar[1]{15},\irrepbar{15})+(\irrep{10},\irrep{27},\irrep{1})+(\irrep{10},\irrep{1},\irrep{27})+(\irrep{27},\irrep{10},\irrep{1})+(\irrep{27},\irrepbar{10},\irrep{1})+(\irrep{27},\irrep{1},\irrep{10})+(\irrep{27},\irrep{1},\irrepbar{10})+(\irrepbar{10},\irrep{27},\irrep{1})+(\irrepbar{10},\irrep{1},\irrep{27})+(\irrep{1},\irrep{10},\irrep{27})+(\irrep{1},\irrep{27},\irrep{10})+(\irrep{1},\irrep{27},\irrepbar{10})+(\irrep{1},\irrepbar{10},\irrep{27})+3(\irrepbar{24},\irrep{15},\irrep{3})+2(\irrepbar{24},\irrep{3},\irrep{15})+2(\irrepbar{15},\irrep{24},\irrep{3})+3(\irrepbar{15},\irrep{3},\irrep{24})+2(\irrepbar{3},\irrep{15},\irrep{24})+3(\irrepbar{3},\irrep{24},\irrep{15})+3(\irrep{15},\irrepbar{24},\irrepbar{3})+2(\irrep{15},\irrepbar{3},\irrepbar{24})+2(\irrep{24},\irrepbar{15},\irrepbar{3})+3(\irrep{24},\irrepbar{3},\irrepbar{15})+2(\irrep{3},\irrepbar{24},\irrepbar{15})+3(\irrep{3},\irrepbar{15},\irrepbar{24})+7(\irrep{27},\irrep{8},\irrep{8})+7(\irrep{8},\irrep{27},
\irrep{8})+7(\irrep{8},\irrep{8},\irrep{27})+(\irrep{35},\irrep{8},\irrep{1})+(\irrepbar{35},\irrep{1},\irrep{8})+(\irrep{8},\irrep{35},\irrep{1})+(\irrep{8},\irrep{1},\irrepbar{35})+(\irrep{1},\irrepbar{35},\irrep{8})+(\irrep{1},\irrep{8},\irrep{35})+4(\irrepbar{15},\irrep{15},\irrep{15})+4(\irrep{15},\irrepbar{15},\irrepbar{15})+2(\irrep{10},\irrep{27},\irrep{8})+(\irrep{10},\irrep{8},\irrep{27})+2(\irrep{27},\irrep{10},\irrep{8})+(\irrep{27},\irrep{8},\irrep{10})+2(\irrep{27},\irrep{8},\irrepbar{10})+(\irrep{27},\irrepbar{10},\irrep{8})+(\irrep{8},\irrep{10},\irrep{27})+2(\irrep{8},\irrep{27},\irrep{10})+(\irrep{8},\irrep{27},\irrepbar{10})+2(\irrep{8},\irrepbar{10},\irrep{27})+(\irrepbar{10},\irrep{27},\irrep{8})+2(\irrepbar{10},\irrep{8},\irrep{27})+2(\irrepbar{24},\irrep{15},\irrepbar{6})+(\irrepbar{24},\irrepbar{6},\irrep{15})+(\irrep{6},\irrep{15},\irrep{24})+2(\irrep{6},\irrep{24},\irrep{15})+(\irrepbar{15},\irrep{24},\irrepbar{6})+2(\irrepbar{15},\irrepbar{6},\irrep{24})+2(\irrep{15},\irrepbar{24},\irrep{6})+(\irrep{15},\irrep{6},\irrepbar{24})+2(\irrep{24},\irrep{6},\irrepbar{15})+(\irrep{24},\irrepbar{15},\irrep{6})+(\irrepbar{6},\irrepbar{24},\irrepbar{15})+2(\irrepbar{6},\irrepbar{15},\irrepbar{24})+(\irrep{35},\irrepbar{10},\irrep{1})+(\irrep{10},\irrep{1},\irrepbar{35})+(\irrepbar{35},\irrep{1},\irrep{10})+(\irrepbar{10},\irrep{35},\irrep{1})+(\irrep{1},\irrep{10},\irrep{35})+(\irrep{1},\irrepbar{35},\irrepbar{10})+(\irrep{10},\irrep{27},\irrep{10})+(\irrep{27},\irrep{10},\irrepbar{10})+(\irrepbar{10},\irrepbar{10},\irrep{27})+(\irrepbar{42},\irrep{3},\irrep{3})+(\irrepbar{3},\irrep{42},\irrep{3})+(\irrepbar{3},\irrep{3},\irrep{42})+(\irrep{42},\irrepbar{3},\irrepbar{3})+(\irrep{3},\irrepbar{42},\irrepbar{3})+(\irrep{3},\irrepbar{3},\irrepbar{42})+(\irrepbar{24},\irrep{24},\irrep{3})+(\irrepbar{24},\irrep{3},\irrep{24})+(\irrepbar{3},\irrep{24},\irrep{24})+(\irrep{24},\irrepbar{24},\irrepbar{3})+(\irrep{24},\irrepbar{3},\irrepbar{24})+(\irrep{3},\irrepbar{24},\irrepbar{24})+(\irrep{35},\irrep{8},\irrep{8})+(\irrepbar{35},\irrep{8},\irrep{8})+(\irrep{8},\irrep{35},\irrep{8})+(\irrep{8},\irrepbar{35},\irrep{8})+(\irrep{8},\irrep{8},\irrep{35})+(\irrep{8},\irrep{8},\irrepbar{35})+(\irrepbar{42},\irrep{3},\irrepbar{6})+(\irrepbar{42},\irrepbar{6},\irrep{3})+(\irrep{6},\irrep{42},\irrep{3})+(\irrep{6},\irrep{3},\irrep{42})+(\irrepbar{3},\irrep{42},\irrepbar{6})+(\irrepbar{3},\irrepbar{6},\irrep{42})+(\irrep{42},\irrep{6},\irrepbar{3})+(\irrep{42},\irrepbar{3},\irrep{6})+(\irrep{3},\irrepbar{42},\irrep{6})+(\irrep{3},\irrep{6},\irrepbar{42})+(\irrepbar{6},\irrepbar{42},\irrepbar{3})+(\irrepbar{6},\irrepbar{3},\irrepbar{42})+(\irrepbar{24},\irrep{15},\irrep{15})+(\irrepbar{15},\irrep{15},\irrep{24})+(\irrepbar{15},\irrep{24},\irrep{15})+(\irrep{15},\irrepbar{24},\irrepbar{15})+(\irrep{15},\irrepbar{15},\irrepbar{24})+(\irrep{24},\irrepbar{15},\irrepbar{15})+(\irrep{27},\irrep{27},\irrep{1})+(\irrep{27},\irrep{1},\irrep{27})+(\irrep{1},\irrep{27},\irrep{27})+(\irrepbar{42},\irrep{3},\irrep{15})+(\irrepbar{15},\irrep{42},\irrep{3})+(\irrepbar{3},\irrep{15},\irrep{42})+(\irrep{42},\irrepbar{15},\irrepbar{3})+(\irrep{15},\irrepbar{3},\irrepbar{42})+(\irrep{3},\irrepbar{42},\irrepbar{15})+(\irrep{27},\irrep{27},\irrep{8})+(\irrep{27},\irrep{8},\irrep{27})+(\irrep{8},\irrep{27},\irrep{27})+(\irrepbar{42},\irrep{15},\irrepbar{6})+(\irrep{6},\irrep{42},\irrep{15})+(\irrepbar{15},\irrepbar{6},\irrep{42})+(\irrep{42},\irrep{6},\irrepbar{15})+(\irrep{15},\irrepbar{42},\irrep{6})+(\irrepbar{6},\irrepbar{15},\irrepbar{42})\}$
\end{mathout}

\pagebreak

\vspace{-10pt}
\section{\LaTeX\ Package}
\label{LaTeXPackage}
LieART comes with a \LaTeX\ package (\texttt{lieart.sty} in the subdirectory \texttt{latex/}) that defines commands to display irreps, roots and weights properly (see Table~\ref{tab:LaTeXCommands}),
which are displayed by LieART using the \texttt{LaTeXForm} on an appropriate expression, e.g.:
\begin{mathin}
DecomposeProduct[Irrep[SU3][8],Irrep[SU3][8]]//LaTeXForm
\end{mathin}
\begin{mathout}
\verb#$\irrep{1}+2(\irrep{8})+\irrep{10}+\irrepbar{10}+\irrep{27}$#
\end{mathout}

\newcommand{\bsl}{\textbackslash}
\begin{table}[!h]
\begin{center}
\begin{tabularx}{\textwidth}{llX}
    \toprule
    \textbf{Command Example} & \textbf{Output} & \textbf{Description}\\
    \midrule
     \texttt{\bsl irrep\{10\}} & \irrep{10} & dimensional name of irrep\\
     \texttt{\bsl irrepbar\{10\}} & \irrepbar{10} & dimensional name of conjugated irrep\\
     \texttt{\bsl irrep[2]\{175\}} & \irrep[2]{175} & number of primes as optional parameter\\
     \texttt{\bsl irrepsub\{8\}\{s\}} & \irrepsub{8}{s} & irrep with subscript, e.g., irreps of \SO8 \\
     \texttt{\bsl irrepbarsub\{10\}\{a\}} & \irrepbarsub{10}{a} & conjugated irrep with subscript, e.g., for labeling antisymmetric product\\
     \texttt{\bsl dynkin\{0,1,0,0\}} & \dynkin{0,1,0,0} & Dynkin label of irrep\\
     \texttt{\bsl dynkincomma\{0,10,0,0\}} & \dynkincomma{0,10,0,0} & for Dynkin labels with at least one digit larger then 9\newline
                                                                      (also available as \texttt{\bsl root}, \texttt{\bsl rootorthogonal},\newline
                                                                      \texttt{\bsl weightalpha} and \texttt{\bsl weightorthogonal}\newline for negative integers) \\
     \texttt{\bsl weight\{0,1,0,{-}1\}} & \weight{0,1,0,{-1}} & weight in $\omega$-basis\\
     \texttt{\bsl rootomega\{{-}1,2,{-}1,0\}} & \rootomega{{-}1,2,{-}1,0} & root in $\omega$-basis\\
    \bottomrule
\end{tabularx}
\caption{\label{tab:LaTeXCommands}\LaTeX\ commands defined in supplemental style file \texttt{lieart.sty}}
\end{center}
\end{table}
\vspace{-20pt}

\enlargethispage{2ex}

\vspace{-2ex}
\section{Conclusions and Outlook}
\label{ConclusionsAndOutlook}

We have programmed the Mathematica application LieART, which brings Lie-algebra and representation-theory related computations to Mathematica.
It provides functions for the decomposition of tensor products and branching rules of irreducible representations, which are of high interest
in particle physics, especially unified model building. LieART exploits the Weyl reflection group in most of its applications, making it fast
and memory efficient. The user interface focuses on usability, allowing one to enter irreducible representations by their dimensional name and
giving results in textbook style. We have reproduced and extended existing tabulated data on irreducible representations, their tensor products
and branching rules.

In future versions we plan to add more branching rules to LieART. Currently, only a selection of common branching rules used in the tables are implemented.
We consider the tables given in the appendix as dynamical: They are included in LieART as Mathematica notebooks and can easily modified and extended by the user.
Tables for algebras of high rank and/or higher dimensional irreducible representations have large CPU time and high memory consumption. Nevertheless, we
plan to extend the tables even further and make them available online in a standard format (pdf and/or html).

\vspace{-2ex}
\section{Acknowledgments}

We thank Carl Albright for many useful discussions that led to the development 
of LieART. We also thank Tanja Feger for checking the tables against those found 
in \cite{Slansky}. We thank Savdeep Sethi and Bruce Westbury for runtime 
comparisons with \texttt{LiE} \cite{LiE}, which led to the implementation of 
Klymik's formula for tensor products. We thank Florian Hartmann, Constantin 
Sluka and Giulia Ferlito for reporting bugs in the initial version.

Most of the work was performed while RPF was affiliated with the Department of Physics and Astronomy, 
Vanderbilt University, Nashville, and his work was supported by a fellowship within the Postdoc-Programme 
of the German Academic Exchange Service (DAAD). The work of RPF and TWK was supported by US
DOE grant \# E-FG05-85ER40226. The work of TWK was also supported by DoE grant \# DE-SC0011981.

\newpage

\appendix
\colorlet{tableoverheadcolor}{gray!37.5}
\colorlet{tableheadcolor}{gray!40}
\colorlet{tablerowcolor}{gray!20}

\rowcolors{4}{}{tablerowcolor}

\renewcommand*{\appendixname}{}

\section{Tables}
\setcounter{table}{0}
We present here tables of properties of irreps, such as Dynkin labels,
dimensional names, indices, congruency classes and the number of singlets in
various subalgebra branchings in Section \ref{ssec:IrrepProperties}, as well as
tables of tensor products in Section \ref{ssec:TensorProducts} and subalgebra
branching rules in Section \ref{ssec:BranchingRules} for many classical and all
exceptional Lie algebras. In presentation style, selection of irreps and
subalgebra branching we closely follow \cite{Slansky}, which has been the
definitive reference for unified model building since its publication. The
tables were created by the supplemental package \texttt{Tables.m}, which uses LieART
for the computation. The tables can also be found as Mathematica notebooks in the
LieART documentation integrated into the Mathematica documentation center as
``Representation Properties'', ``Tensor Products'' and ``Branching Rules'' under
the section ``Tables'' on the LieART documentation home. Since LieART comes with
the functions that generate the tables, the user may extend them to the limit of
his or her computer power.

\begin{table}[!!h]
\begin{center}
\begin{tabular}{l|ll|ll|ll}
    \toprule\rowcolor{tableheadcolor}
    &\multicolumn{2}{>{\columncolor{tableheadcolor}}l|}{\textbf{Irrep Properties}} & \multicolumn{2}{>{\columncolor{tableheadcolor}}l|}{\textbf{Tensor Products}} & \multicolumn{2}{>{\columncolor{tableheadcolor}}l}{\textbf{Branching Rules}}\\
    \rowcolor{tableheadcolor}\textbf{Algebra} & \textbf{Number} & \textbf{Page} & \textbf{Number} & \textbf{Page} & \textbf{Number} & \textbf{Page}\\
    \midrule
    \SU2    & \ref{tab:SU2Irreps}  & \pageref{tab:SU2Irreps}  & \ref{tab:SU2TensorProducts}  & \pageref{tab:SU2TensorProducts}  & \ref{tab:SU2BranchingRules}  & \pageref{tab:SU2BranchingRules} \\
    \SU3    & \ref{tab:SU3Irreps}  & \pageref{tab:SU3Irreps}  & \ref{tab:SU3TensorProducts}  & \pageref{tab:SU3TensorProducts}  & \ref{tab:SU3BranchingRules}  & \pageref{tab:SU3BranchingRules} \\
    \SU4    & \ref{tab:SU4Irreps}  & \pageref{tab:SU4Irreps}  & \ref{tab:SU4TensorProducts}  & \pageref{tab:SU4TensorProducts}  & \ref{tab:SU4BranchingRules}  & \pageref{tab:SU4BranchingRules} \\
    \SU5    & \ref{tab:SU5Irreps}  & \pageref{tab:SU5Irreps}  & \ref{tab:SU5TensorProducts}  & \pageref{tab:SU5TensorProducts}  & \ref{tab:SU5BranchingRules}  & \pageref{tab:SU5BranchingRules} \\
    \SU6    & \ref{tab:SU6Irreps}  & \pageref{tab:SU6Irreps}  & \ref{tab:SU6TensorProducts}  & \pageref{tab:SU6TensorProducts}  & \ref{tab:SU6BranchingRules}  & \pageref{tab:SU6BranchingRules} \\
    \SU7    & \ref{tab:SU7Irreps}  & \pageref{tab:SU7Irreps}  & \ref{tab:SU7TensorProducts}  & \pageref{tab:SU7TensorProducts}  & \ref{tab:SU7BranchingRules}  & \pageref{tab:SU7BranchingRules} \\
    \SU8    & \ref{tab:SU8Irreps}  & \pageref{tab:SU8Irreps}  & \ref{tab:SU8TensorProducts}  & \pageref{tab:SU8TensorProducts}  & \ref{tab:SU8BranchingRules}  & \pageref{tab:SU8BranchingRules} \\
    \SU9    & \ref{tab:SU9Irreps}  & \pageref{tab:SU9Irreps}  & \ref{tab:SU9TensorProducts}  & \pageref{tab:SU9TensorProducts}  & \ref{tab:SU9BranchingRules}  & \pageref{tab:SU9BranchingRules} \\
    \SU{10} & \ref{tab:SU10Irreps} & \pageref{tab:SU10Irreps} & \ref{tab:SU10TensorProducts} & \pageref{tab:SU10TensorProducts} & \ref{tab:SU10BranchingRules} & \pageref{tab:SU10BranchingRules}\\
    \SU{11} & \ref{tab:SU11Irreps} & \pageref{tab:SU11Irreps} & \ref{tab:SU11TensorProducts} & \pageref{tab:SU11TensorProducts} & \ref{tab:SU11BranchingRules} & \pageref{tab:SU11BranchingRules}\\
    \SU{12} & \ref{tab:SU12Irreps} & \pageref{tab:SU12Irreps} & \ref{tab:SU12TensorProducts} & \pageref{tab:SU12TensorProducts} & \ref{tab:SU12BranchingRules} & \pageref{tab:SU12BranchingRules}\\
    \midrule
    \SO7    & \ref{tab:SO7Irreps}  & \pageref{tab:SO7Irreps}  & \ref{tab:SO7TensorProducts}  & \pageref{tab:SO7TensorProducts}  & \ref{tab:SO7BranchingRules}  & \pageref{tab:SO7BranchingRules} \\
    \SO8    & \ref{tab:SO8Irreps}  & \pageref{tab:SO8Irreps}  & \ref{tab:SO8TensorProducts}  & \pageref{tab:SO8TensorProducts}  & \ref{tab:SO8BranchingRules}  & \pageref{tab:SO8BranchingRules} \\
    \SO9    & \ref{tab:SO9Irreps}  & \pageref{tab:SO9Irreps}  & \ref{tab:SO9TensorProducts}  & \pageref{tab:SO9TensorProducts}  & \ref{tab:SO9BranchingRules}  & \pageref{tab:SO9BranchingRules} \\
    \SO{10} & \ref{tab:SO10Irreps} & \pageref{tab:SO10Irreps} & \ref{tab:SO10TensorProducts} & \pageref{tab:SO10TensorProducts} & \ref{tab:SO10BranchingRules} & \pageref{tab:SO10BranchingRules}\\
    \SO{11} & \ref{tab:SO11Irreps} & \pageref{tab:SO11Irreps} & \ref{tab:SO11TensorProducts} & \pageref{tab:SO11TensorProducts} & --                           &  --                             \\
    \SO{12} & \ref{tab:SO12Irreps} & \pageref{tab:SO12Irreps} & \ref{tab:SO12TensorProducts} & \pageref{tab:SO12TensorProducts} & --                           &  --                             \\
    \SO{13} & \ref{tab:SO13Irreps} & \pageref{tab:SO13Irreps} & \ref{tab:SO13TensorProducts} & \pageref{tab:SO13TensorProducts} & --                           &  --                             \\
    \SO{14} & \ref{tab:SO14Irreps} & \pageref{tab:SO14Irreps} & \ref{tab:SO14TensorProducts} & \pageref{tab:SO14TensorProducts} & \ref{tab:SO14BranchingRules} & \pageref{tab:SO14BranchingRules}\\
    \SO{18} & \ref{tab:SO18Irreps} & \pageref{tab:SO18Irreps} & \ref{tab:SO18TensorProducts} & \pageref{tab:SO18TensorProducts} & \ref{tab:SO18BranchingRules} & \pageref{tab:SO18BranchingRules}\\
    \SO{22} & \ref{tab:SO22Irreps} & \pageref{tab:SO22Irreps} & \ref{tab:SO22TensorProducts} & \pageref{tab:SO22TensorProducts} & \ref{tab:SO22BranchingRules} & \pageref{tab:SO22BranchingRules}\\
    \SO{26} & \ref{tab:SO26Irreps} & \pageref{tab:SO26Irreps} & \ref{tab:SO26TensorProducts} & \pageref{tab:SO26TensorProducts} & \ref{tab:SO26BranchingRules} & \pageref{tab:SO26BranchingRules}\\
    \midrule
    \Sp4    & \ref{tab:Sp4Irreps}  & \pageref{tab:Sp4Irreps}  & \ref{tab:Sp4TensorProducts}  & \pageref{tab:Sp4TensorProducts}  & --                           & --                              \\
    \Sp6    & \ref{tab:Sp6Irreps}  & \pageref{tab:Sp6Irreps}  & \ref{tab:Sp6TensorProducts}  & \pageref{tab:Sp6TensorProducts}  & --                           & --                              \\
    \Sp8    & \ref{tab:Sp8Irreps}  & \pageref{tab:Sp8Irreps}  & \ref{tab:Sp8TensorProducts}  & \pageref{tab:Sp8TensorProducts}  & --                           & --                              \\
    \Sp{10} & \ref{tab:Sp10Irreps} & \pageref{tab:Sp10Irreps} & \ref{tab:Sp10TensorProducts} & \pageref{tab:Sp10TensorProducts} & --                           & --                              \\
    \Sp{12} & \ref{tab:Sp12Irreps} & \pageref{tab:Sp12Irreps} & \ref{tab:Sp12TensorProducts} & \pageref{tab:Sp12TensorProducts} & --                           & --                              \\
    \midrule
    \E6     & \ref{tab:E6Irreps}   & \pageref{tab:E6Irreps}   & \ref{tab:E6TensorProducts}   & \pageref{tab:E6TensorProducts}   & \ref{tab:E6BranchingRules}   & \pageref{tab:E6BranchingRules} \\
    \E7     & \ref{tab:E7Irreps}   & \pageref{tab:E7Irreps}   & \ref{tab:E7TensorProducts}   & \pageref{tab:E7TensorProducts}   & \ref{tab:E7BranchingRules}   & \pageref{tab:E7BranchingRules} \\
    \E8     & \ref{tab:E8Irreps}   & \pageref{tab:E8Irreps}   & \ref{tab:E8TensorProducts}   & \pageref{tab:E8TensorProducts}   & \ref{tab:E8BranchingRules}   & \pageref{tab:E8BranchingRules} \\
    \F4     & \ref{tab:F4Irreps}   & \pageref{tab:F4Irreps}   & \ref{tab:F4TensorProducts}   & \pageref{tab:F4TensorProducts}   & --                           & --                              \\
    \G2     & \ref{tab:G2Irreps}   & \pageref{tab:G2Irreps}   & \ref{tab:G2TensorProducts}   & \pageref{tab:G2TensorProducts}   & --                           & --                              \\
    \bottomrule
\end{tabular}
\end{center}
\caption{Table of tables}
\end{table}
\newpage

\subsection{Properties of Irreducible Representations}
\label{ssec:IrrepProperties}

{
	\rowcolors{2}{tablerowcolor}{}
    \newcommand\starred[1]{#1\makebox[0pt][l]{${}^\ast$}}
    \renewcommand{\irrep}[2][0]{\ensuremath{\irrepbase{#2}\makebox[0pt][l]{$^{\primes{#1}{\prime\hspace{-1pt}}}$}}}
    \renewcommand{\irrepbar}[2][0]{\ensuremath{\irrepbarbase{#2}}\makebox[0pt][l]{$^{\primes{#1}{\prime\hspace{-1pt}}}$}}
    \renewcommand{\irrepsub}[3][0]{\ensuremath{\irrep[#1]{#2}\makebox[0pt][l]{$_\text{#3}$}}}
    \renewcommand{\irrepbarsub}[3][0]{\ensuremath{\irrepbar[#1]{#2}}\makebox[0pt][l]{$_\text{#3}$}}%
    \subsubsection{\SU{N}}
\enlargethispage{10pt}
\rowcolors{2}{}{tablerowcolor}
\begin{longtable}{lrrc}
\rowcolor{white}
\caption{SU(2) Irreps\label{tab:SU2Irreps}}\\
\toprule
\rowcolor{tableheadcolor}
Dynkin & Dimension & l & \\
\rowcolor{tableheadcolor}label & (name) & (index) & Duality\\
\midrule
\endfirsthead
\caption[]{SU(2) Irreps (continued)}\\
\toprule
\rowcolor{tableheadcolor}
Dynkin & Dimension & l & \\
\rowcolor{tableheadcolor}label & (name) & (index) & Duality\\
\midrule
\endhead
\endfoot
\bottomrule
\endlastfoot
\dynkin{1} & \irrep{2} & 1 & 1\\
\dynkin{2} & \irrep{3} & 4 & 0\\
\dynkin{3} & \irrep{4} & 10 & 1\\
\dynkin{4} & \irrep{5} & 20 & 0\\
\dynkin{5} & \irrep{6} & 35 & 1\\
\dynkin{6} & \irrep{7} & 56 & 0\\
\dynkin{7} & \irrep{8} & 84 & 1\\
\dynkin{8} & \irrep{9} & 120 & 0\\
\end{longtable}
\newpage

\rowcolors{2}{}{tablerowcolor}
\begin{longtable}{lrrcc}
\rowcolor{white}
\caption{\label{tab:SU3Irreps}SU(3) Irreps}\\
\toprule
\rowcolor{tableheadcolor}
Dynkin & Dimension & l &  & SU(2)\\
\rowcolor{tableheadcolor}label & (name) & (index) & Triality & singlets\\
\midrule
\endfirsthead
\rowcolor{white}
\caption[]{SU(3) Irreps (continued)}\\
\toprule
\rowcolor{tableheadcolor}
Dynkin & Dimension & l &  & SU(2)\\
\rowcolor{tableheadcolor}label & (name) & (index) & Triality & singlets\\
\midrule
\endhead
\rowcolor{white}\multicolumn{5}{l}{\footnotesize $^\ast$SU(2)${\times}$U(1) singlet.}
\endfoot
\bottomrule
\rowcolor{white}\multicolumn{5}{l}{\footnotesize $^\ast$SU(2)${\times}$U(1) singlet.}
\endlastfoot
\dynkin{1, 0} & \irrep{3} & 1 & 1 & 1\\
\dynkin{2, 0} & \irrep{6} & 5 & 2 & 1\\
\dynkin{1, 1} & \irrep{8} & 6 & 0 & \starred{1}\\
\dynkin{3, 0} & \irrep{10} & 15 & 0 & 1\\
\dynkin{2, 1} & \irrep{15} & 20 & 1 & 1\\
\dynkin{4, 0} & \irrep[1]{15} & 35 & 1 & 1\\
\dynkin{0, 5} & \irrep{21} & 70 & 1 & 1\\
\dynkin{1, 3} & \irrep{24} & 50 & 1 & 1\\
\end{longtable}
\newpage

\rowcolors{2}{}{tablerowcolor}
\begin{longtable}{lrrcc}
\caption{\label{tab:SU4Irreps}SU(4) Irreps}\\
\toprule
\rowcolor{tableheadcolor}
Dynkin & Dimension & l &  & SU(3)\\
\rowcolor{tableheadcolor}label & (name) & (index) & Quadrality & singlets\\
\midrule
\endfirsthead
\rowcolor{white}
\caption[]{SU(4) Irreps (continued)}\\
\toprule
\rowcolor{tableheadcolor}
Dynkin & Dimension & l &  & SU(3)\\
\rowcolor{tableheadcolor}label & (name) & (index) & Quadrality & singlets\\
\midrule
\endhead
\rowcolor{white}\multicolumn{5}{l}{\footnotesize $^\ast$SU(3)${\times}$U(1) singlet.}
\endfoot
\bottomrule
\rowcolor{white}\multicolumn{5}{l}{\footnotesize $^\ast$SU(3)${\times}$U(1) singlet.}
\endlastfoot
\dynkin{1, 0, 0} & \irrep{4} & 1 & 1 & 1\\
\dynkin{0, 1, 0} & \irrep{6} & 2 & 2 & 0\\
\dynkin{2, 0, 0} & \irrep{10} & 6 & 2 & 1\\
\dynkin{1, 0, 1} & \irrep{15} & 8 & 0 & \starred{1}\\
\dynkin{0, 1, 1} & \irrep{20} & 13 & 1 & 0\\
\dynkin{0, 2, 0} & \irrep[1]{20} & 16 & 0 & 0\\
\dynkin{0, 0, 3} & \irrep[2]{20} & 21 & 1 & 1\\
\dynkin{4, 0, 0} & \irrep{35} & 56 & 0 & 1\\
\end{longtable}
\newpage

\rowcolors{2}{}{tablerowcolor}
\begin{longtable}{lrrccc}
\caption{\label{tab:SU5Irreps}SU(5) Irreps}\\
\toprule
\rowcolor{tableheadcolor}
Dynkin & Dimension & l &  & SU(4) & SU(3)${\times}$SU(2)\\
\rowcolor{tableheadcolor}label & (name) & (index) & Quintality & singlets & singlets\\
\midrule
\endfirsthead
\rowcolor{white}
\caption[]{SU(5) Irreps (continued)}\\
\toprule
\rowcolor{tableheadcolor}
Dynkin & Dimension & l &  & SU(4) & SU(3)${\times}$SU(2)\\
\rowcolor{tableheadcolor}label & (name) & (index) & Quintality & singlets & singlets\\
\midrule
\endhead
\rowcolor{white}\multicolumn{6}{l}{\footnotesize $^\ast$SU(4)${\times}$U(1) and SU(3)${\times}$SU(2)${\times}$U(1) singlets resp.}
\endfoot
\bottomrule
\rowcolor{white}\multicolumn{6}{l}{\footnotesize $^\ast$SU(4)${\times}$U(1) and SU(3)${\times}$SU(2)${\times}$U(1) singlets resp.}
\endlastfoot
\dynkin{1, 0, 0, 0} & \irrep{5} & 1 & 1 & 1 & 0\\
\dynkin{0, 1, 0, 0} & \irrep{10} & 3 & 2 & 0 & 1\\
\dynkin{2, 0, 0, 0} & \irrep{15} & 7 & 2 & 1 & 0\\
\dynkin{1, 0, 0, 1} & \irrep{24} & 10 & 0 & \starred{1} & \starred{1}\\
\dynkin{0, 0, 0, 3} & \irrep{35} & 28 & 2 & 1 & 0\\
\dynkin{0, 0, 1, 1} & \irrep{40} & 22 & 2 & 0 & 0\\
\dynkin{0, 1, 0, 1} & \irrep{45} & 24 & 1 & 0 & 0\\
\dynkin{0, 0, 2, 0} & \irrep{50} & 35 & 1 & 0 & 1\\
\end{longtable}
\newpage

\rowcolors{2}{}{tablerowcolor}
\begin{longtable}{lrrcccc}
\caption{\label{tab:SU6Irreps}SU(6) Irreps}\\
\toprule
\rowcolor{tableheadcolor}
Dynkin & Dimension & l &  & SU(5) & SU(4)${\times}$SU(2) & SU(3)${\times}$SU(3)\\
\rowcolor{tableheadcolor}label & (name) & (index) & Sextality & singlets & singlets & singlets\\
\midrule
\endfirsthead
\rowcolor{white}
\caption[]{SU(6) Irreps (continued)}\\
\toprule
\rowcolor{tableheadcolor}
Dynkin & Dimension & l &  & SU(5) & SU(4)${\times}$SU(2) & SU(3)${\times}$SU(3)\\
\rowcolor{tableheadcolor}label & (name) & (index) & Sextality & singlets & singlets & singlets\\
\midrule
\endhead
\rowcolor{white}\multicolumn{7}{l}{\footnotesize $^\ast$SU(5)${\times}$U(1) and SU(4)${\times}$SU(2)${\times}$U(1) and SU(3)${\times}$SU(3)${\times}$U(1) singlets resp.}
\endfoot
\bottomrule
\rowcolor{white}\multicolumn{7}{l}{\footnotesize $^\ast$SU(5)${\times}$U(1) and SU(4)${\times}$SU(2)${\times}$U(1) and SU(3)${\times}$SU(3)${\times}$U(1) singlets resp.}
\endlastfoot
\dynkin{1, 0, 0, 0, 0} & \irrep{6} & 1 & 1 & 1 & 0 & 0\\
\dynkin{0, 1, 0, 0, 0} & \irrep{15} & 4 & 2 & 0 & 1 & 0\\
\dynkin{0, 0, 1, 0, 0} & \irrep{20} & 6 & 3 & 0 & 0 & 2\\
\dynkin{2, 0, 0, 0, 0} & \irrep{21} & 8 & 2 & 1 & 0 & 0\\
\dynkin{1, 0, 0, 0, 1} & \irrep{35} & 12 & 0 & \starred{1} & \starred{1} & \starred{1}\\
\dynkin{3, 0, 0, 0, 0} & \irrep{56} & 36 & 3 & 1 & 0 & 0\\
\dynkin{1, 1, 0, 0, 0} & \irrep{70} & 33 & 3 & 0 & 0 & 0\\
\dynkin{0, 1, 0, 0, 1} & \irrep{84} & 38 & 1 & 0 & 0 & 0\\
\end{longtable}
\newpage

\rowcolors{2}{}{tablerowcolor}
\begin{longtable}{lrrcccc}
\caption{\label{tab:SU7Irreps}SU(7) Irreps}\\
\toprule
\rowcolor{tableheadcolor}
Dynkin & Dimension & l &  & SU(6) & SU(5)${\times}$SU(2) & SU(4)${\times}$SU(3)\\
\rowcolor{tableheadcolor}label & (name) & (index) & Septality & singlets & singlets & singlets\\
\midrule
\endfirsthead
\rowcolor{white}
\caption[]{SU(7) Irreps (continued)}\\
\toprule
\rowcolor{tableheadcolor}
Dynkin & Dimension & l &  & SU(6) & SU(5)${\times}$SU(2) & SU(4)${\times}$SU(3)\\
\rowcolor{tableheadcolor}label & (name) & (index) & Septality & singlets & singlets & singlets\\
\midrule
\endhead
\rowcolor{white}\multicolumn{7}{l}{\footnotesize $^\ast$SU(6)${\times}$U(1) and SU(5)${\times}$SU(2)${\times}$U(1) and SU(4)${\times}$SU(3)${\times}$U(1) singlets resp.}
\endfoot
\bottomrule
\rowcolor{white}\multicolumn{7}{l}{\footnotesize $^\ast$SU(6)${\times}$U(1) and SU(5)${\times}$SU(2)${\times}$U(1) and SU(4)${\times}$SU(3)${\times}$U(1) singlets resp.}
\endlastfoot
\dynkin{1, 0, 0, 0, 0, 0} & \irrep{7} & 1 & 1 & 1 & 0 & 0\\
\dynkin{0, 1, 0, 0, 0, 0} & \irrep{21} & 5 & 2 & 0 & 1 & 0\\
\dynkin{2, 0, 0, 0, 0, 0} & \irrep{28} & 9 & 2 & 1 & 0 & 0\\
\dynkin{0, 0, 1, 0, 0, 0} & \irrep{35} & 10 & 3 & 0 & 0 & 1\\
\dynkin{1, 0, 0, 0, 0, 1} & \irrep{48} & 14 & 0 & \starred{1} & \starred{1} & \starred{1}\\
\dynkin{3, 0, 0, 0, 0, 0} & \irrep{84} & 45 & 3 & 1 & 0 & 0\\
\dynkin{1, 1, 0, 0, 0, 0} & \irrep{112} & 46 & 3 & 0 & 0 & 0\\
\dynkin{0, 1, 0, 0, 0, 1} & \irrep{140} & 55 & 1 & 0 & 0 & 0\\
\end{longtable}
\newpage

\setlength{\tabcolsep}{5.5pt}
\rowcolors{2}{}{tablerowcolor}
\begin{longtable}{lrrccccc}
\caption{\label{tab:SU8Irreps}SU(8) Irreps}\\
\toprule
\rowcolor{tableheadcolor}
Dynkin & Dimension & l &  & SU(7) & SU(6)${\times}$SU(2) & SU(5)${\times}$SU(3) & SU(4)${\times}$SU(4)\\
\rowcolor{tableheadcolor}label & (name) & (index) & Octality & singlets & singlets & singlets & singlets\\
\midrule
\endfirsthead
\rowcolor{white}
\caption[]{SU(8) Irreps (continued)}\\
\toprule
\rowcolor{tableheadcolor}
Dynkin & Dimension & l &  & SU(7) & SU(6)${\times}$SU(2) & SU(5)${\times}$SU(3) & SU(4)${\times}$SU(4)\\
\rowcolor{tableheadcolor}label & (name) & (index) & Octality & singlets & singlets & singlets & singlets\\
\midrule
\endhead
\rowcolor{white}\multicolumn{8}{l}{\footnotesize $^\ast$SU(7)${\times}$U(1) and SU(6)${\times}$SU(2)${\times}$U(1) and SU(5)${\times}$SU(3)${\times}$U(1) and SU(4)${\times}$SU(4)${\times}$U(1) singlets resp.}
\endfoot
\bottomrule
\rowcolor{white}\multicolumn{8}{l}{\footnotesize $^\ast$SU(7)${\times}$U(1) and SU(6)${\times}$SU(2)${\times}$U(1) and SU(5)${\times}$SU(3)${\times}$U(1) and SU(4)${\times}$SU(4)${\times}$U(1) singlets resp.}
\endlastfoot
\dynkin{1, 0, 0, 0, 0, 0, 0} & \irrep{8} & 1 & 1 & 1 & 0 & 0 & 0\\
\dynkin{0, 1, 0, 0, 0, 0, 0} & \irrep{28} & 6 & 2 & 0 & 1 & 0 & 0\\
\dynkin{2, 0, 0, 0, 0, 0, 0} & \irrep{36} & 10 & 2 & 1 & 0 & 0 & 0\\
\dynkin{0, 0, 1, 0, 0, 0, 0} & \irrep{56} & 15 & 3 & 0 & 0 & 1 & 0\\
\dynkin{1, 0, 0, 0, 0, 0, 1} & \irrep{63} & 16 & 0 & \starred{1} & \starred{1} & \starred{1} & \starred{1}\\
\dynkin{0, 0, 0, 1, 0, 0, 0} & \irrep{70} & 20 & 4 & 0 & 0 & 0 & 2\\
\dynkin{3, 0, 0, 0, 0, 0, 0} & \irrep{120} & 55 & 3 & 1 & 0 & 0 & 0\\
\dynkin{1, 1, 0, 0, 0, 0, 0} & \irrep{168} & 61 & 3 & 0 & 0 & 0 & 0\\
\end{longtable}
\newpage

\setlength{\tabcolsep}{5pt}
\enlargethispage{20pt}
{
\rowcolors{2}{}{tablerowcolor}
\begin{longtable}{lrrccccc}
\caption{\label{tab:SU9Irreps}SU(9) Irreps}\\
\toprule
\rowcolor{tableheadcolor}
Dynkin & Dimension & l &  & SU(8) & SU(7)${\times}$SU(2) & SU(6)${\times}$SU(3) & SU(5)${\times}$SU(4)\\
\rowcolor{tableheadcolor}label & (name) & (index) & Nonality & singlets & singlets & singlets & singlets\\
\midrule
\endfirsthead
\rowcolor{white}
\caption[]{SU(9) Irreps (continued)}\\
\toprule
\rowcolor{tableheadcolor}
Dynkin & Dimension & l &  & SU(8) & SU(7)${\times}$SU(2) & SU(6)${\times}$SU(3) & SU(5)${\times}$SU(4)\\
\rowcolor{tableheadcolor}label & (name) & (index) & Nonality & singlets & singlets & singlets & singlets\\
\midrule
\endhead
\rowcolor{white}\multicolumn{8}{l}{\footnotesize $^\ast$SU(8)${\times}$U(1) and SU(7)${\times}$SU(2)${\times}$U(1) and SU(6)${\times}$SU(3)${\times}$U(1) and SU(5)${\times}$SU(4)${\times}$U(1) singlets resp.}
\endfoot
\bottomrule
\rowcolor{white}\multicolumn{8}{l}{\footnotesize $^\ast$SU(8)${\times}$U(1) and SU(7)${\times}$SU(2)${\times}$U(1) and SU(6)${\times}$SU(3)${\times}$U(1) and SU(5)${\times}$SU(4)${\times}$U(1) singlets resp.}
\endlastfoot
\dynkin{1, 0, 0, 0, 0, 0, 0, 0} & \irrep{9} & 1 & 1 & 1 & 0 & 0 & 0\\
\dynkin{0, 1, 0, 0, 0, 0, 0, 0} & \irrep{36} & 7 & 2 & 0 & 1 & 0 & 0\\
\dynkin{2, 0, 0, 0, 0, 0, 0, 0} & \irrep{45} & 11 & 2 & 1 & 0 & 0 & 0\\
\dynkin{1, 0, 0, 0, 0, 0, 0, 1} & \irrep{80} & 18 & 0 & \starred{1} & \starred{1} & \starred{1} & \starred{1}\\
\dynkin{0, 0, 1, 0, 0, 0, 0, 0} & \irrep{84} & 21 & 3 & 0 & 0 & 1 & 0\\
\dynkin{0, 0, 0, 1, 0, 0, 0, 0} & \irrep{126} & 35 & 4 & 0 & 0 & 0 & 1\\
\dynkin{3, 0, 0, 0, 0, 0, 0, 0} & \irrep{165} & 66 & 3 & 1 & 0 & 0 & 0\\
\dynkin{1, 1, 0, 0, 0, 0, 0, 0} & \irrep{240} & 78 & 3 & 0 & 0 & 0 & 0\\
\end{longtable}
\vspace{-20pt}

\setlength{\tabcolsep}{1pt}
\rowcolors{2}{}{tablerowcolor}
\begin{longtable}{lrrcccccc}
\caption{\label{tab:SU10Irreps}SU(10) Irreps}\\
\toprule
\rowcolor{tableheadcolor}
Dynkin & Dimension & l &  & SU(9) & SU(8)${\times}$SU(2) & SU(7)${\times}$SU(3) & SU(6)${\times}$SU(4) & SU(5)${\times}$SU(5)\\
\rowcolor{tableheadcolor}label & (name) & (index) & Decality & singlets & singlets & singlets & singlets & singlets\\
\midrule
\endfirsthead
\rowcolor{white}
\caption[]{SU(10) Irreps (continued)}\\
\toprule
\rowcolor{tableheadcolor}
Dynkin & Dimension & l &  & SU(9) & SU(8)${\times}$SU(2) & SU(7)${\times}$SU(3) & SU(6)${\times}$SU(4) & SU(5)${\times}$SU(5)\\
\rowcolor{tableheadcolor}label & (name) & (index) & Decality & singlets & singlets & singlets & singlets & singlets\\
\midrule
\endhead
\rowcolor{white}\multicolumn{9}{>{\raggedright\arraybackslash}p{0.8\textwidth}}{\footnotesize $^\ast$SU(9)${\times}$U(1) and SU(8)${\times}$SU(2)${\times}$U(1) and SU(7)${\times}$SU(3)${\times}$U(1) and SU(6)${\times}$SU(4)${\times}$U(1) and SU(5)${\times}$SU(5)${\times}$U(1) singlets resp.}
\endfoot
\bottomrule
\rowcolor{white}\multicolumn{9}{>{\raggedright\arraybackslash}p{0.8\textwidth}}{\footnotesize $^\ast$SU(9)${\times}$U(1) and SU(8)${\times}$SU(2)${\times}$U(1) and SU(7)${\times}$SU(3)${\times}$U(1) and SU(6)${\times}$SU(4)${\times}$U(1) and SU(5)${\times}$SU(5)${\times}$U(1) singlets resp.}
\endlastfoot
\dynkin{1, 0, 0, 0, 0, 0, 0, 0, 0} & \irrep{10} & 1 & 1 & 1 & 0 & 0 & 0 & 0\\
\dynkin{0, 1, 0, 0, 0, 0, 0, 0, 0} & \irrep{45} & 8 & 2 & 0 & 1 & 0 & 0 & 0\\
\dynkin{2, 0, 0, 0, 0, 0, 0, 0, 0} & \irrep{55} & 12 & 2 & 1 & 0 & 0 & 0 & 0\\
\dynkin{1, 0, 0, 0, 0, 0, 0, 0, 1} & \irrep{99} & 20 & 0 & \starred{1} & \starred{1} & \starred{1} & \starred{1} & \starred{1}\\
\dynkin{0, 0, 1, 0, 0, 0, 0, 0, 0} & \irrep{120} & 28 & 3 & 0 & 0 & 1 & 0 & 0\\
\dynkin{0, 0, 0, 1, 0, 0, 0, 0, 0} & \irrep{210} & 56 & 4 & 0 & 0 & 0 & 1 & 0\\
\dynkin{3, 0, 0, 0, 0, 0, 0, 0, 0} & \irrep{220} & 78 & 3 & 1 & 0 & 0 & 0 & 0\\
\dynkin{0, 0, 0, 0, 1, 0, 0, 0, 0} & \irrep{252} & 70 & 5 & 0 & 0 & 0 & 0 & 2\\
\end{longtable}
}
\newpage

\setlength{\tabcolsep}{0.5pt}
\rowcolors{2}{}{tablerowcolor}
\begin{longtable}{lrrcccccc}
\caption{\label{tab:SU11Irreps}SU(11) Irreps}\\
\toprule
\rowcolor{tableheadcolor}
Dynkin & Dimension & l &  & SU(10) & SU(9)${\times}$SU(2) & SU(8)${\times}$SU(3) & SU(7)${\times}$SU(4) & SU(6)${\times}$SU(5)\\
\rowcolor{tableheadcolor}label & (name) & (index) & Decality & singlets & singlets & singlets & singlets & singlets\\
\midrule
\endfirsthead
\rowcolor{white}
\caption[]{SU(11) Irreps (continued)}\\
\toprule
\rowcolor{tableheadcolor}
Dynkin & Dimension & l &  & SU(10) & SU(9)${\times}$SU(2) & SU(8)${\times}$SU(3) & SU(7)${\times}$SU(4) & SU(6)${\times}$SU(5)\\
\rowcolor{tableheadcolor}label & (name) & (index) & Decality & singlets & singlets & singlets & singlets & singlets\\
\midrule
\endhead
\rowcolor{white}\multicolumn{9}{>{\raggedright\arraybackslash}p{0.8\textwidth}}{\footnotesize $^\ast$SU(10)${\times}$U(1) and SU(9)${\times}$SU(2)${\times}$U(1) and SU(8)${\times}$SU(3)${\times}$U(1) and SU(7)${\times}$SU(4)${\times}$U(1) and SU(6)${\times}$SU(5)${\times}$U(1) singlets resp.}
\endfoot
\bottomrule
\rowcolor{white}\multicolumn{9}{>{\raggedright\arraybackslash}p{0.8\textwidth}}{\footnotesize $^\ast$SU(10)${\times}$U(1) and SU(9)${\times}$SU(2)${\times}$U(1) and SU(8)${\times}$SU(3)${\times}$U(1) and SU(7)${\times}$SU(4)${\times}$U(1) and SU(6)${\times}$SU(5)${\times}$U(1) singlets resp.}
\endlastfoot
\dynkin{1, 0, 0, 0, 0, 0, 0, 0, 0, 0} & \irrep{11} & 1 & 1 & 1 & 0 & 0 & 0 & 0\\
\dynkin{0, 1, 0, 0, 0, 0, 0, 0, 0, 0} & \irrep{55} & 9 & 2 & 0 & 1 & 0 & 0 & 0\\
\dynkin{2, 0, 0, 0, 0, 0, 0, 0, 0, 0} & \irrep{66} & 13 & 2 & 1 & 0 & 0 & 0 & 0\\
\dynkin{1, 0, 0, 0, 0, 0, 0, 0, 0, 1} & \irrep{120} & 22 & 0 & \starred{1} & \starred{1} & \starred{1} & \starred{1} & \starred{1}\\
\dynkin{0, 0, 1, 0, 0, 0, 0, 0, 0, 0} & \irrep{165} & 36 & 3 & 0 & 0 & 1 & 0 & 0\\
\dynkin{3, 0, 0, 0, 0, 0, 0, 0, 0, 0} & \irrep{286} & 91 & 3 & 1 & 0 & 0 & 0 & 0\\
\dynkin{0, 0, 0, 1, 0, 0, 0, 0, 0, 0} & \irrep{330} & 84 & 4 & 0 & 0 & 0 & 1 & 0\\
\dynkin{1, 1, 0, 0, 0, 0, 0, 0, 0, 0} & \irrep{440} & 118 & 3 & 0 & 0 & 0 & 0 & 0\\
\end{longtable}

\rowcolors{2}{}{tablerowcolor}
\begin{longtable}{lrrcccccc}
\caption{\label{tab:SU12Irreps}SU(12) Irreps}\\
\toprule
\rowcolor{tableheadcolor}
Dynkin & Dimension & l &  & SU(11) & SU(10)${\times}$SU(2) & SU(9)${\times}$SU(3) & SU(7)${\times}$SU(5) & SU(6)${\times}$SU(6)\\
\rowcolor{tableheadcolor}label & (name) & (index) & Decality & singlets & singlets & singlets & singlets & singlets\\
\midrule
\endfirsthead
\rowcolor{white}
\caption[]{SU(12) Irreps (continued)}\\
\toprule
\rowcolor{tableheadcolor}
Dynkin & Dimension & l &  & SU(11) & SU(10)${\times}$SU(2) & SU(9)${\times}$SU(3) & SU(7)${\times}$SU(5) & SU(6)${\times}$SU(6)\\
\rowcolor{tableheadcolor}label & (name) & (index) & Decality & singlets & singlets & singlets & singlets & singlets\\
\midrule
\endhead
\rowcolor{white}\multicolumn{9}{>{\raggedright\arraybackslash}p{0.8\textwidth}}{\footnotesize $^\ast$SU(11)${\times}$U(1) and SU(10)${\times}$SU(2)${\times}$U(1) and SU(9)${\times}$SU(3)${\times}$U(1) and SU(7)${\times}$SU(5)${\times}$U(1) and SU(6)${\times}$SU(6)${\times}$U(1) singlets resp.}
\endfoot
\bottomrule
\rowcolor{white}\multicolumn{9}{>{\raggedright\arraybackslash}p{0.8\textwidth}}{\footnotesize $^\ast$SU(11)${\times}$U(1) and SU(10)${\times}$SU(2)${\times}$U(1) and SU(9)${\times}$SU(3)${\times}$U(1) and SU(7)${\times}$SU(5)${\times}$U(1) and SU(6)${\times}$SU(6)${\times}$U(1) singlets resp.}
\endlastfoot
\dynkin{1, 0, 0, 0, 0, 0, 0, 0, 0, 0, 0} & \irrep{12} & 1 & 1 & 1 & 0 & 0 & 0 & 0\\
\dynkin{0, 1, 0, 0, 0, 0, 0, 0, 0, 0, 0} & \irrep{66} & 10 & 2 & 0 & 1 & 0 & 0 & 0\\
\dynkin{2, 0, 0, 0, 0, 0, 0, 0, 0, 0, 0} & \irrep{78} & 14 & 2 & 1 & 0 & 0 & 0 & 0\\
\dynkin{1, 0, 0, 0, 0, 0, 0, 0, 0, 0, 1} & \irrep{143} & 24 & 0 & \starred{1} & \starred{1} & \starred{1} & \starred{1} & \starred{1}\\
\dynkin{0, 0, 1, 0, 0, 0, 0, 0, 0, 0, 0} & \irrep{220} & 45 & 3 & 0 & 0 & 1 & 0 & 0\\
\dynkin{3, 0, 0, 0, 0, 0, 0, 0, 0, 0, 0} & \irrep{364} & 105 & 3 & 1 & 0 & 0 & 0 & 0\\
\dynkin{0, 0, 0, 1, 0, 0, 0, 0, 0, 0, 0} & \irrep{495} & 120 & 4 & 0 & 0 & 0 & 0 & 0\\
\dynkin{1, 1, 0, 0, 0, 0, 0, 0, 0, 0, 0} & \irrep{572} & 141 & 3 & 0 & 0 & 0 & 0 & 0\\
\end{longtable}
\newpage

\subsubsection{\SO{N}}

\setlength{\tabcolsep}{7pt}

\rowcolors{2}{}{tablerowcolor}
\begin{longtable}{lrrcc}
\caption{\label{tab:SO7Irreps}SO(7) Irreps}\\
\toprule
\rowcolor{tableheadcolor}
Dynkin & Dimension & l/2 & Congruency & SU(4)\\
\rowcolor{tableheadcolor}label & (name) & (index) & class & singlets\\
\midrule
\endfirsthead
\rowcolor{white}
\caption[]{SO(7) Irreps (continued)}\\
\toprule
\rowcolor{tableheadcolor}
Dynkin & Dimension & l/2 & Congruency & SU(4)\\
\rowcolor{tableheadcolor}label & (name) & (index) & class & singlets\\
\midrule
\endhead
\endfoot
\bottomrule
\endlastfoot
\dynkin{1, 0, 0} & \irrep{7} & 1 & 0 & 1\\
\dynkin{0, 0, 1} & \irrep{8} & 1 & 1 & 0\\
\dynkin{0, 1, 0} & \irrep{21} & 5 & 0 & 0\\
\dynkin{2, 0, 0} & \irrep{27} & 9 & 0 & 1\\
\dynkin{0, 0, 2} & \irrep{35} & 10 & 0 & 0\\
\dynkin{1, 0, 1} & \irrep{48} & 14 & 1 & 0\\
\dynkin{3, 0, 0} & \irrep{77} & 44 & 0 & 1\\
\dynkin{1, 1, 0} & \irrep{105} & 45 & 0 & 0\\
\end{longtable}
\newpage

\rowcolors{2}{}{tablerowcolor}
\begin{longtable}{lrrcc}
\caption{\label{tab:SO8Irreps}SO(8) Irreps}\\
\toprule
\rowcolor{tableheadcolor}
Dynkin & Dimension & l/2 & Congruency & SO(7)\\
\rowcolor{tableheadcolor}label & (name) & (index) & class & singlets\\
\midrule
\endfirsthead
\rowcolor{white}
\caption[]{SO(8) Irreps (continued)}\\
\toprule
\rowcolor{tableheadcolor}
Dynkin & Dimension & l/2 & Congruency & SO(7)\\
\rowcolor{tableheadcolor}label & (name) & (index) & class & singlets\\
\midrule
\endhead
\endfoot
\bottomrule
\endlastfoot
\dynkin{0, 0, 0, 1} & \irrepsub{8}{s} & 1 & (10) & 1\\
\dynkin{1, 0, 0, 0} & \irrepsub{8}{v} & 1 & (02) & 0\\
\dynkin{0, 0, 1, 0} & \irrepsub{8}{c} & 1 & (12) & 0\\
\dynkin{0, 1, 0, 0} & \irrep{28} & 6 & (00) & 0\\
\dynkin{2, 0, 0, 0} & \irrepsub{35}{v} & 10 & (00) & 0\\
\dynkin{0, 0, 2, 0} & \irrepsub{35}{c} & 10 & (00) & 0\\
\dynkin{0, 0, 0, 2} & \irrepsub{35}{s} & 10 & (00) & 1\\
\dynkin{1, 0, 1, 0} & \irrepsub{56}{s} & 15 & (10) & 0\\
\end{longtable}
\newpage

\rowcolors{2}{}{tablerowcolor}
\begin{longtable}{lrrccc}
\caption{\label{tab:SO9Irreps}SO(9) Irreps}\\
\toprule
\rowcolor{tableheadcolor}
Dynkin & Dimension & l/2 & Congruency & SO(8) & SU(4)${\times}$SU(2)\\
\rowcolor{tableheadcolor}label & (name) & (index) & class & singlets & singlets\\
\midrule
\endfirsthead
\rowcolor{white}
\caption[]{SO(9) Irreps (continued)}\\
\toprule
\rowcolor{tableheadcolor}
Dynkin & Dimension & l/2 & Congruency & SO(8) & SU(4)${\times}$SU(2)\\
\rowcolor{tableheadcolor}label & (name) & (index) & class & singlets & singlets\\
\midrule
\endhead
\endfoot
\bottomrule
\endlastfoot
\dynkin{1, 0, 0, 0} & \irrep{9} & 1 & 0 & 1 & 0\\
\dynkin{0, 0, 0, 1} & \irrep{16} & 2 & 1 & 0 & 0\\
\dynkin{0, 1, 0, 0} & \irrep{36} & 7 & 0 & 0 & 0\\
\dynkin{2, 0, 0, 0} & \irrep{44} & 11 & 0 & 1 & 1\\
\dynkin{0, 0, 1, 0} & \irrep{84} & 21 & 0 & 0 & 1\\
\dynkin{0, 0, 0, 2} & \irrep{126} & 35 & 0 & 0 & 0\\
\dynkin{1, 0, 0, 1} & \irrep{128} & 32 & 1 & 0 & 0\\
\dynkin{3, 0, 0, 0} & \irrep{156} & 65 & 0 & 1 & 0\\
\end{longtable}
\newpage

{
\setlength{\tabcolsep}{3pt}
\rowcolors{2}{}{tablerowcolor}
\begin{longtable}{lrrccccc}
\caption{\label{tab:SO10Irreps}SO(10) Irreps}\\
\toprule
\rowcolor{tableheadcolor}
Dynkin & Dimension & l/2 & Congruency & SU(5) & SU(2)${\times}$SU(2)${\times}$SU(4) & SO(9) & SU(2)${\times}$SO(7)\\
\rowcolor{tableheadcolor}label & (name) & (index) & class & singlets & singlets & singlets & singlets\\
\midrule
\endfirsthead
\rowcolor{white}
\caption[]{SO(10) Irreps (continued)}\\
\toprule
\rowcolor{tableheadcolor}
Dynkin & Dimension & l/2 & Congruency & SU(5) & SU(2)${\times}$SU(2)${\times}$SU(4) & SO(9) & SU(2)${\times}$SO(7)\\
\rowcolor{tableheadcolor}label & (name) & (index) & class & singlets & singlets & singlets & singlets\\
\midrule
\endhead
\rowcolor{white}\multicolumn{8}{l}{\footnotesize $^\ast$SU(5)${\times}$U(1) singlets resp.}
\endfoot
\bottomrule
\rowcolor{white}\multicolumn{8}{l}{\footnotesize $^\ast$SU(5)${\times}$U(1) singlets resp.}
\endlastfoot
\dynkin{1, 0, 0, 0, 0} & \irrep{10} & 1 & (02) & 0 & 0 & 1 & 0\\
\dynkin{0, 0, 0, 0, 1} & \irrep{16} & 2 & (11) & 1 & 0 & 0 & 0\\
\dynkin{0, 1, 0, 0, 0} & \irrep{45} & 8 & (00) & \starred{1} & 0 & 0 & 0\\
\dynkin{2, 0, 0, 0, 0} & \irrep{54} & 12 & (00) & 0 & 1 & 1 & 1\\
\dynkin{0, 0, 1, 0, 0} & \irrep{120} & 28 & (02) & 0 & 0 & 0 & 1\\
\dynkin{0, 0, 0, 2, 0} & \irrep{126} & 35 & (02) & 1 & 0 & 0 & 0\\
\dynkin{1, 0, 0, 1, 0} & \irrep{144} & 34 & (11) & 0 & 0 & 0 & 0\\
\dynkin{0, 0, 0, 1, 1} & \irrep{210} & 56 & (00) & \starred{1} & 1 & 0 & 0\\
\end{longtable}
}
\newpage

\rowcolors{2}{}{tablerowcolor}
\begin{longtable}{lrrc}
\caption{\label{tab:SO11Irreps}SO(11) Irreps}\\
\toprule
\rowcolor{tableheadcolor}
Dynkin & Dimension & l/2 & Congruency\\
\rowcolor{tableheadcolor}label & (name) & (index) & class\\
\midrule
\endfirsthead
\rowcolor{white}
\caption[]{SO(11) Irreps (continued)}\\
\toprule
\rowcolor{tableheadcolor}
Dynkin & Dimension & l/2 & Congruency\\
\rowcolor{tableheadcolor}label & (name) & (index) & class\\
\midrule
\endhead
\endfoot
\bottomrule
\endlastfoot
\dynkin{1, 0, 0, 0, 0} & \irrep{11} & 1 & 0\\
\dynkin{0, 0, 0, 0, 1} & \irrep{32} & 4 & 1\\
\dynkin{0, 1, 0, 0, 0} & \irrep{55} & 9 & 0\\
\dynkin{2, 0, 0, 0, 0} & \irrep{65} & 13 & 0\\
\dynkin{0, 0, 1, 0, 0} & \irrep{165} & 36 & 0\\
\dynkin{3, 0, 0, 0, 0} & \irrep{275} & 90 & 0\\
\dynkin{1, 0, 0, 0, 1} & \irrep{320} & 72 & 1\\
\dynkin{0, 0, 0, 1, 0} & \irrep{330} & 84 & 0\\
\end{longtable}
\newpage

\rowcolors{2}{}{tablerowcolor}
\begin{longtable}{lrrc}
\caption{\label{tab:SO12Irreps}SO(12) Irreps}\\
\toprule
\rowcolor{tableheadcolor}
Dynkin & Dimension & l/2 & Congruency\\
\rowcolor{tableheadcolor}label & (name) & (index) & class\\
\midrule
\endfirsthead
\rowcolor{white}
\caption[]{SO(12) Irreps (continued)}\\
\toprule
\rowcolor{tableheadcolor}
Dynkin & Dimension & l/2 & Congruency\\
\rowcolor{tableheadcolor}label & (name) & (index) & class\\
\midrule
\endhead
\endfoot
\bottomrule
\endlastfoot
\dynkin{1, 0, 0, 0, 0, 0} & \irrep{12} & 1 & (02)\\
\dynkin{0, 0, 0, 0, 1, 0} & \irrep{32} & 4 & (10)\\
\dynkin{0, 1, 0, 0, 0, 0} & \irrep{66} & 10 & (00)\\
\dynkin{2, 0, 0, 0, 0, 0} & \irrep{77} & 14 & (00)\\
\dynkin{0, 0, 1, 0, 0, 0} & \irrep{220} & 45 & (02)\\
\dynkin{1, 0, 0, 0, 0, 1} & \irrep{352} & 76 & (10)\\
\dynkin{3, 0, 0, 0, 0, 0} & \irrep[1]{352} & 104 & (02)\\
\dynkin{0, 0, 0, 0, 2, 0} & \irrep{462} & 126 & (00)\\
\end{longtable}
\newpage

\rowcolors{2}{}{tablerowcolor}
\begin{longtable}{lrrc}
\caption{\label{tab:SO13Irreps}SO(13) Irreps}\\
\toprule
\rowcolor{tableheadcolor}
Dynkin & Dimension & l/2 & Congruency\\
\rowcolor{tableheadcolor}label & (name) & (index) & class\\
\midrule
\endfirsthead
\rowcolor{white}
\caption[]{SO(13) Irreps (continued)}\\
\toprule
\rowcolor{tableheadcolor}
Dynkin & Dimension & l/2 & Congruency\\
\rowcolor{tableheadcolor}label & (name) & (index) & class\\
\midrule
\endhead
\endfoot
\bottomrule
\endlastfoot
\dynkin{1, 0, 0, 0, 0, 0} & \irrep{13} & 1 & 0\\
\dynkin{0, 0, 0, 0, 0, 1} & \irrep{64} & 8 & 1\\
\dynkin{0, 1, 0, 0, 0, 0} & \irrep{78} & 11 & 0\\
\dynkin{2, 0, 0, 0, 0, 0} & \irrep{90} & 15 & 0\\
\dynkin{0, 0, 1, 0, 0, 0} & \irrep{286} & 55 & 0\\
\dynkin{3, 0, 0, 0, 0, 0} & \irrep{442} & 119 & 0\\
\dynkin{1, 1, 0, 0, 0, 0} & \irrep{715} & 165 & 0\\
\dynkin{1, 0, 0, 0, 0, 1} & \irrep{768} & 160 & 1\\
\dynkin{0, 0, 0, 0, 1, 0} & \irrep{1287} & 330 & 0\\
\dynkin{0, 0, 0, 0, 0, 2} & \irrep{1716} & 462 & 0\\
\dynkin{4, 0, 0, 0, 0, 0} & \irrep{1729} & 665 & 0\\
\dynkin{0, 2, 0, 0, 0, 0} & \irrep{2275} & 700 & 0\\
\dynkin{1, 0, 1, 0, 0, 0} & \irrep{2925} & 825 & 0\\
\dynkin{2, 1, 0, 0, 0, 0} & \irrep{3927} & 1309 & 0\\
\dynkin{0, 1, 0, 0, 0, 1} & \irrep{4160} & 1160 & 1\\
\dynkin{2, 0, 0, 0, 0, 1} & \irrep{4992} & 1520 & 1\\
\dynkin{5, 0, 0, 0, 0, 0} & \irrep{5733} & 2940 & 0\\
\dynkin{1, 0, 0, 1, 0, 0} & \irrep{7722} & 2475 & 0\\
\dynkin{0, 1, 1, 0, 0, 0} & \irrep{12285} & 4410 & 0\\
\dynkin{0, 0, 1, 0, 0, 1} & \irrep{13312} & 4480 & 1\\
\dynkin{1, 0, 0, 0, 1, 0} & \irrep{14300} & 4950 & 0\\
\dynkin{3, 1, 0, 0, 0, 0} & \irrep{16302} & 7315 & 0\\
\dynkin{1, 2, 0, 0, 0, 0} & \irrep{16575} & 6800 & 0\\
\dynkin{6, 0, 0, 0, 0, 0} & \irrep{16744} & 10948 & 0\\
\dynkin{2, 0, 1, 0, 0, 0} & \irrep{17017} & 6545 & 0\\
\end{longtable}
\vspace{-10pt}

\rowcolors{2}{}{tablerowcolor}
\begin{longtable}{lrrcc}
\caption{\label{tab:SO14Irreps}SO(14) Irreps}\\
\toprule
\rowcolor{tableheadcolor}
Dynkin & Dimension & l/2 & Congruency & SU(2)${\times}$SU(2)${\times}$SO(10)\\
\rowcolor{tableheadcolor}label & (name) & (index) & class & singlets\\
\midrule
\endfirsthead
\rowcolor{white}
\caption[]{SO(14) Irreps (continued)}\\
\toprule
\rowcolor{tableheadcolor}
Dynkin & Dimension & l/2 & Congruency & SU(2)${\times}$SU(2)${\times}$SO(10)\\
\rowcolor{tableheadcolor}label & (name) & (index) & class & singlets\\
\midrule
\endhead
\endfoot
\bottomrule
\endlastfoot
\dynkin{1, 0, 0, 0, 0, 0, 0} & \irrep{14} & 1 & (02) & 0\\
\dynkin{0, 0, 0, 0, 0, 1, 0} & \irrep{64} & 8 & (11) & 0\\
\dynkin{0, 1, 0, 0, 0, 0, 0} & \irrep{91} & 12 & (00) & 0\\
\dynkin{2, 0, 0, 0, 0, 0, 0} & \irrep{104} & 16 & (00) & 1\\
\dynkin{0, 0, 1, 0, 0, 0, 0} & \irrep{364} & 66 & (02) & 0\\
\dynkin{3, 0, 0, 0, 0, 0, 0} & \irrep{546} & 135 & (02) & 0\\
\dynkin{1, 0, 0, 0, 0, 0, 1} & \irrep{832} & 168 & (11) & 0\\
\dynkin{1, 1, 0, 0, 0, 0, 0} & \irrep{896} & 192 & (02) & 0\\
\dynkin{0, 0, 0, 1, 0, 0, 0} & \irrep{1001} & 220 & (00) & 1\\
\dynkin{0, 0, 0, 0, 0, 2, 0} & \irrep{1716} & 462 & (02) & 0\\
\dynkin{0, 0, 0, 0, 1, 0, 0} & \irrep{2002} & 495 & (02) & 0\\
\dynkin{4, 0, 0, 0, 0, 0, 0} & \irrep{2275} & 800 & (00) & 1\\
\dynkin{0, 0, 0, 0, 0, 1, 1} & \irrep{3003} & 792 & (00) & 0\\
\dynkin{0, 2, 0, 0, 0, 0, 0} & \irrep{3080} & 880 & (00) & 1\\
\dynkin{1, 0, 1, 0, 0, 0, 0} & \irrep{4004} & 1056 & (00) & 0\\
\dynkin{0, 1, 0, 0, 0, 1, 0} & \irrep{4928} & 1320 & (11) & 0\\
\dynkin{2, 1, 0, 0, 0, 0, 0} & \irrep{5265} & 1620 & (00) & 0\\
\dynkin{2, 0, 0, 0, 0, 1, 0} & \irrep{5824} & 1688 & (11) & 0\\
\end{longtable}
\newpage

\rowcolors{2}{}{tablerowcolor}
\begin{longtable}{lrrcc}
\caption{\label{tab:SO18Irreps}SO(18) Irreps}\\
\toprule
\rowcolor{tableheadcolor}
Dynkin & Dimension & l/2 & Congruency & SO(8)${\times}$SO(10)\\
\rowcolor{tableheadcolor}label & (name) & (index) & class & singlets\\
\midrule
\endfirsthead
\rowcolor{white}
\caption[]{SO(18) Irreps (continued)}\\
\toprule
\rowcolor{tableheadcolor}
Dynkin & Dimension & l/2 & Congruency & SO(8)${\times}$SO(10)\\
\rowcolor{tableheadcolor}label & (name) & (index) & class & singlets\\
\midrule
\endhead
\endfoot
\bottomrule
\endlastfoot
\dynkin{1, 0, 0, 0, 0, 0, 0, 0, 0} & \irrep{18} & 1 & (02) & 0\\
\dynkin{0, 1, 0, 0, 0, 0, 0, 0, 0} & \irrep{153} & 16 & (00) & 0\\
\dynkin{2, 0, 0, 0, 0, 0, 0, 0, 0} & \irrep{170} & 20 & (00) & 1\\
\dynkin{0, 0, 0, 0, 0, 0, 0, 0, 1} & \irrep{256} & 32 & (11) & 0\\
\dynkin{0, 0, 1, 0, 0, 0, 0, 0, 0} & \irrep{816} & 120 & (02) & 0\\
\dynkin{3, 0, 0, 0, 0, 0, 0, 0, 0} & \irrep{1122} & 209 & (02) & 0\\
\dynkin{1, 1, 0, 0, 0, 0, 0, 0, 0} & \irrep{1920} & 320 & (02) & 0\\
\dynkin{0, 0, 0, 1, 0, 0, 0, 0, 0} & \irrep{3060} & 560 & (00) & 0\\
\dynkin{1, 0, 0, 0, 0, 0, 0, 1, 0} & \irrep{4352} & 800 & (11) & 0\\
\end{longtable}

\rowcolors{2}{}{tablerowcolor}
\begin{longtable}{lrrcc}
\caption{\label{tab:SO22Irreps}SO(22) Irreps}\\
\toprule
\rowcolor{tableheadcolor}
Dynkin & Dimension & l/2 & Congruency & SO(12)${\times}$SO(10)\\
\rowcolor{tableheadcolor}label & (name) & (index) & class & singlets\\
\midrule
\endfirsthead
\rowcolor{white}
\caption[]{SO(22) Irreps (continued)}\\
\toprule
\rowcolor{tableheadcolor}
Dynkin & Dimension & l/2 & Congruency & SO(12)${\times}$SO(10)\\
\rowcolor{tableheadcolor}label & (name) & (index) & class & singlets\\
\midrule
\endhead
\endfoot
\bottomrule
\endlastfoot
\dynkin{1, 0, 0, 0, 0, 0, 0, 0, 0, 0, 0} & \irrep{22} & 1 & (02) & 0\\
\dynkin{0, 1, 0, 0, 0, 0, 0, 0, 0, 0, 0} & \irrep{231} & 20 & (00) & 0\\
\dynkin{2, 0, 0, 0, 0, 0, 0, 0, 0, 0, 0} & \irrep{252} & 24 & (00) & 1\\
\dynkin{0, 0, 0, 0, 0, 0, 0, 0, 0, 1, 0} & \irrep{1024} & 128 & (11) & 0\\
\dynkin{0, 0, 1, 0, 0, 0, 0, 0, 0, 0, 0} & \irrep{1540} & 190 & (02) & 0\\
\dynkin{1, 1, 0, 0, 0, 0, 0, 0, 0, 0, 0} & \irrep{3520} & 480 & (02) & 0\\
\dynkin{0, 0, 0, 1, 0, 0, 0, 0, 0, 0, 0} & \irrep{7315} & 1140 & (00) & 0\\
\end{longtable}

\rowcolors{2}{}{tablerowcolor}
\begin{longtable}{lrrcc}
\caption{\label{tab:SO26Irreps}SO(26) Irreps}\\
\toprule
\rowcolor{tableheadcolor}
Dynkin & Dimension & l/2 & Congruency & SO(16)${\times}$SO(10)\\
\rowcolor{tableheadcolor}label & (name) & (index) & class & singlets\\
\midrule
\endfirsthead
\rowcolor{white}
\caption[]{SO(26) Irreps (continued)}\\
\toprule
\rowcolor{tableheadcolor}
Dynkin & Dimension & l/2 & Congruency & SO(16)${\times}$SO(10)\\
\rowcolor{tableheadcolor}label & (name) & (index) & class & singlets\\
\midrule
\endhead
\endfoot
\bottomrule
\endlastfoot
\dynkin{1, 0, 0, 0, 0, 0, 0, 0, 0, 0, 0, 0, 0} & \irrep{26} & 1 & (02) & 0\\
\dynkin{0, 1, 0, 0, 0, 0, 0, 0, 0, 0, 0, 0, 0} & \irrep{325} & 24 & (00) & 0\\
\dynkin{0, 0, 1, 0, 0, 0, 0, 0, 0, 0, 0, 0, 0} & \irrep{2600} & 276 & (02) & 0\\
\dynkin{0, 0, 0, 0, 0, 0, 0, 0, 0, 0, 0, 0, 1} & \irrep{4096} & 512 & (11) & 0\\
\dynkin{1, 1, 0, 0, 0, 0, 0, 0, 0, 0, 0, 0, 0} & \irrep{5824} & 672 & (02) & 0\\
\dynkin{0, 0, 0, 1, 0, 0, 0, 0, 0, 0, 0, 0, 0} & \irrep{14950} & 2024 & (00) & 0\\
\dynkin{1, 0, 1, 0, 0, 0, 0, 0, 0, 0, 0, 0, 0} & \irrep{52325} & 7728 & (00) & 0\\
\dynkin{0, 0, 0, 0, 1, 0, 0, 0, 0, 0, 0, 0, 0} & \irrep{65780} & 10626 & (02) & 0\\
\dynkin{1, 0, 0, 0, 0, 0, 0, 0, 0, 0, 0, 1, 0} & \irrep{102400} & 16896 & (11) & 0\\
\dynkin{0, 0, 0, 0, 0, 1, 0, 0, 0, 0, 0, 0, 0} & \irrep{230230} & 42504 & (00) & 0\\
\dynkin{1, 0, 0, 1, 0, 0, 0, 0, 0, 0, 0, 0, 0} & \irrep{320320} & 56672 & (02) & 0\\
\dynkin{0, 1, 1, 0, 0, 0, 0, 0, 0, 0, 0, 0, 0} & \irrep{450450} & 83853 & (02) & 0\\
\dynkin{0, 0, 0, 0, 0, 0, 1, 0, 0, 0, 0, 0, 0} & \irrep{657800} & 134596 & (02) & 0\\
\end{longtable}
\newpage

\subsubsection{\Sp{N}}

\rowcolors{2}{}{tablerowcolor}
\begin{longtable}{lrrc}
\caption{\label{tab:Sp4Irreps}Sp(4) Irreps}\\
\toprule
\rowcolor{tableheadcolor}
Dynkin & Dimension & l & Congruency\\
\rowcolor{tableheadcolor}label & (name) & (index) & class\\
\midrule
\endfirsthead
\rowcolor{white}
\caption[]{Sp(4) Irreps (continued)}\\
\toprule
\rowcolor{tableheadcolor}
Dynkin & Dimension & l & Congruency\\
\rowcolor{tableheadcolor}label & (name) & (index) & class\\
\midrule
\endhead
\endfoot
\bottomrule
\endlastfoot
\dynkin{1, 0} & \irrep{4} & 1 & 1\\
\dynkin{0, 1} & \irrep{5} & 2 & 0\\
\dynkin{2, 0} & \irrep{10} & 6 & 0\\
\dynkin{0, 2} & \irrep{14} & 14 & 0\\
\dynkin{1, 1} & \irrep{16} & 12 & 1\\
\dynkin{3, 0} & \irrep{20} & 21 & 1\\
\dynkin{0, 3} & \irrep{30} & 54 & 0\\
\dynkin{2, 1} & \irrep{35} & 42 & 0\\
\end{longtable}
\newpage

\rowcolors{2}{}{tablerowcolor}
\begin{longtable}{lrrc}
\caption{\label{tab:Sp6Irreps}Sp(6) Irreps}\\
\toprule
\rowcolor{tableheadcolor}
Dynkin & Dimension & l & Congruency\\
\rowcolor{tableheadcolor}label & (name) & (index) & class\\
\midrule
\endfirsthead
\rowcolor{white}
\caption[]{Sp(6) Irreps (continued)}\\
\toprule
\rowcolor{tableheadcolor}
Dynkin & Dimension & l & Congruency\\
\rowcolor{tableheadcolor}label & (name) & (index) & class\\
\midrule
\endhead
\endfoot
\bottomrule
\endlastfoot
\dynkin{1, 0, 0} & \irrep{6} & 1 & 1\\
\dynkin{0, 1, 0} & \irrep{14} & 4 & 0\\
\dynkin{0, 0, 1} & \irrep[1]{14} & 5 & 1\\
\dynkin{2, 0, 0} & \irrep{21} & 8 & 0\\
\dynkin{3, 0, 0} & \irrep{56} & 36 & 1\\
\dynkin{1, 1, 0} & \irrep{64} & 32 & 1\\
\dynkin{1, 0, 1} & \irrep{70} & 40 & 0\\
\dynkin{0, 0, 2} & \irrep{84} & 72 & 0\\
\end{longtable}
\newpage

\rowcolors{2}{}{tablerowcolor}
\begin{longtable}{lrrc}
\caption{\label{tab:Sp8Irreps}Sp(8) Irreps}\\
\toprule
\rowcolor{tableheadcolor}
Dynkin & Dimension & l & Congruency\\
\rowcolor{tableheadcolor}label & (name) & (index) & class\\
\midrule
\endfirsthead
\rowcolor{white}
\caption[]{Sp(8) Irreps (continued)}\\
\toprule
\rowcolor{tableheadcolor}
Dynkin & Dimension & l & Congruency\\
\rowcolor{tableheadcolor}label & (name) & (index) & class\\
\midrule
\endhead
\endfoot
\bottomrule
\endlastfoot
\dynkin{1, 0, 0, 0} & \irrep{8} & 1 & 1\\
\dynkin{0, 1, 0, 0} & \irrep{27} & 6 & 0\\
\dynkin{2, 0, 0, 0} & \irrep{36} & 10 & 0\\
\dynkin{0, 0, 0, 1} & \irrep{42} & 14 & 0\\
\dynkin{0, 0, 1, 0} & \irrep{48} & 14 & 1\\
\dynkin{3, 0, 0, 0} & \irrep{120} & 55 & 1\\
\dynkin{1, 1, 0, 0} & \irrep{160} & 60 & 1\\
\dynkin{1, 0, 0, 1} & \irrep{288} & 140 & 1\\
\end{longtable}
\newpage

\rowcolors{2}{}{tablerowcolor}
\begin{longtable}{lrrc}
\caption{\label{tab:Sp10Irreps}Sp(10) Irreps}\\
\toprule
\rowcolor{tableheadcolor}
Dynkin & Dimension & l & Congruency\\
\rowcolor{tableheadcolor}label & (name) & (index) & class\\
\midrule
\endfirsthead
\rowcolor{white}
\caption[]{Sp(10) Irreps (continued)}\\
\toprule
\rowcolor{tableheadcolor}
Dynkin & Dimension & l & Congruency\\
\rowcolor{tableheadcolor}label & (name) & (index) & class\\
\midrule
\endhead
\endfoot
\bottomrule
\endlastfoot
\dynkin{1, 0, 0, 0, 0} & \irrep{10} & 1 & 1\\
\dynkin{0, 1, 0, 0, 0} & \irrep{44} & 8 & 0\\
\dynkin{2, 0, 0, 0, 0} & \irrep{55} & 12 & 0\\
\dynkin{0, 0, 1, 0, 0} & \irrep{110} & 27 & 1\\
\dynkin{0, 0, 0, 0, 1} & \irrep{132} & 42 & 1\\
\dynkin{0, 0, 0, 1, 0} & \irrep{165} & 48 & 0\\
\dynkin{3, 0, 0, 0, 0} & \irrep{220} & 78 & 1\\
\dynkin{1, 1, 0, 0, 0} & \irrep{320} & 96 & 1\\
\end{longtable}
\newpage

\rowcolors{2}{}{tablerowcolor}
\begin{longtable}{lrrc}
\caption{\label{tab:Sp12Irreps}Sp(12) Irreps}\\
\toprule
\rowcolor{tableheadcolor}
Dynkin & Dimension & l & Congruency\\
\rowcolor{tableheadcolor}label & (name) & (index) & class\\
\midrule
\endfirsthead
\rowcolor{white}
\caption[]{Sp(12) Irreps (continued)}\\
\toprule
\rowcolor{tableheadcolor}
Dynkin & Dimension & l & Congruency\\
\rowcolor{tableheadcolor}label & (name) & (index) & class\\
\midrule
\endhead
\endfoot
\bottomrule
\endlastfoot
\dynkin{1, 0, 0, 0, 0, 0} & \irrep{12} & 1 & 1\\
\dynkin{0, 1, 0, 0, 0, 0} & \irrep{65} & 10 & 0\\
\dynkin{2, 0, 0, 0, 0, 0} & \irrep{78} & 14 & 0\\
\dynkin{0, 0, 1, 0, 0, 0} & \irrep{208} & 44 & 1\\
\dynkin{3, 0, 0, 0, 0, 0} & \irrep{364} & 105 & 1\\
\dynkin{0, 0, 0, 1, 0, 0} & \irrep{429} & 110 & 0\\
\dynkin{0, 0, 0, 0, 0, 1} & \irrep[1]{429} & 132 & 0\\
\dynkin{1, 1, 0, 0, 0, 0} & \irrep{560} & 140 & 1\\
\dynkin{0, 0, 0, 0, 1, 0} & \irrep{572} & 165 & 1\\
\dynkin{4, 0, 0, 0, 0, 0} & \irrep{1365} & 560 & 0\\
\dynkin{0, 2, 0, 0, 0, 0} & \irrep{1650} & 550 & 0\\
\dynkin{1, 0, 1, 0, 0, 0} & \irrep{2002} & 616 & 0\\
\dynkin{2, 1, 0, 0, 0, 0} & \irrep{2925} & 1050 & 0\\
\dynkin{1, 0, 0, 1, 0, 0} & \irrep{4368} & 1540 & 1\\
\dynkin{5, 0, 0, 0, 0, 0} & \irrep[1]{4368} & 2380 & 1\\
\dynkin{1, 0, 0, 0, 0, 1} & \irrep{4576} & 1848 & 1\\
\dynkin{1, 0, 0, 0, 1, 0} & \irrep{6006} & 2310 & 0\\
\dynkin{0, 1, 1, 0, 0, 0} & \irrep{7800} & 3050 & 1\\
\dynkin{2, 0, 1, 0, 0, 0} & \irrep{11088} & 4620 & 1\\
\dynkin{1, 2, 0, 0, 0, 0} & \irrep{11440} & 5060 & 1\\
\dynkin{3, 1, 0, 0, 0, 0} & \irrep{11648} & 5600 & 1\\
\dynkin{6, 0, 0, 0, 0, 0} & \irrep{12376} & 8568 & 0\\
\dynkin{0, 0, 2, 0, 0, 0} & \irrep{13650} & 6300 & 0\\
\dynkin{0, 1, 0, 1, 0, 0} & \irrep{18954} & 8262 & 0\\
\dynkin{0, 1, 0, 0, 0, 1} & \irrep{21450} & 10450 & 0\\
\dynkin{0, 3, 0, 0, 0, 0} & \irrep{24310} & 13090 & 0\\
\end{longtable}

\rowcolors{2}{}{tablerowcolor}
\begin{longtable}{lrrc}
\caption{\label{tab:Sp14Irreps}Sp(14) Irreps}\\
\toprule
\rowcolor{tableheadcolor}
Dynkin & Dimension & l & Congruency\\
\rowcolor{tableheadcolor}label & (name) & (index) & class\\
\midrule
\endfirsthead
\rowcolor{white}
\caption[]{Sp(14) Irreps (continued)}\\
\toprule
\rowcolor{tableheadcolor}
Dynkin & Dimension & l & Congruency\\
\rowcolor{tableheadcolor}label & (name) & (index) & class\\
\midrule
\endhead
\endfoot
\bottomrule
\endlastfoot
\dynkin{1, 0, 0, 0, 0, 0, 0} & \irrep{14} & 1 & 1\\
\dynkin{0, 1, 0, 0, 0, 0, 0} & \irrep{90} & 12 & 0\\
\dynkin{2, 0, 0, 0, 0, 0, 0} & \irrep{105} & 16 & 0\\
\dynkin{0, 0, 1, 0, 0, 0, 0} & \irrep{350} & 65 & 1\\
\dynkin{3, 0, 0, 0, 0, 0, 0} & \irrep{560} & 136 & 1\\
\dynkin{1, 1, 0, 0, 0, 0, 0} & \irrep{896} & 192 & 1\\
\dynkin{0, 0, 0, 1, 0, 0, 0} & \irrep{910} & 208 & 0\\
\dynkin{0, 0, 0, 0, 0, 0, 1} & \irrep{1430} & 429 & 1\\
\dynkin{0, 0, 0, 0, 1, 0, 0} & \irrep{1638} & 429 & 1\\
\dynkin{0, 0, 0, 0, 0, 1, 0} & \irrep{2002} & 572 & 0\\
\dynkin{4, 0, 0, 0, 0, 0, 0} & \irrep{2380} & 816 & 0\\
\dynkin{0, 2, 0, 0, 0, 0, 0} & \irrep{3094} & 884 & 0\\
\dynkin{1, 0, 1, 0, 0, 0, 0} & \irrep{3900} & 1040 & 0\\
\dynkin{2, 1, 0, 0, 0, 0, 0} & \irrep{5355} & 1632 & 0\\
\dynkin{5, 0, 0, 0, 0, 0, 0} & \irrep{8568} & 3876 & 1\\
\end{longtable}
\newpage

\subsubsection{Exceptional Algebras}

\enlargethispage{10pt}
\rowcolors{2}{}{tablerowcolor}
\begin{longtable}{lrrcccc}
\caption{\label{tab:E6Irreps}\E6 Irreps}\\
\toprule
\rowcolor{tableheadcolor}
Dynkin & Dimension & l/6 &  & SO(10) & SU(6)${\times}$SU(2) & SU(3)${\times}$SU(3)${\times}$SU(3)\\
\rowcolor{tableheadcolor}label & (name) & (index) & Triality & singlets & singlets & singlets\\
\midrule
\endfirsthead
\rowcolor{white}
\caption[]{$E_6$ Irreps (continued)}\\
\toprule
\rowcolor{tableheadcolor}
Dynkin & Dimension & l/6 &  & SO(10) & SU(6)${\times}$SU(2) & SU(3)${\times}$SU(3)${\times}$SU(3)\\
\rowcolor{tableheadcolor}label & (name) & (index) & Triality & singlets & singlets & singlets\\
\midrule
\endhead
\rowcolor{white}\multicolumn{7}{l}{\footnotesize $^\ast$SO(10)${\times}$U(1) singlets resp.}
\endfoot
\bottomrule
\rowcolor{white}\multicolumn{7}{l}{\footnotesize $^\ast$SO(10)${\times}$U(1) singlets resp.}
\endlastfoot
\dynkin{1, 0, 0, 0, 0, 0} & \irrep{27} & 1 & 1 & 1 & 0 & 0\\
\dynkin{0, 0, 0, 0, 0, 1} & \irrep{78} & 4 & 0 & \starred{1} & 0 & 0\\
\dynkin{0, 0, 0, 1, 0, 0} & \irrep{351} & 25 & 1 & 0 & 0 & 0\\
\dynkin{0, 0, 0, 0, 2, 0} & \irrep[1]{351} & 28 & 1 & 1 & 0 & 0\\
\dynkin{1, 0, 0, 0, 1, 0} & \irrep{650} & 50 & 0 & \starred{1} & 1 & 2\\
\dynkin{1, 0, 0, 0, 0, 1} & \irrep{1728} & 160 & 1 & 1 & 0 & 0\\
\dynkin{0, 0, 0, 0, 0, 2} & \irrep{2430} & 270 & 0 & \starred{1} & 1 & 1\\
\dynkin{0, 0, 1, 0, 0, 0} & \irrep{2925} & 300 & 0 & 0 & 0 & 1\\
\end{longtable}
\newpage

\rowcolors{2}{}{tablerowcolor}
\begin{longtable}{lrrc}
\caption{\label{tab:E7Irreps}\E7 Irreps}\\
\toprule
\rowcolor{tableheadcolor}
Dynkin & Dimension & l/12 & Congruency\\
\rowcolor{tableheadcolor}label & (name) & (index) & class\\
\midrule
\endfirsthead
\rowcolor{white}
\caption[]{$E_7$ Irreps (continued)}\\
\toprule
\rowcolor{tableheadcolor}
Dynkin & Dimension & l/12 & Congruency\\
\rowcolor{tableheadcolor}label & (name) & (index) & class\\
\midrule
\endhead
\endfoot
\bottomrule
\endlastfoot
\dynkin{0, 0, 0, 0, 0, 1, 0} & \irrep{56} & 1 & 1\\
\dynkin{1, 0, 0, 0, 0, 0, 0} & \irrep{133} & 3 & 0\\
\dynkin{0, 0, 0, 0, 0, 0, 1} & \irrep{912} & 30 & 1\\
\dynkin{0, 0, 0, 0, 0, 2, 0} & \irrep{1463} & 55 & 0\\
\dynkin{0, 0, 0, 0, 1, 0, 0} & \irrep{1539} & 54 & 0\\
\dynkin{1, 0, 0, 0, 0, 1, 0} & \irrep{6480} & 270 & 1\\
\dynkin{2, 0, 0, 0, 0, 0, 0} & \irrep{7371} & 351 & 0\\
\dynkin{0, 1, 0, 0, 0, 0, 0} & \irrep{8645} & 390 & 0\\
\end{longtable}
\vspace{-10pt}

\rowcolors{2}{}{tablerowcolor}
\begin{longtable}{lrrc}
\caption{\label{tab:E8Irreps}\E8 Irreps}\\
\toprule
\rowcolor{tableheadcolor}
Dynkin & Dimension & l/60 & Congruency\\
\rowcolor{tableheadcolor}label & (name) & (index) & class\\
\midrule
\endfirsthead
\rowcolor{white}
\caption[]{$E_8$ Irreps (continued)}\\
\toprule
\rowcolor{tableheadcolor}
Dynkin & Dimension & l/60 & Congruency\\
\rowcolor{tableheadcolor}label & (name) & (index) & class\\
\midrule
\endhead
\endfoot
\bottomrule
\endlastfoot
\dynkin{0, 0, 0, 0, 0, 0, 1, 0} & \irrep{248} & 1 & 0\\
\dynkin{1, 0, 0, 0, 0, 0, 0, 0} & \irrep{3875} & 25 & 0\\
\dynkin{0, 0, 0, 0, 0, 0, 2, 0} & \irrep{27000} & 225 & 0\\
\dynkin{0, 0, 0, 0, 0, 1, 0, 0} & \irrep{30380} & 245 & 0\\
\dynkin{0, 0, 0, 0, 0, 0, 0, 1} & \irrep{147250} & 1425 & 0\\
\dynkin{1, 0, 0, 0, 0, 0, 1, 0} & \irrep{779247} & 8379 & 0\\
\dynkin{0, 0, 0, 0, 0, 0, 3, 0} & \irrep{1763125} & 22750 & 0\\
\dynkin{0, 0, 0, 0, 1, 0, 0, 0} & \irrep{2450240} & 29640 & 0\\
\dynkin{0, 0, 0, 0, 0, 1, 1, 0} & \irrep{4096000} & 51200 & 0\\
\dynkin{2, 0, 0, 0, 0, 0, 0, 0} & \irrep{4881384} & 65610 & 0\\
\dynkin{0, 1, 0, 0, 0, 0, 0, 0} & \irrep{6696000} & 88200 & 0\\
\dynkin{0, 0, 0, 0, 0, 0, 1, 1} & \irrep{26411008} & 372736 & 0\\
\dynkin{1, 0, 0, 0, 0, 0, 2, 0} & \irrep{70680000} & 1083000 & 0\\
\dynkin{1, 0, 0, 0, 0, 1, 0, 0} & \irrep{76271625} & 1148175 & 0\\
\end{longtable}
\newpage

\rowcolors{2}{}{tablerowcolor}
\begin{longtable}{lrrc}
\caption{\label{tab:F4Irreps}\F4 Irreps}\\
\toprule
\rowcolor{tableheadcolor}
Dynkin & Dimension & l/6 & Congruency\\
\rowcolor{tableheadcolor}label & (name) & (index) & class\\
\midrule
\endfirsthead
\rowcolor{white}
\caption[]{$F_4$ Irreps (continued)}\\
\toprule
\rowcolor{tableheadcolor}
Dynkin & Dimension & l/6 & Congruency\\
\rowcolor{tableheadcolor}label & (name) & (index) & class\\
\midrule
\endhead
\endfoot
\bottomrule
\endlastfoot
\dynkin{0, 0, 0, 1} & \irrep{26} & 1 & 0\\
\dynkin{1, 0, 0, 0} & \irrep{52} & 3 & 0\\
\dynkin{0, 0, 1, 0} & \irrep{273} & 21 & 0\\
\dynkin{0, 0, 0, 2} & \irrep{324} & 27 & 0\\
\dynkin{1, 0, 0, 1} & \irrep{1053} & 108 & 0\\
\dynkin{2, 0, 0, 0} & \irrep[1]{1053} & 135 & 0\\
\dynkin{0, 1, 0, 0} & \irrep{1274} & 147 & 0\\
\dynkin{0, 0, 0, 3} & \irrep{2652} & 357 & 0\\
\end{longtable}
\newpage

\rowcolors{2}{}{tablerowcolor}
\begin{longtable}{lrrc}
\caption{\label{tab:G2Irreps}\G2 Irreps}\\
\toprule
\rowcolor{tableheadcolor}
Dynkin & Dimension & l/2 & Congruency\\
\rowcolor{tableheadcolor}label & (name) & (index) & class\\
\midrule
\endfirsthead
\rowcolor{white}
\caption[]{$G_2$ Irreps (continued)}\\
\toprule
\rowcolor{tableheadcolor}
Dynkin & Dimension & l/2 & Congruency\\
\rowcolor{tableheadcolor}label & (name) & (index) & class\\
\midrule
\endhead
\endfoot
\bottomrule
\endlastfoot
\dynkin{1, 0} & \irrep{7} & 1 & 0\\
\dynkin{0, 1} & \irrep{14} & 4 & 0\\
\dynkin{2, 0} & \irrep{27} & 9 & 0\\
\dynkin{1, 1} & \irrep{64} & 32 & 0\\
\dynkin{3, 0} & \irrep{77} & 44 & 0\\
\dynkin{0, 2} & \irrep[1]{77} & 55 & 0\\
\dynkin{4, 0} & \irrep{182} & 156 & 0\\
\dynkin{2, 1} & \irrep{189} & 144 & 0\\
\end{longtable}
\newpage

}
\newpage

\subsection{Tensor Products}
\label{ssec:TensorProducts}

\setlength\extrarowheight{1.3pt}
\renewcommand{\tabcolsep}{2pt}
\subsubsection{\SU{N}}

\vspace{-10pt}
\enlargethispage{10pt}
\rowcolors{2}{tablerowcolor}{}\begin{longtable}{rcl}
\rowcolor{white}\caption{\label{tab:SU2TensorProducts}SU(2) Tensor Products}\\
\toprule
\endfirsthead
\rowcolor{white}\caption[]{SU(2) Tensor Products (continued)}\\
\endhead
\bottomrule
\endlastfoot
$\irrep{2}\times\irrep{2}$ & = & $\irrep{1}+\irrep{3}$\\
$\irrep{3}\times\irrep{2}$ & = & $\irrep{2}+\irrep{4}$\\
$\irrep{3}\times\irrep{3}$ & = & $\irrep{1}+\irrep{3}+\irrep{5}$\\
$\irrep{4}\times\irrep{2}$ & = & $\irrep{3}+\irrep{5}$\\
$\irrep{4}\times\irrep{3}$ & = & $\irrep{2}+\irrep{4}+\irrep{6}$\\
$\irrep{4}\times\irrep{4}$ & = & $\irrep{1}+\irrep{3}+\irrep{5}+\irrep{7}$\\
$\irrep{5}\times\irrep{2}$ & = & $\irrep{4}+\irrep{6}$\\
$\irrep{5}\times\irrep{3}$ & = & $\irrep{3}+\irrep{5}+\irrep{7}$\\
\end{longtable}
\newpage

\rowcolors{2}{tablerowcolor}{}\begin{longtable}{rcl}
\rowcolor{white}\caption{\label{tab:SU3TensorProducts}SU(3) Tensor Products}\\
\toprule
\endfirsthead
\rowcolor{white}\caption[]{SU(3) Tensor Products (continued)}\\
\endhead
\bottomrule
\endlastfoot
$\irrepbar{3}\times\irrep{3}$ & = & $\irrep{1}+\irrep{8}$\\
$\irrep{3}\times\irrep{3}$ & = & $\irrepbar{3}+\irrep{6}$\\
$\irrepbar{6}\times\irrep{3}$ & = & $\irrepbar{3}+\irrepbar{15}$\\
$\irrep{6}\times\irrep{3}$ & = & $\irrep{8}+\irrep{10}$\\
$\irrepbar{6}\times\irrepbar{6}$ & = & $\irrep{6}+\irrepbar{15}+\irrepbar[1]{15}$\\
$\irrep{6}\times\irrepbar{6}$ & = & $\irrep{1}+\irrep{8}+\irrep{27}$\\
$\irrep{8}\times\irrep{3}$ & = & $\irrep{3}+\irrepbar{6}+\irrep{15}$\\
$\irrep{8}\times\irrepbar{6}$ & = & $\irrep{3}+\irrepbar{6}+\irrep{15}+\irrep{24}$\\
\end{longtable}
\newpage

\enlargethispage{10pt}
\rowcolors{2}{tablerowcolor}{}\begin{longtable}{rcl}
\rowcolor{white}\caption{\label{tab:SU4TensorProducts}SU(4) Tensor Products}\\
\toprule
\endfirsthead
\rowcolor{white}\caption[]{SU(4) Tensor Products (continued)}\\
\endhead
\bottomrule
\endlastfoot
$\irrepbar{4}\times\irrep{4}$ & = & $\irrep{1}+\irrep{15}$\\
$\irrep{4}\times\irrep{4}$ & = & $\irrep{6}+\irrep{10}$\\
$\irrep{6}\times\irrep{4}$ & = & $\irrepbar{4}+\irrepbar{20}$\\
$\irrep{6}\times\irrep{6}$ & = & $\irrep{1}+\irrep{15}+\irrep[1]{20}$\\
$\irrepbar{10}\times\irrep{4}$ & = & $\irrepbar{4}+\irrepbar{36}$\\
$\irrep{10}\times\irrep{4}$ & = & $\irrepbar{20}+\irrepbar[2]{20}$\\
$\irrep{10}\times\irrep{6}$ & = & $\irrep{15}+\irrep{45}$\\
$\irrepbar{10}\times\irrep{10}$ & = & $\irrep{1}+\irrep{15}+\irrep{84}$\\
\end{longtable}
\newpage

\rowcolors{2}{tablerowcolor}{}\begin{longtable}{rcl}
\rowcolor{white}\caption{\label{tab:SU5TensorProducts}SU(5) Tensor Products}\\
\toprule
\endfirsthead
\rowcolor{white}\caption[]{SU(5) Tensor Products (continued)}\\
\endhead
\bottomrule
\endlastfoot
$\irrepbar{5}\times\irrep{5}$ & = & $\irrep{1}+\irrep{24}$\\
$\irrep{5}\times\irrep{5}$ & = & $\irrep{10}+\irrep{15}$\\
$\irrepbar{10}\times\irrep{5}$ & = & $\irrepbar{5}+\irrepbar{45}$\\
$\irrep{10}\times\irrep{5}$ & = & $\irrepbar{10}+\irrepbar{40}$\\
$\irrepbar{10}\times\irrep{10}$ & = & $\irrep{1}+\irrep{24}+\irrep{75}$\\
$\irrep{10}\times\irrep{10}$ & = & $\irrepbar{5}+\irrepbar{45}+\irrepbar{50}$\\
$\irrepbar{15}\times\irrep{5}$ & = & $\irrepbar{5}+\irrepbar{70}$\\
$\irrep{15}\times\irrep{5}$ & = & $\irrepbar{35}+\irrepbar{40}$\\
\end{longtable}
\newpage

\rowcolors{2}{tablerowcolor}{}\begin{longtable}{rcl}
\rowcolor{white}\caption{\label{tab:SU6TensorProducts}SU(6) Tensor Products}\\
\toprule
\endfirsthead
\rowcolor{white}\caption[]{SU(6) Tensor Products (continued)}\\
\endhead
\bottomrule
\endlastfoot
$\irrepbar{6}\times\irrep{6}$ & = & $\irrep{1}+\irrep{35}$\\
$\irrep{6}\times\irrep{6}$ & = & $\irrep{15}+\irrep{21}$\\
$\irrepbar{15}\times\irrep{6}$ & = & $\irrepbar{6}+\irrepbar{84}$\\
$\irrep{15}\times\irrep{6}$ & = & $\irrep{20}+\irrep{70}$\\
$\irrepbar{15}\times\irrep{15}$ & = & $\irrep{1}+\irrep{35}+\irrep{189}$\\
$\irrep{15}\times\irrep{15}$ & = & $\irrepbar{15}+\irrepbar{105}+\irrepbar[1]{105}$\\
$\irrep{20}\times\irrep{6}$ & = & $\irrepbar{15}+\irrepbar{105}$\\
$\irrep{20}\times\irrep{15}$ & = & $\irrepbar{6}+\irrepbar{84}+\irrepbar{210}$\\
\end{longtable}
\newpage

\rowcolors{2}{tablerowcolor}{}\begin{longtable}{rcl}
\rowcolor{white}\caption{\label{tab:SU7TensorProducts}SU(7) Tensor Products}\\
\toprule
\endfirsthead
\rowcolor{white}\caption[]{SU(7) Tensor Products (continued)}\\
\endhead
\bottomrule
\endlastfoot
$\irrepbar{7}\times\irrep{7}$ & = & $\irrep{1}+\irrep{48}$\\
$\irrep{7}\times\irrep{7}$ & = & $\irrep{21}+\irrep{28}$\\
$\irrepbar{21}\times\irrep{7}$ & = & $\irrepbar{7}+\irrepbar{140}$\\
$\irrep{21}\times\irrep{7}$ & = & $\irrep{35}+\irrep{112}$\\
$\irrepbar{21}\times\irrep{21}$ & = & $\irrep{1}+\irrep{48}+\irrep{392}$\\
$\irrep{21}\times\irrep{21}$ & = & $\irrepbar{35}+\irrepbar{196}+\irrepbar{210}$\\
$\irrepbar{28}\times\irrep{7}$ & = & $\irrepbar{7}+\irrepbar{189}$\\
$\irrep{28}\times\irrep{7}$ & = & $\irrep{84}+\irrep{112}$\\
\end{longtable}
\newpage

\rowcolors{2}{tablerowcolor}{}\begin{longtable}{rcl}
\rowcolor{white}\caption{\label{tab:SU8TensorProducts}SU(8) Tensor Products}\\
\toprule
\endfirsthead
\rowcolor{white}\caption[]{SU(8) Tensor Products (continued)}\\
\endhead
\bottomrule
\endlastfoot
$\irrepbar{8}\times\irrep{8}$ & = & $\irrep{1}+\irrep{63}$\\
$\irrep{8}\times\irrep{8}$ & = & $\irrep{28}+\irrep{36}$\\
$\irrepbar{28}\times\irrep{8}$ & = & $\irrepbar{8}+\irrepbar{216}$\\
$\irrep{28}\times\irrep{8}$ & = & $\irrep{56}+\irrep{168}$\\
$\irrepbar{28}\times\irrep{28}$ & = & $\irrep{1}+\irrep{63}+\irrep{720}$\\
$\irrep{28}\times\irrep{28}$ & = & $\irrep{70}+\irrep{336}+\irrep{378}$\\
$\irrepbar{36}\times\irrep{8}$ & = & $\irrepbar{8}+\irrepbar{280}$\\
$\irrep{36}\times\irrep{8}$ & = & $\irrep{120}+\irrep{168}$\\
\end{longtable}%
\newpage

\subsubsection{\SO{N}}

\rowcolors{2}{tablerowcolor}{}%
\newpage

\subsubsection{\Sp{N}}
\enlargethispage{10pt}
\rowcolors{2}{tablerowcolor}{}%
\newpage

\subsubsection{Exceptional Algebras}

\rowcolors{2}{tablerowcolor}{}%
}

\newpage

\subsection{Branching Rules}
\label{ssec:BranchingRules}

\subsubsection{\SU{N}}

\rowcolors{2}{tablerowcolor}{}\rowcolors{2}{tablerowcolor}{}%
\newpage

\subsubsection{\SO{N}}

\rowcolors{2}{tablerowcolor}{}\begin{longtable}{rcp{0.9\textwidth}}
\rowcolor{white}\caption{\label{tab:SO7BranchingRules}SO(7) Branching Rules}\\
\endfirsthead
\rowcolor{white}\caption[]{SO(7) Branching Rules (continued)}\\
\endhead
\endfoot
\toprule
\rowcolor{tableheadcolor}
SO(7)& $\to$ &SU(4)\\
\midrule
\irrep{7} & = & $\irrep{1}+\irrep{6}$\\
\irrep{8} & = & $\irrep{4}+\irrepbar{4}$\\
\irrep{21} & = & $\irrep{6}+\irrep{15}$\\
\irrep{27} & = & $\irrep{1}+\irrep{6}+\irrep[1]{20}$\\
\irrep{35} & = & $\irrep{10}+\irrepbar{10}+\irrep{15}$\\
\irrep{48} & = & $\irrep{4}+\irrepbar{4}+\irrep{20}+\irrepbar{20}$\\
\irrep{77} & = & $\irrep{1}+\irrep{6}+\irrep[1]{20}+\irrep{50}$\\
\bottomrule
\end{longtable}%
\newpage

\subsubsection{Exceptional Algebras}

{\setlength\extrarowheight{1.1pt}
\enlargethispage{15pt}
\rowcolors{2}{tablerowcolor}{}%

\bibliographystyle{elsarticle-num}
\bibliography{LieART}

\begin{thebibliography}{10}
\expandafter\ifx\csname url\endcsname\relax
  \def\url#1{\texttt{#1}}\fi
\expandafter\ifx\csname urlprefix\endcsname\relax\def\urlprefix{URL }\fi
\expandafter\ifx\csname href\endcsname\relax
  \def\href#1#2{#2} \def\path#1{#1}\fi

\bibitem{Dynkin:1957um}
E.~Dynkin, {Semisimple subalgebras of semisimple Lie algebras},
  Trans.Am.Math.Soc. 6 (1957) 111.

\bibitem{Dynkin:1957dm}
E.~Dynkin, {Maximal subgroups of the classical groups}, Trans.Am.Math.Soc. 6
  (1957) 245.

\bibitem{Georgi:1974sy}
H.~Georgi, S.~Glashow, {Unity of All Elementary Particle Forces},
  Phys.Rev.Lett. 32 (1974) 438--441.
\newblock \href {http://dx.doi.org/10.1103/PhysRevLett.32.438}
  {\path{doi:10.1103/PhysRevLett.32.438}}.

\bibitem{Georgi:1974xy}
{H. Georgi. Particles And Fields: Williamsburg 1974. AIP Conference Proceedings
  No. 23 - C. E. Carlson (eds.)}.

\bibitem{Fritzsch:1974nn}
H.~Fritzsch, P.~Minkowski, {Unified Interactions of Leptons and Hadrons},
  Annals Phys. 93 (1975) 193--266.
\newblock \href {http://dx.doi.org/10.1016/0003-4916(75)90211-0}
  {\path{doi:10.1016/0003-4916(75)90211-0}}.

\bibitem{Gursey:1975ki}
F.~Gursey, P.~Ramond, P.~Sikivie, {A Universal Gauge Theory Model Based on E6},
  Phys.Lett. B60 (1976) 177.
\newblock \href {http://dx.doi.org/10.1016/0370-2693(76)90417-2}
  {\path{doi:10.1016/0370-2693(76)90417-2}}.

\bibitem{Slansky}
R.~Slansky, {Group Theory for Unified Model Building}, Phys.Rept. 79 (1981)
  1--128.
\newblock \href {http://dx.doi.org/10.1016/0370-1573(81)90092-2}
  {\path{doi:10.1016/0370-1573(81)90092-2}}.

\bibitem{McKay:99021}
W.~G. McKay, J.~Patera, {Tables of dimensions, indices, and branching rules for
  representations of simple Lie algebras}, Lecture Notes in Pure and Applied
  Mathematics, Dekker, New York, NY, 1981.

\bibitem{Georgi:1982jb}
H.~Georgi, {Lie Algebras in Particle Physics. from Isospin to Unified
  Theories}, Front.Phys. 54 (1982) 1--255.

\bibitem{Ramond:2010zz}
P.~Ramond, {Group theory: A physicist's survey}, Cambridge University Press,
  Cambridge, UK, 2010.

\bibitem{cahn1984semi}
R.~Cahn, {Semi-simple lie algebras and their representations}, Vol.~59 of
  Front.Phys., Benjamin Cummings, Menlo Park, CA, 1984.

\bibitem{LiE}
M.~van Leeuwen, A.~Cohen, B.~Lisser,
  \href{{http://young.sp2mi.univ-poitiers.fr/~marc/LiE/}}{{LiE, A Package for
  Lie Group Computations}}, Computer Algebra Nederland, Amsterdam, 1992.
\newline\urlprefix\url{{http://young.sp2mi.univ-poitiers.fr/~marc/LiE/}}

\bibitem{Schur}
B.~G. Wybourne, \href{{http://smc.vnet.net/Schur.html}}{Schur} (2002).
\newline\urlprefix\url{{http://smc.vnet.net/Schur.html}}

\bibitem{Simplie}
T.~Nutma, \href{{http://code.google.com/p/simplie/}}{Simplie} (2009).
\newline\urlprefix\url{{http://code.google.com/p/simplie/}}

\bibitem{Nazarov:2011mv}
A.~Nazarov, {Affine.m - Mathematica package for computations in representation
  theory of finite-dimensional and affine Lie algebras}\href
  {http://arxiv.org/abs/1107.4681} {\path{arXiv:1107.4681}}.

\bibitem{Fonseca:2011sy}
R.~M. Fonseca, Calculating the renormalisation group equations of a susy model
  with susyno, Computer Physics Communications 183~(10) (2012) 2298 -- 2306.
\newblock \href {http://arxiv.org/abs/1106.5016} {\path{arXiv:1106.5016}},
  \href {http://dx.doi.org/10.1016/j.cpc.2012.05.017}
  {\path{doi:10.1016/j.cpc.2012.05.017}}.

\bibitem{Albright:2012zt}
C.~H. Albright, R.~P. Feger, T.~W. Kephart, {An explicit SU(12) family and
  flavor unification model with natural fermion masses and mixings}, Phys.Rev.
  D86 (2012) 015012.
\newblock \href {http://arxiv.org/abs/1204.5471} {\path{arXiv:1204.5471}},
  \href {http://dx.doi.org/10.1103/PhysRevD.86.015012}
  {\path{doi:10.1103/PhysRevD.86.015012}}.

\bibitem{klimyk_orbit_2006}
A.~Klimyk, J.~Patera, {Orbit Functions}, SIGMA 2 (2006) 6--66.
\newblock \href {http://arxiv.org/abs/math-ph/0601037}
  {\path{arXiv:math-ph/0601037}}, \href
  {http://dx.doi.org/10.3842/SIGMA.2006.006}
  {\path{doi:10.3842/SIGMA.2006.006}}.

\bibitem{Moody:1982}
R.~V. Moody, J.~Patera, {Fast Recursion Formula for Weight Multiplicities},
  Bull.Amer.Math.Soc.(N.S.) 7~(1) (1982) 237--242.

\bibitem{lemire_congruence_1980}
F.~Lemire, J.~Patera, {Congruence number, a generalization of {SU(3)}
  triality}, J.Math.Phys. 21~(8) (1980) 2026.
\newblock \href {http://dx.doi.org/10.1063/1.524711}
  {\path{doi:10.1063/1.524711}}.

\bibitem{klimyk_translation}
A.~U. Klimyk, {Decomposition of the direct product of irreducible
  representations of a semisimple Lie algebra into irreducible
  representations}, American Mathematical Society Translations: Series 2 76
  (1967) 63.

\bibitem{Humphreys:1980dw}
J.~Humphreys, {Introduction to Lie Algebras and Representation Theory},
  Springer, New York, 1972.

\bibitem{larouche_branching_2009}
M.~Larouche, M.~Nesterenko, J.~Patera, {Branching rules for the Weyl group
  orbits of the Lie algebra A(n)}, J. Phys. A: Math. Theor. 42~(48) (2009)
  485203.
\newblock \href {http://arxiv.org/abs/0909.2337} {\path{arXiv:0909.2337}},
  \href {http://dx.doi.org/10.1088/1751-8113/42/48/485203}
  {\path{doi:10.1088/1751-8113/42/48/485203}}.

\bibitem{larouche_branching_2011}
M.~Larouche, J.~Patera, {Branching rules for Weyl group orbits of simple Lie
  algebras B(n), C(n) and D(n)}, J. Phys. A: Math. Theor. 44~(11) (2011)
  115203.
\newblock \href {http://arxiv.org/abs/1101.6043} {\path{arXiv:1101.6043}},
  \href {http://dx.doi.org/10.1088/1751-8113/44/11/115203}
  {\path{doi:10.1088/1751-8113/44/11/115203}}.

\end{thebibliography}

\end{document}